\documentclass[12pt]{article}

\usepackage{amsthm,amsmath,amsfonts,amssymb}
\usepackage[authoryear]{natbib}
\usepackage{hyperref}
\hypersetup{
	colorlinks,
	citecolor=blue,
	urlcolor=blue,
	linkcolor=blue
}
\usepackage[margin=1in]{geometry}
\usepackage{graphicx}
\usepackage{xcolor}         
\usepackage{tikz}
\usepackage{tikz-layers}
\usetikzlibrary{calc, shapes, positioning}
\usepackage{array}
\usepackage{type1cm,type1ec}
\RequirePackage{mathtools}
\usepackage{float}
\usepackage[section]{placeins}
\usepackage[ruled, vlined]{algorithm2e}
\usepackage{enumitem}
\usepackage{xr}
\theoremstyle{plain}

\newtheorem{lemma}{Lemma}

\newtheorem{proposition}{Proposition}
\theoremstyle{remark}

\newtheorem{remark}{Remark}

\title{Simultaneous clustering and estimation of additive shape invariant models for recurrent event data}

\author{Zitong Zhang\footnote{zztzhang@ucdavis.edu, Department of Statistics, University of California, Davis} and Shizhe Chen\footnote{szdchen@ucdavis.edu, Department of Statistics, University of California, Davis}}
\date{} 

\begin{document}

\maketitle


\begin{abstract}
Technological advancements have enabled the recording of spiking activities from large neuron ensembles, presenting  an exciting yet  challenging opportunity for statistical analysis. This project considers the challenges from a common type of neuroscience experiments, where randomized interventions are applied over the course of each trial. The objective is to identify groups of neurons with unique stimulation responses and estimate these responses. The observed data, however, comprise superpositions of neural responses to all stimuli, which is further complicated by varying response latencies across neurons. We introduce a novel additive shape invariant model that is capable of simultaneously accommodating multiple clusters, additive components, and unknown time-shifts. We establish conditions for the identifiability of model parameters, offering guidance for the design of future experiments. We examine the properties of the proposed algorithm through simulation studies, and apply the proposed method on neural data collected in mice.

\end{abstract}




\section{Introduction}
\label{sec:introduction}

\begin{figure}[ht]
\centering
\begin{tikzpicture}
    \node[below right,inner sep=0] (image_11) at (0,0) 
        {\includegraphics[width=0.23\textwidth, trim=0 0 0 0, clip]
            {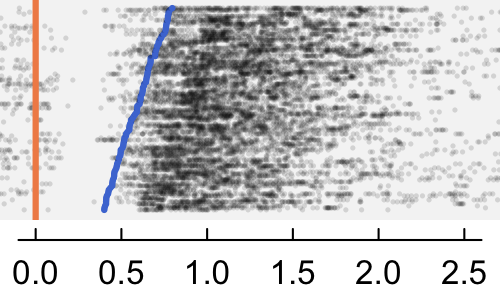}
        }; 
    \begin{scope}[
        shift={($(image_11.south west)$)},
        x={($0.1*(image_11.south east)$)},
        y={($0.1*(image_11.north west)$)}]
        \node[above, rotate=90] at ($(0,6.7)$) { Trials};
        \node[above] at ($(1.8,10)$) { Neuron 1};
    \end{scope}  
    \node[below right = 0 and 0.2 of image_11.north east,inner sep=0] (image_21) at (image_11.north east) 
        {\includegraphics[width=0.23\textwidth, trim=0 0 0 0, clip]
            {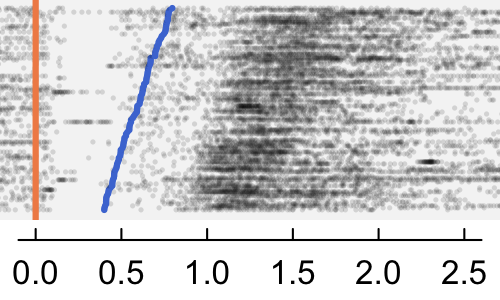}
        }; 
    \begin{scope}[
        shift={($(image_21.south west)$)},
        x={($0.1*(image_21.south east)$)},
        y={($0.1*(image_21.north west)$)}]
        \node[above] at ($(1.8,10)$) { Neuron 2};
    \end{scope} 
    \node[below right = 0 and 0.2 of image_21.north east,inner sep=0] (image_22) at (image_21.north east) 
        {\includegraphics[width=0.23\textwidth, trim=0 0 0 0, clip]
            {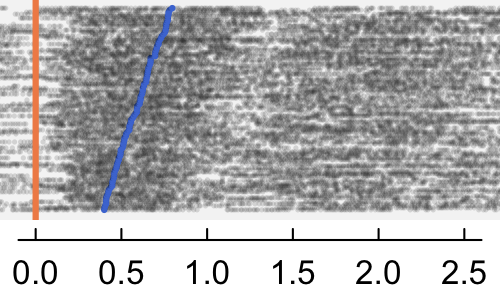}
        };
    \begin{scope}[
        shift={($(image_22.south west)$)},
        x={($0.1*(image_22.south east)$)},
        y={($0.1*(image_22.north west)$)}]
        \node[above] at ($(1.8,10)$) { Neuron 3};
    \end{scope}  
    \node[below right = -0.55 and 0.2 of image_22.north east,inner sep=0] (image_legend) at (image_22.north east) 
        {\includegraphics[width=0.23\textwidth, trim=0 0 0 0, clip]
            {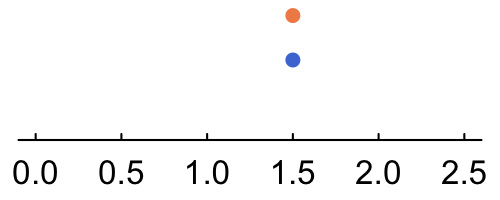}
        }; 
    \begin{scope}[
        shift={($(image_legend.south west)$)},
        x={($0.1*(image_legend.south east)$)},
        y={($0.1*(image_legend.north west)$)}]
        \node[right] at ($(5.8,9.3)$) {\scriptsize Visual};
        \node[right] at ($(5.8,7.1)$) {\scriptsize Auditory};
    \end{scope}  
    \node[below right = 0 and 0.2 of image_22.north east,inner sep=0] (image_23) at (image_22.north east) 
        {\includegraphics[width=0.23\textwidth, trim=0 0 0 0, clip]
            {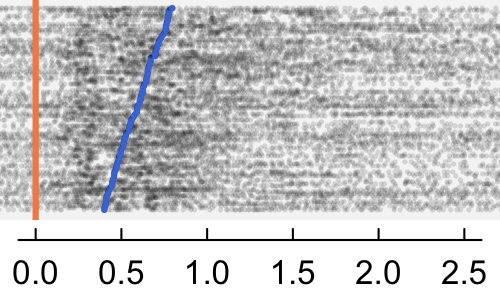}
        };
    \begin{scope}[
        shift={($(image_23.south west)$)},
        x={($0.1*(image_23.south east)$)},
        y={($0.1*(image_23.north west)$)}]
        \node[above] at ($(1.8,10)$) { Neuron 4};
    \end{scope} 
    \begin{scope}[
        shift={($(image_11.south west)$)},
        x={($0.1*(image_23.south east)$)},
        y={($0.1*(image_11.north west)$)}]
        \node[below] at ($(5.0,-0.3)$) { Time from visual stimulus onset (s)};
    \end{scope}  
\end{tikzpicture}
\caption{ 
Activities of four example neurons from  \cite{Steinmetz2019}. 
The four neurons are all from the midbrain region of the mouse brain, where their firings are shown as black dots.
Each panel corresponds to a single neuron, where the x-axis represents time since visual stimulus onset, and the y-axis represents experiment trials.
Trials are aligned by the visual stimulus onset time shown as orange dots , and ordered by the auditory stimulus onset time shown as blue dots.  
}
\label{fig:real_data_motivation}
\end{figure}
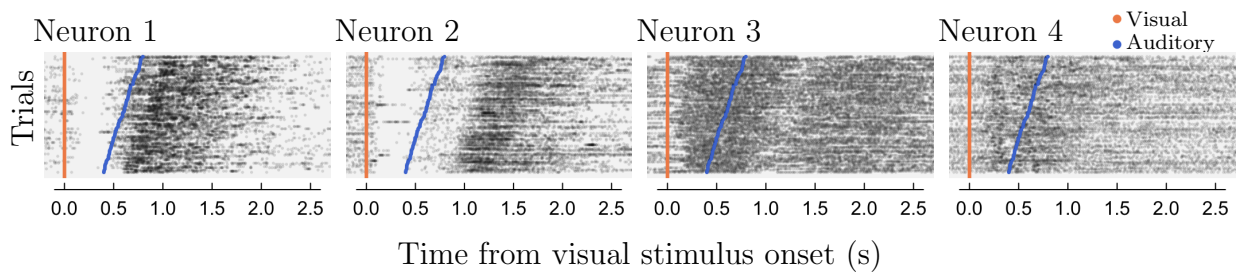

Recent technological advancements have greatly improved our ability to record neural activities with high spatiotemporal resolution and over large volumes. 
Recording device such as Neuropixels~\citep{jun2017fully,steinmetz2018challenges,steinmetz2021neuropixels} enable scientists to record neurons across multiple brain areas while  subjects engage in behavioral tasks.
The growing amount of neural data facilitates the exploration of how neural firing patterns encode information, but also presents challenges to existing statistical methods. 
In this paper, we consider the problem of identifying subgroups of neurons that share similar responses to stimuli. 
In a typical experiment, a subject will be exposed to a series of stimuli (e.g., visual cue, auditory cue, reward) over the course of a trial, which means that the recorded spike train is a superposition of neural responses to these stimuli~\citep{benucci2009coding,capilla2011steady,orhan2015neural}. 
It is challenging to disentangle the response to each stimulus given the stochastic nature of neural activities. 
Moreover, even if two neurons share the same responses to a stimulus, they might respond in different time due to the difference in their response latency~\citep{oram2002temporal,levakova2015review,lee2020neural}, making the estimation of shared neural responses more challenging. 
Lastly, even within the same area of the brain, we cannot assume that neurons share the same responses~\citep{molyneaux2007neuronal,lake2016neuronal,cembrowski2019heterogeneity}.
In light of these intricacies, the primary objective of this study is to develop statistical methods for recurrent events to simultaneously address three key tasks: (i) \emph{decompose} the neural firing activities into their constituent components, (ii) \emph{align} neural firing patterns across  neurons, and (iii) \emph{cluster} neurons based on their similar firing patterns.
As a concrete example, we consider the representative experiment by \cite{Steinmetz2019}, where mice were exposed to a series of visual and auditory stimuli, and thousands of neurons were recorded throughout the experiment.
We display four neurons  selected by our method in Figure~\ref{fig:real_data_motivation} that displayed distinct firing patterns to the stimuli. 

However, existing methods are inadequate in addressing these challenges at hand.
Shape invariant models~\citep{beath2007infant,Bigot2010,vimond2010efficient,Bigot2013,Bigot2013a,Bontemps2014} have been studied extensively to align a collection of curves with identical means up to unknown time shifts.
The combination of shape invariant models and mixture models has led to the development of techniques for simultaneously aligning and clustering curves~\citep{Chudova2003,Gaffney2004,Liu2009,Lu2019}.
In a related vein,~\citet{Sangalli2010a} propose an approach to simultaneously align and cluster curves that share a common shape but are subject to non-linear distortions in temporal alignment. 
Nonetheless, all of these methods are limited in their capacity to decompose curves comprising multiple components. 

A class of problems related to the decomposing superimposed curves has been studied in the context of functional principal component analysis~\citep{Yao2005,morris2006wavelet,di2009multilevel,crainiceanu2009generalized}. 
Functional principal component analysis aims to deconstruct functional data into orthogonal components that effectively account for the variation within the dataset. 
Researchers have applied the functional principal component analysis to analyze point processes through the modeling of intensity functions~\citep{wu2013functional,xu2020semi}.
Furthermore, the functional principal component analysis models have been combined with mixture models to simultaneously cluster and decompose curves~\citep[][]{Chiou2007,bouveyron2011model,jacques2013funclust, yin2021row}.
However, the orthogonal components derived from these methods may not match the components of our primary interest. Indeed, the congruence between the outcomes of these methods and our proposed approach is limited to specific scenarios (see Section~\ref{sec:connection_with_prior_studies}).
Besides, unknown time shifts have not been studied in functional principal component analysis. 

In this paper, we propose an additive shape invariant {mixture} model capable of simultaneously decompose, align, and cluster  recurrent event data, such as neural firings.
The rest of this paper is organized as follows.
In Section~\ref{sec:additive_sim}, we introduce an additive shape invariant model for simultaneous decomposition and alignment. 
Section~\ref{sec:simultaneous_clustering} expands upon the additive shape invariant model by integrating simultaneous clustering. 
In Section~\ref{sec:algorithm}, we present a computationally efficient algorithm.
The performance of the proposed method is assessed through simulation experiments in Section~\ref{sec:simulation}. 
In Section~\ref{sec:real_data},
we use the proposed method to study neural firing patterns in mice.
Finally, we discuss potential future research directions in Section~\ref{sec:discussion}.


\section{Additive shape invariant model}
\label{sec:additive_sim}
\subsection{Model}

Consider a set of \emph{recurrent events} from repeated measurements of $n$ subjects $\mathcal{O} \equiv \big\{ \left\{t_{i,r,j} \right\}_{j = 1, \ldots, N_{i,r}(T)}: 0 < t_{i,r,1} < \cdots < t_{i,r,N_{i,r}(T)} < T, i = 1, \ldots,n, r = 1,\ldots,R \big\},$ where $t_{i,r,j}$ denotes the time of the $j$-th event of the $i$-th subject in the $r$-th observation, $T$ denotes the duration of each observation, $N_{i,r}(T)$ denotes the total number of events associated with the $i$-th subject and the $r$-th observation. 
For $i\in[n]$ and $r\in[R]$, we adopt the definition of \emph{counting process} \cite[see, e.g.][]{Daley2003} as $N_{i,r}(t) \equiv \sum_{j = 1}^{N_{i,r}(T)} \textbf{1}(t_{i,r,j} \leq t)$ for $t \in [0,T]$.
We use the \emph{intensity}  function, i.e.,  $\lambda_{i,r}(t) \equiv \mathbb{E}\{\mathrm{d}N_{i,r}(t)/\mathrm{d}t\}$, to characterize each counting process. 

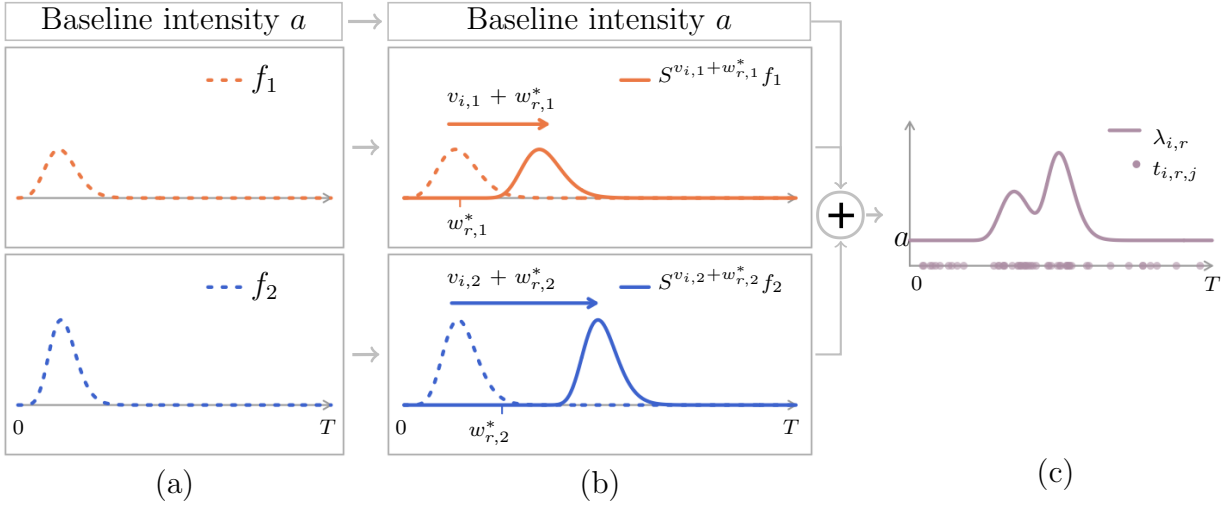
\begin{figure}
\centering
\begin{tikzpicture}
    \node[below right,inner sep=0] (image_21) at (0,0) 
    {\includegraphics[width=0.277\textwidth, trim=0 0 0 0, clip]
        {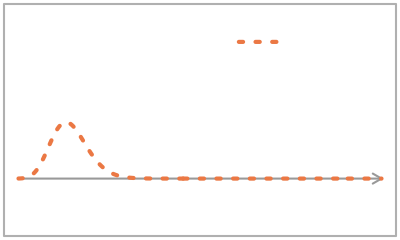}
    };  
    \begin{scope}[
      shift={(image_21.south west)},
      x={($0.1*(image_21.south east)$)},
      y={($0.1*(image_21.north west)$)}]
      \node[right] at ($(6.9,8.3)$) { $f_1$};
    \end{scope}   
    \node[below right,inner sep=0] (image_31) at (image_21.south west) 
    {\includegraphics[width=0.277\textwidth, trim=0 0 0 0, clip]
        {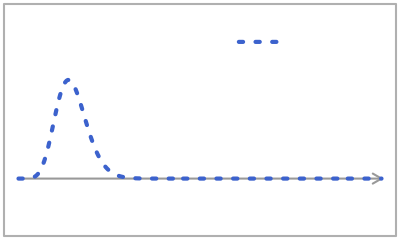}
    };
    \begin{scope}[
      shift={(image_31.south west)},
      x={($0.1*(image_31.south east)$)},
      y={($0.1*(image_31.north west)$)}]
      \node[right] at ($(6.9,8.3)$) { $f_2$};
      \node[right] at ($(0.0,1.5)$) {\scriptsize $0$};
      \node[left] at ($(10.0,1.5)$) {\scriptsize $T$};
      \node[below] at ($(5,0)$) { (a)};
    \end{scope}   
    \node[below right = 0 and 0.5 of image_21.north east,inner sep=0] (image_22) at (image_21.north east) 
    {\includegraphics[width=0.346\textwidth, trim=0 0 0 0, clip]
        {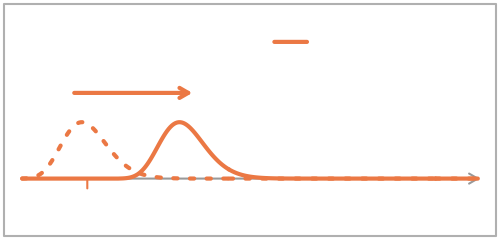}
    };    
    \begin{scope}[
      shift={(image_22.south west)},
      x={($0.1*(image_22.south east)$)},
      y={($0.1*(image_22.north west)$)}]
      \node[right] at ($(6.1,8.6)$) {\scriptsize  $S^{v_{i,1}+w^*_{r,1}}f_1$};
      \node[right] at ($(1.2,7.3)$) {\scriptsize $v_{i,1}$ + $w^*_{r,1}$};
      \node[right] at ($(1.2,1.2)$) {\scriptsize $w^*_{r,1}$};
    \end{scope}
    \node[draw = lightgray, rectangle, minimum width = 0.271\textwidth, minimum height = 0.5cm, semithick, above right,inner sep=0, xshift=1pt, yshift=1pt] (image_11) at (image_21.north west) 
    { Baseline intensity $a$};   
    \node[draw = lightgray, rectangle, minimum width = 0.34\textwidth, minimum height = 0.5cm, semithick, above right,inner sep=0, xshift=1pt, yshift=1pt] (image_12) at (image_22.north west) 
    { Baseline intensity $a$};
    \node[below right,inner sep=0] (image_32) at (image_22.south west) 
    {\includegraphics[width=0.346\textwidth, trim=0 0 0 0, clip]
        {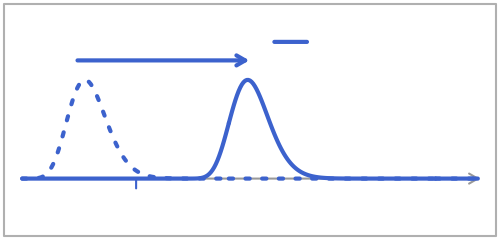}
    };    
    \begin{scope}[
      shift={(image_32.south west)},
      x={($0.1*(image_32.south east)$)},
      y={($0.1*(image_32.north west)$)}]
      \node[right] at ($(6.1,8.6)$) {\scriptsize  $S^{v_{i,2}+w^*_{r,2}}f_2$};
      \node[right] at ($(1.2,8.6)$) {\scriptsize $v_{i,2}$ + $w^*_{r,2}$};
      \node[right] at ($(1.7,1.2)$) {\scriptsize $w^*_{r,2}$};
      \node[right] at ($(0.0,1.5)$) {\scriptsize $0$};
      \node[right] at ($(9.0,1.5)$) {\scriptsize $T$};
      \node[below] at ($(5,0)$) { (b)};
    \end{scope}   
    \node[right ,inner sep=0, yshift=-0.9cm, xshift=0.9cm] (image_13) at ($(image_22.east)$) 
    {\includegraphics[width=0.277\textwidth, trim=0 0 0 0, clip]
        {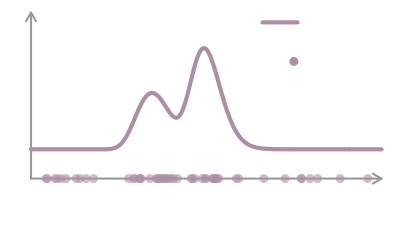}
    }; 
    \begin{scope}[
      shift={(image_13.south west)},
      x={($0.1*(image_13.south east)$)},
      y={($0.1*(image_13.north west)$)}]
      \node[right] at ($(7.5,8.6)$) {\scriptsize $\lambda_{i,r}$};
      \node[right] at ($(7.5,7.1)$) {\scriptsize $t_{i,r,j}$};
      \node[right] at ($(0.0,3.82)$) { $a$};      
      \node[right] at ($(0.50,1.7)$) {\scriptsize $0$};
      \node[right] at ($(9.0,1.7)$) {\scriptsize $T$};
      \node[below] at ($(5,-6.2)$) { (c)};
    \end{scope}   
    
    \draw[->, very thick, lightgray, shorten <=2pt, shorten >=2pt] (image_11.east) -- (image_12.west);
    \draw[->, very thick, lightgray, shorten <=2pt, shorten >=2pt] (image_21.east) -- (image_22.west);
    \draw[->, very thick, lightgray, shorten <=2pt, shorten >=2pt] (image_31.east) -- (image_32.west);
   \node[right,inner sep=0, yshift=-0.9cm] (plus_sign) at (image_22.east) 
   {\includegraphics[width=0.04\textwidth, trim=0 0 0 0, clip]
        {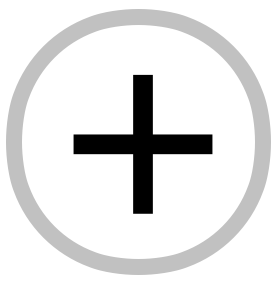}
    }; 
   \draw[->, thick, lightgray, shorten <=0pt, shorten >=0pt] (image_12.east) -- (plus_sign |- image_12) -- (plus_sign.north);
   \draw[->, thick, lightgray, shorten <=0pt, shorten >=0pt] (image_22.east) -- (plus_sign |- image_22) -- (plus_sign.north);
   \draw[->, thick, lightgray, shorten <=0pt, shorten >=0pt] (image_32.east) -- (plus_sign |- image_32) -- (plus_sign.south);
   \draw[->, very thick, lightgray, shorten <=0pt, shorten >=0pt] (plus_sign.east) -- (image_13.west);
\end{tikzpicture}
\caption{ 
Graphical representation of the additive shape invariant model with two components. 
Panel (a) shows the intensity components shared across subjects and observations.
Panel (b) shows the intensity components  associated with subject $i$ and observation $r$ where the intensity is shifted by $v_{i,m}+w^*_{r,m}$. 
Panel (c) shows the expected intensity of subject $i$ in observation $r$, and one realization of the point process $\{t_{i,r,j}: j=1,\cdots,N_{i,r}(T)\}$. 
}
\label{fig:model}
\end{figure}

We start with a model that tackles the first two of the aforementioned tasks: decomposition and alignment.
To be specific, we assume that the true underlying intensities are formed through the superposition of multiple components, and these intensity components exhibit uniformity across subjects, differing only in temporal shifts.
We arrive at the following model
\begin{align}
\label{model:simplified_2}
\begin{split}
\lambda_{i,r}(t)
= a_{} + \sum_{m\in[M]} S^{v_{i,m}+w^*_{r,m}} f_{m}(t),
\end{split}
\end{align}
where
$a \in [0,\infty)$ represents the \emph{baseline intensity},
$M \in \mathbb{N}^+$ is the number of components,
$v_{i,m} \in [0,V]$ represents the \emph{subject-specific time shift} associated with the $i$-th subject and the $m$-th component,
$w^*_{r,m} \in [0,W]$ represents the {known} \emph{observation-specific time shift} of the $m$-th component in the $r$-th observation,
$S^v$ is the shift operator defined as $S^vx(t) \equiv x(t-v)$,
and $f_m(t) \in \mathcal{F}$ represents the $m$-th \emph{intensity component}.
Here $V,W \in (0,T)$, and $\mathcal{F} \equiv \{f \in L^2(\mathbb{R}): f(t)=0 \text{ for } t\in \mathbb{R}\setminus(0, T_0) \}$ with $T_0 \in (0,T)$.  
For further clarity, a graphical representation of the model in~\eqref{model:simplified_2} is provided in Figure~\ref{fig:model}.

To provide a concrete example for Model~\ref{model:simplified_2}, consider the experiment in \cite{Steinmetz2019} where $N_{i,r}(t)$ represents the recorded spike train of the $i$-th neuron in the $r$-th trial, and $\lambda_{i,r}(t)$ denotes the corresponding firing rate (see for instance Figure~\ref{fig:real_data_motivation}).
The terms $a$ and $\{f_m(\cdot): m=1,2\}$ can be interpreted as the spontaneous firing rate and the neural response elicited by the visual gratings and auditory cue, respectively.
The subject-specific time shift $v_{i,m}$ corresponds to the response latency of the $i$-th neuron in response to the $m$-th stimulus. 
The observation-specific time shift $w^*_{r,m}$ corresponds to the occurrence time of the $m$-th stimulus in the $r$-th trial, which is randomly generated by the experimenters.

\subsection{Identifiability}
\label{sec:identifiability}
We denote the collection of model parameters as $\boldsymbol{\theta}_0 \equiv ({a}, \mathbf{f}, \mathbf{v}) \in \Theta_0$, where $\mathbf{f} \equiv (f_{m})_{m\in[M]}$, $\mathbf{v} \equiv (v_{i,m})_{i\in[n], m\in[M]}$, and 
$\Theta_0 \equiv 
\{ \hspace*{-2pt} ({a}, \mathbf{f}, \mathbf{v})  :
{a} \in [0,\infty),  \mathbf{f} \in \mathcal{F}^{M},
\mathbf{v} \in [0,V]^{n \times M} 
\}$.
In addition, for any $N_{i,r}(t)$, we denote its conditional intensity given the observation-specific time shifts as $\lambda_{\boldsymbol{\theta}_0,i}(t,\mathbf{w}^*) \equiv \mathbb{E}_{\boldsymbol{\theta}_0}\{\mathrm{d}N_{i,r}(t) / \mathrm{d}t \mid \mathbf{w}^*_r = \mathbf{w}^*\}$, where $\mathbf{w}^*_r \equiv (w^*_{r,m})_{m\in[M]}$.
We further denote $\boldsymbol{\lambda}_{\boldsymbol{\theta}_0}(t,\mathbf{w}^*) \equiv (\lambda_{\boldsymbol{\theta}_0,i}(t,\mathbf{w}^*))_{i\in[n]}$.
In this context, \emph{identifiability} pertains to the \emph{injectivity} of the mapping $\boldsymbol{\theta}_0 \mapsto \boldsymbol{\lambda}_{\boldsymbol{\theta}_0}(\cdot,\cdot)$ for $\boldsymbol{\theta}_0 \in \Theta_0 $.

With these notations, we can formally  present the identifiability results in Proposition~\ref{prop:identifiability_simplified}. 
\begin{proposition}
\label{prop:identifiability_simplified}
Suppose that the following assumptions hold.
\begin{enumerate}[label=(A\arabic*),ref=A\arabic*]
    \item \label{assump:no_censorship_s}
    $T \geq T_0 + V + W$.
    \item \label{assump:time_shifts_different_s}
    The matrix {$\mathbb{E}[\overline{\boldsymbol{\eta}^*(\xi)} {\boldsymbol{\eta}^*(\xi)}^\top]$ } is invertible {for $\xi \in \mathbb{R} \setminus \{0\}$},
    where ${{\boldsymbol{\eta}^*(\xi)}} \equiv (\eta_m)_{m\in[M]}$, $\eta_m \equiv \exp\{-\operatorname{j} 2 \pi \xi w^*_m\}$,{~$\operatorname{j}$ denotes the imaginary unit, and $\overline{\boldsymbol{\eta}^*(\xi)}$ denotes the complex conjugate of ${\boldsymbol{\eta}^*(\xi)}$.
    }
\end{enumerate}
Let $({a}^*, \mathbf{f}^*, \mathbf{v}^*) \in \Theta_0$ denote the  true parameters.
Then, we can verify the following statements hold.
\begin{enumerate}[label=(P\arabic*), ref=P\arabic*]
   \item \label{prop_res:f_s} 
   For $m\in[M]$, the intensity component ${f}_m^*$ is identifiable up to a time shift.
   \item \label{prop_res:v_s}
   For $m\in[M]$ such that $f^*_{m}$ is non-zero on a set of positive measure,
   the subject-specific time shifts $\{v^*_{i,m}: {i \in [n]} \}$ are identifiable up to an additive constant.
   \item \label{prop_res:a_s} 
   The baseline intensity ${a}^*$ is identifiable. 
\end{enumerate}
\end{proposition}

Assumption~\ref{assump:no_censorship_s} posits that the observation duration sufficiently extends to avoid censorship. 
Given~\ref{assump:no_censorship_s}, the model described in~\eqref{model:simplified_2} can be formulated in the form of a linear regression model  in the frequency domain, where the response variables are determined by $N_{i,r}$'s, the explanatory variables depend on $w^*_{r,m}$'s, and the regression coefficients are functions of $f_m$'s and $v_{i,m}$'s.
Within the context of linear regression, the significance of Assumption~\ref{assump:time_shifts_different_s} becomes straightforward.
Specifically, Assumption~\ref{assump:time_shifts_different_s}  effectively assumes no collinearity among explanatory variables.
We note that Assumption~\ref{assump:time_shifts_different_s} is satisfied if either (i) the observation-specific time shifts are randomized, i.e., $\{w^*_m \in \mathbb{C}: m= 1, 2, \ldots M\}$ or (ii) the gaps between  the observation-specific time shifts are randomized,  i.e., $w^*_m = w^*_{m-1} + \delta_{m-1}$ for $m=2,\ldots,M$, where $\{\delta_m \in \mathbb{C}: m\in[M-1]\}$  are independent random variables with non-zero variance. 
Detailed proof of Proposition~\ref{prop:identifiability_simplified} is presented in Section~\ref*{ssec:proof_prop1} of the Supplementary Materials.

\subsection{Connections with existing models}
\label{sec:connection_with_prior_studies}

In special scenarios, the model presented in~\eqref{model:simplified_2} has connections with three distinct branches of research.
Firstly, considering a special case where $M = 1$ and $R=1$,
the model in~\eqref{model:simplified_2} can be simplified as follows:
\begin{align}
\label{eq:lambda_i1}
\lambda_{i,1}(t)
& = a + S^{v_{i,1}+w^*_{1,1}} f_{1}(t) \equiv S^{v_{i,1}} g(t),
\end{align}
where $g(t) \equiv a + S^{w^*_{1,1}}f_{1}(t)$.
This model has been investigated in the realm of \emph{shape invariant models}~\citep[see, e.g.,][]{beath2007infant,Bigot2010,vimond2010efficient,Bigot2013,Bigot2013a,Bontemps2014}.
Consequently, the proposed model can be seen as a generalization of shape invariant models to incorporate additive  components.
Various assumptions have been proposed to ensure the identifiability of shape invariant models, including availability of repeated observations~\citep{Bigot2013a}, partial knowledge of time shifts~\citep{Bigot2010,Bigot2013}, or the knowledge of the noise distribution~\citep{Bontemps2014}.
For the proposed additive shape invariant model, where $M$ is allowed to exceed 2, a combination of repeated observations and partial knowledge of time shifts suffices to ensure the model identifiability, as elucidated in Proposition~\ref{prop:identifiability_simplified}.

Secondly, considering the scenario where the intensity components $\{S^{v_{i,m}+w^*_{r,m}}f_{m}(t): m\in[M]\}$ have non-overlapping supports for all $i\in[n]$ and $r \in [R]$, model~\eqref{model:simplified_2} can be expressed as:
\begin{align}
\label{eq:lambda_ir_g_h}
\lambda_{i,r}(t)
& = a + \sum_{m\in[M]} S^{v_{i,m}+w^*_{r,m}} f_{m}(t) \equiv g\{h_{i,r}(t)\},
\end{align}
where $g(s) \equiv a + \sum_{m \in [M]} f_{m}(s)$ and $h_{i,r}(t) \equiv \sum_{m\in[M]} \{t - (v_{i,m}+w^*_{r,m})\} \times \mathbf{1}[f_{m} \{ t - (v_{i,m}+w^*_{r,m} )\} \neq 0]$.
The reformulated model in~\eqref{eq:lambda_ir_g_h} means that $\lambda_{i,r}$'s are identical subject to unknown \emph{time warping functions} (i.e., $h_{i,r}(t)$'s).
The task of estimating time warping functions has been explored in the domain of \emph{curve registration} \citep{kneip1992statistical,ramsay1998curve,james2007curve,telesca2008bayesian,cheng2016bayesian}.

Lastly, considering the scenario where the variances of $v_{i,m}$'s and $w^*_{r,m}$'s are both close to zero,  model~\eqref{model:simplified_2} can be approximated using Taylor expansion as:
\begin{align}
\label{model:conn_with_fpca}
\begin{split}
\lambda_{i, r}(t) 
& \approx 
\mu(t) + \sum_{m \in[M]} \zeta_{i,r,m} ~\psi_m(t)
\end{split}
\end{align}
where $\mu(t) \equiv a + \sum_{m \in[M]} f_{m}(t-\mathbb{E}u_{i,r,m})$, 
$u_{i,r,m} \equiv v_{i,m}+w^*_{r,m}$, 
$\zeta_{i,r,m} \equiv -(u_{i,r,m}-\mathbb{E}u_{i,r,m}) \| Df_{m}(t-\mathbb{E}u_{i,r,m}) \|_t$, 
$Df_m$ denotes the first order derivative of $f_m$,
and $\psi_m(t) \equiv Df_{m}(t-\mathbb{E}u_{i,r,m}) \| Df_{m}(t-\mathbb{E}u_{i,r,m}) \|_t^{-1}$.
We defer the derivation of this approximation to Section~\ref*{ssec:connection_FPCA} of the Supplementary Materials.
When functions $\{S^{\mathbb{E}u_{i,r,m}}f_{m}(t): m\in[M]\}$ exhibit non-overlapping supports, the approximate model in~\eqref{model:conn_with_fpca} corresponds to the models of \emph{functional principal component analysis} (FPCA) \citep{Yao2005,morris2006wavelet,di2009multilevel,crainiceanu2009generalized,xu2020semi}.

\subsection{Estimation}
\label{sec:method_simplified}
We consider the case that $N_{i,r}(t)$'s are Poisson processes.
To effectively estimate the parameters in~\eqref{model:simplified_2}, we exploit the two sources of stochasticity of the Poisson processes.
Firstly, the number of events over $[0,T]$ for any Poisson process is a random variable that follows a Poisson distribution,
in other words, $N_{i,r}(T) \sim \mathrm{Poisson}(\Lambda_{i,r})$, where $\Lambda_{i,r} \equiv \int_0^T \lambda_{i,r}(t) \mathrm{d}t$ represents the \emph{expected event count}.
When Assumption~\ref{assump:no_censorship_s} holds, we can derive from the model in~\eqref{model:simplified_2} that $\Lambda_{i,r} = aT + \sum_{m\in[M]}\int_0^T f_{m}(t)\mathrm{d}t$, denoted as $\Lambda$ for simplicity.
Secondly, conditioning $N_{i,r}(T)$, the event times $\{t_{i,r,j}: j=1,\ldots,N_{i,r}(T)\}$ can be regarded as independent and identically distributed random variables.
The probability density function of these event times, or \emph{event time distribution}, is characterized by $\lambda_{i,r}(t)\Lambda_{}^{-1} = a \Lambda^{-1} + \sum_{m\in[M]} S^{v_{i,m}+w^*_{r,m}} f_m(t) \Lambda^{-1}$.

Accordingly, we estimate $a,\mathbf{f},\mathbf{v}$ through the following reparameterization.
Letting $a' \equiv a \Lambda^{-1}$and $\mathbf{f}' \equiv (f_m \Lambda^{-1})_{m\in[M]}$, we have the following optimization problem:
\begin{align}
\label{eq:optimization_problem_simplified}
\begin{split}
\hat{{a}}', \hat{\mathbf{f}}', \hat{\mathbf{v}}
& \equiv 
\underset{ {a}', \mathbf{f}', \mathbf{v} }{\arg\min}
~L_1( {a}', \mathbf{f}', \mathbf{v} ) 
\\ 
& \equiv 
\underset{ {a}', \mathbf{f}', \mathbf{v} }{\arg\min}
\sum_{i\in[n], r\in [R]} 
\hspace*{-10pt} \beta_{i,r}~
\frac{1}{T} \Bigg\|
\frac{y_{i,r}(t)}{N_{i,r}(T)} 
- \bigg\{ a' + \sum_{m\in[M]} S^{v_{i,m}+w^*_{r,m}} f'_{m}(t) \bigg\}  \Bigg\|_t^2,
\end{split}
\end{align}
where $\|\cdot\|_t$ denotes the $L^2$-norm with respect to $t$, $\beta_{i,r}\equiv N_{i,r}(T)$, $y_{i,r}(t) \equiv \{N_{i,r}(t+\Delta t) - N_{i,r}(t)\} \Delta t^{-1}$, and $\Delta t$ represents a infinitesimally small value.
In~\eqref{eq:optimization_problem_simplified}, the objective function $L_1( {a}', \mathbf{f}', \mathbf{v} )$ measures the discrepancy between the empirical and estimated distributions of event timings, where $y_{i,r}(t) N_{i,r}(T)^{-1}$ and $\{ a' + \sum_{m\in[M]} S^{v_{i,m}+w^*_{r,m}} f'_{m}(t) \}$ serve as the \emph{empirical distribution} and the \emph{estimated distribution}  of $\{t_{i,r,j}: j=1,\ldots,N_{i,r}(T)\}$, respectively.
The term $\beta_{i,r}$ serves as the weight of the counting process $N_{i,r}(t)$. For instance, setting $\beta_{i,r}=N_{i,r}(T)$ means equal weights for all events.  
Finally,  we estimate $\Lambda$ using the empirical mean
\begin{align}
\label{eq:optimization_problem_Lambda_hat}
\begin{split}
\hat{\Lambda}
& \equiv \left( {nR} \right)^{-1} \sum_{i \in [n], r \in [R]} N_{i,r}(T).
\end{split}
\end{align}
The parameters $a,\mathbf{f}$ can be estimated  from $\hat{\Lambda}, \hat{a}', \hat{\mathbf{f}}'$ by $\hat{a} \equiv \hat{a}' \hat{\Lambda}$ and $\hat{\mathbf{f}} \equiv \hat{\mathbf{f}}' \hat{\Lambda}$.

\section{Additive shape invariant mixture model}
\label{sec:simultaneous_clustering}

\subsection{Model}
We extend model~\eqref{model:simplified_2} to simultaneously perform  decomposition, alignment, and clustering. 
Assume that the $n$ subjects can be classified into $K$ distinct clusters, that is, $[n]=\cup_{k=1}^K \mathcal{C}_k$, where $\mathcal{C}_1, \ldots, \mathcal{C}_K$ represent mutually exclusive subsets of $[n]$. 
These clusters are delineated based on the similarity of intensity components across subjects.
Specifically, we introduce the following model:
\begin{align}
\label{model:full}
\begin{split}
\lambda_{i,r}(t)
= a_{z_i} + \sum_{m\in[M]} S^{v_{i,m}+w^*_{r,m}} f_{z_{i},m}(t),
\end{split}
\end{align}
where $z_i \in [K]$ represents the \emph{cluster membership} of subject $i$ such that $z_i = k$ if $i \in \mathcal{C}_k$.
{We refer to the model in~\eqref{model:full} as the \emph{additive shape invariant mixture model}, or ASIMM for short.}
Conditioning on each cluster, the {additive shape invariant mixture} model in~\eqref{model:full} simplifies to the additive shape invariant model in~\eqref{model:simplified_2}.
Similar to the connection between model in~\eqref{model:simplified_2} and FPCA, the {additive shape invariant mixture} model in~\eqref{model:full} has a close connection with clustering methods based on FPCA~\citep[][]{Chiou2007,bouveyron2011model,jacques2013funclust,yin2021row}.

In the context of neural data~\citep{Steinmetz2019}, $\mathcal{C}_k$'s can be interpreted as functional groups of neurons, wherein neurons exhibit similar firing patterns.
The {additive shape invariant mixture model} in~\eqref{model:full} enables us to simultaneously identify functional groups of neurons (i.e., $\mathcal{C}_k$'s), discern representative neural firing patterns (i.e., $f_{k,m}$'s), and estimate individual neural response latencies (i.e., $v_{i,m}$'s).
It is worthwhile to emphasize that the applicability of the proposed method extends beyond neural data analysis. 
For instance, the {additive shape invariant mixture} model in~\eqref{model:full} can be employed in analyzing recurrent consumer activity in response to advertising \citep{xu2014path,zadeh2014modeling,tanaka2016inferring,bues2017mobile}, or studying hospital admission rates following the implementation of societal disease prevention policies \citep{barone2006short,sims2010short,klevens2016paid,evans2021impact}.
Additionally, the proposed model has potential for application on diverse datasets~\citep{tang2023multivariate,schoenberg2023estimating,dempsey2023recurrent,xu2024bias,djorno2024mutually}.

\subsection{Identifiability}
We denote the collection of unknown parameters in~\eqref{model:full} as $\mathbf{z} \equiv (z_i)_{i\in[n]}$, $\mathbf{a} \equiv (a_k)_{k\in[K]}$, $\mathbf{f} \equiv (f_{k,m})_{k\in[K], m\in[M]}$, $\mathbf{v} \equiv (v_{i,m})_{i\in[n], m\in[M]}$.
Let $(\mathbf{z}^*, \boldsymbol{a}^*, \mathbf{f}^*, \mathbf{v}^*) \in \Theta_1$ denote the true parameters, where $\Theta_1 \equiv 
\{ \hspace*{-2pt} (\mathbf{z}, \boldsymbol{a}, \mathbf{f}, \mathbf{v})  :
\mathbf{z} \in [K]^n,
\boldsymbol{a} \in [0,\infty)^K,  \mathbf{f} \in \mathcal{F}^{K \times M},
\mathbf{v} \in [0,V]^{n \times M}\}$.
When conditioning on $\mathbf{z}^*$, Proposition~\ref{prop:identifiability_simplified} establishes the identifiability of $\boldsymbol{a}^*, \mathbf{f}^*, \mathbf{v}^*$.
However, to ensure the identifiability of $\mathbf{z}^*$, an additional assumption regarding the separability of clusters is required. 
The formal presentation of model identifiability is provided in Proposition~\ref{prop:identifiability}.

\begin{proposition}
\label{prop:identifiability}
Suppose that both Assumptions~\ref{assump:no_censorship_s} and~\ref{assump:time_shifts_different_s} hold, and further assume that 
\begin{enumerate}[label=(A\arabic*),ref=A\arabic*,start=3]
    \item \label{assump}
    For any $k,k' \in [K]$ that $k \neq k'$, there exists $m_0\in[M]$ such that  for any $x \in \mathbb{R}$, $\{t\in \mathbb{R}: S^x f^*_{k,m_0}(t) \neq f^*_{k',m_0}(t) \}$ has a positive measure.
\end{enumerate}
 Then, we can verify the following statements hold.
\begin{enumerate}[label=(P\arabic*), ref=P\arabic*,start=4]
   \item \label{prop_res:z} 
   The cluster memberships $\mathbf{z}^*$ are  identifiable up to a permutation of cluster labels.
   \item \label{prop_res:alpha} 
   The baseline values $\boldsymbol{a}^*$ are identifiable up to a permutation of cluster labels. 
   \item \label{prop_res:f} 
   The response components $\mathbf{f}^*$ are identifiable up to a permutation of cluster labels and time shifts.
   \item \label{prop_res:v}
   For $k\in[K],m\in[M]$ such that the set $\{t: f^*_{k,m}(t) \neq 0\}$ is of positive measure,
   the set $(v^*_{i,m})_{i \in \mathcal{C}^*_k}$ is identifiable up to a constant independent of $i$.
\end{enumerate}
\end{proposition}
Assumption~\ref{assump} mandates that each cluster exhibits at least one signature intensity component that is unique to this cluster. Statement~\ref{prop_res:z} directly stems from Assumption~\ref{assump}. 
Statements~\ref{prop_res:alpha},~\ref{prop_res:f}, and~\ref{prop_res:v} can be derived by applying Proposition~\ref{prop:identifiability_simplified} to each individual cluster.
Detailed proof of Proposition~\ref{prop:identifiability} is presented in Section~\ref*{ssec:proof_prop2} of the Supplementary Materials.

\subsection{Estimation}
\label{subsec:optimization_full}
We estimate the parameters in~\eqref{model:full} by generalizing the optimization approach in Section~\ref{sec:method_simplified} to incorporate the cluster structure.
For any $k\in[K]$ and $m\in[M]$, let $\Lambda_{k} \equiv a_{k}T + \sum_{m\in[M]}\int_0^T f_{k,m}(t) \mathrm{d}t$, $a'_{k} \equiv a_{k} \Lambda_{k}^{-1}$, and $f'_{k,m} \equiv f_{k,m} \Lambda_{k}^{-1}$.
Denoting $\mathbf{a}' \equiv (a'_{k})_{k\in[K]}$,  
$\mathbf{f}' \equiv (f'_{k,m})_{k\in[K], m\in[M]}$, 
and $\boldsymbol{\Lambda} \equiv (\Lambda_k)_{k\in[K]}$,
we propose the following optimization problem:
\begin{align}
\begin{split}
\label{eq:optimization_problem}
\hat{\mathbf{z}}, \hat{\mathbf{a}}', \hat{\mathbf{f}}', \hat{\mathbf{v}}, \hat{\boldsymbol{\Lambda}}
& \equiv
\underset{ {\mathbf{z}}, \mathbf{a}', \mathbf{f}', \mathbf{v}, \boldsymbol{\Lambda} }{\arg\min}
\left\{
L_1(\mathbf{z}, \mathbf{a}', \mathbf{f}', \mathbf{v} ) 
+ \gamma ~L_2(\mathbf{z}, \boldsymbol{\Lambda} ) \right\}, 
\end{split}
\end{align}
where 
\begin{align}
\label{eq:L1_zafv}
& L_1(\mathbf{z}, \mathbf{a}', \mathbf{f}', \mathbf{v} ) 
 \equiv 
\sum_{i\in[n], r\in [R]} 
\hspace*{-10pt} \beta_{i,r}~
\frac{1}{T} \Bigg\|
\frac{y_{i,r}(t)}{N_{i,r}(T)} 
- \bigg\{ a'_{z_i} + \sum_{m\in[M]} S^{v_{i,m}+w^*_{r,m}} f'_{z_i,m}(t) \bigg\}  \Bigg\|_t^2,
\\
\label{eq:L2_zLambda}
& L_2(\mathbf{z}, \boldsymbol{\Lambda} ) 
 \equiv 
\sum_{i\in [n], r\in [R]} 
\big| N_{i,r}(T) - \Lambda_{z_i} \big|^2,
\end{align}
and $\gamma \in (0, \infty)$ is a tuning parameter.
In essence, $L_1$ and $L_2$ assess the within-cluster variance of event time distributions and event counts, respectively. 
When the number of clusters is reduced to one (i.e., $K=1$), the definitions of $L_1$ in~\eqref{eq:L1_zafv} is identical to the definition in~\eqref{eq:optimization_problem_simplified}.
The tuning parameter $\gamma$ modulates the relative importance of $L_2$ compared to $L_1$ in the optimization with respect to $\mathbf{z}$.
When $\gamma$ is sufficiently small, the estimator $\hat{\mathbf{z}}$ defined in~\eqref{eq:optimization_problem} is predominantly determined by $L_1$, resulting in a potentially suboptimal value of $L_2$. 
Conversely, when $\gamma$ is sufficiently large, the dominance shifts towards $L_2$, resulting in a $\hat{\mathbf{z}}$ that achieves the minimum value of $L_2$, while $L_1$ may be relegated to suboptimal values.
Subsequent to addressing the optimization problem in~\eqref{eq:optimization_problem}, the estimations of $\mathbf{a}$ and $\mathbf{f}$ can be established via $\hat{a}_k \equiv \hat{a}'_k \hat{\Lambda}_k$ and $\hat{\mathbf{f}}_{k,m} \equiv \hat{\mathbf{f}}'_{k,m} \hat{\Lambda}_k$ for $k\in[K]$, $m\in[M]$.

The optimization problem in~\eqref{eq:optimization_problem} involves two tuning parameters $\gamma$ and $K$. 
To determine these tuning parameters, we employ a heuristic method.
We first establish a preliminary estimate of $K$ using simple methods, such as applying the k-means algorithm on $N_{i,r}(T)$'s and selecting $K$ using the elbow method~\citep{thorndike1953belongs}.
Given this preliminary estimation of $K$, 
we choose the largest $\gamma$ before observing a significant increase in $L_1$.
We provide simulation experiments to justify this heuristic method and demonstrate the robustness of selected $\gamma$ to the change of the preliminary selection of $K$ in Section~\ref*{ssec:heuristic_select_gamma} in the Supplementary Material.
Finally, conditioning on the selected $\gamma$, we refine the value of $K$ by identifying the elbow point on the curve of the overall objective function in~\eqref{eq:optimization_problem} against $K$.


\section{Algorithm} \label{sec:algorithm}
We now present an algorithm for solving the optimization problem~\eqref{eq:optimization_problem}.
This optimization problem aims to minimize the within-cluster variances pertaining to event time distributions and event counts.
To this end, we propose an algorithm that resembles the k-means algorithm that alternates between a \emph{centering step} and a \emph{clustering step} presented as follows. 
\begin{align}
\label{eq:centering_step}
 \text{({centering} step)}~ \quad 
&\hat{\mathbf{a}}', \hat{\mathbf{f}}' 
=  \underset{\mathbf{a}', \mathbf{f}' }{\arg\min} 
~L_1(\hat{\mathbf{z}}, \mathbf{a}', \mathbf{f}', \hat{\mathbf{v}} ),
\quad
\hat{\boldsymbol{\Lambda}}
= \underset{\boldsymbol{\Lambda} }{\arg\min} 
~L_2(\hat{\mathbf{z}}, \boldsymbol{\Lambda} ) ,
\\
\label{eq:clustering_step}
 \text{({clustering} step)} \quad
&\hat{\mathbf{z}}, \hat{\mathbf{v}} 
= \underset{\mathbf{z}, \mathbf{v}}{\arg\min}
~\{L_1(\mathbf{z}, \hat{\mathbf{a}}', \hat{\mathbf{f}}', \mathbf{v} )
+ \gamma ~L_2({\mathbf{z}}, \hat{\boldsymbol{\Lambda}} )\}.
\end{align}
In the centering step~\eqref{eq:centering_step}, the estimators of ${\mathbf{a}'}$, ${\mathbf{f}'}$ and $\boldsymbol{\Lambda}$ are updated conditioned on the values of $\hat{\mathbf{z}}$ and $\hat{\mathbf{v}}$. 
In the clustering step~\eqref{eq:clustering_step}, the estimators of ${\mathbf{z}}$ and ${\mathbf{v}}$ are updated conditioned on the values of $\hat{\mathbf{a}}'$, $\hat{\mathbf{f}}'$ and $\hat{\boldsymbol{\Lambda}}$. 
This alternating scheme facilitates a closed-form solution in the centering step and an effective optimization process in the clustering step.

\subsection{The centering step}
\label{sec:the_centering_step}
The centering step~\eqref{eq:centering_step} involves two optimization problems.
To solve the first optimization problem, we formulate it in the frequency domain as follows:
\begin{align}
\label{eq:L_1_fourier}
\begin{split}
\hat{\mathbf{a}}',\hat{\boldsymbol{\phi}}'
& =  \underset{\mathbf{a}', \boldsymbol{\phi}' }{\arg\min} 
\sum_{i\in [n], r\in [R]} 
\hspace*{-10pt} \beta_{i,r}~
\sum_{l\in \mathbb{Z}} 
\Bigg| \frac{\eta_{i,r,l}}{N_{i,r}(T)}
- \bigg\{ a'_{\hat{z}_i} \mathbf{1}(l=0) 
\hspace*{5pt} + 
\\ & \hspace*{1.5in} 
\sum_{m\in[M]} \exp\big\{- \operatorname{j} 2 \pi l (\hat{v}_{i,m} +w^*_{r,m}) T^{-1} \big\} \phi'_{\hat{z}_i,m,l}
\bigg\}
 \Bigg|^2 ,
\end{split}
\end{align}
where $\boldsymbol{\phi}' \equiv (\phi'_{k,m,l})_{k\in[K],m\in[M],l\in\mathbb{Z}}$, 
$\{\phi'_{k,m,l}: l\in \mathbb{Z}\}$ denotes the Fourier coefficients of $f'_{k,m}(t)$,
$\{\eta_{i,r,l}: l \in \mathbb{Z}\}$ denotes the Fourier coefficients of $y_{i,r}(t)$, and $\operatorname{j}$ denotes the imaginary unit.
Notably, the multiplicative term $\exp( - \operatorname{j} 2 \pi l [\hat{v}_{i,m} +w^*_{r,m}] T^{-1} ) $ serves as the frequency domain counterpart of the shift operator $S^{\hat{v}_{i,m}+w^*_{r,m}}$.
In essence, the Fourier transformation converts the \emph{shift operators} into \emph{multiplication}. 
As a result, the objective function becomes more tractable compared to its counterpart in the original domain.
Indeed,  an analytical solution for $\hat{\boldsymbol{\phi}}'$ can be derived. 
Let $\boldsymbol{\phi}'_{k,*,l} \equiv ({\phi}'_{k,m,l})_{m\in[M]}$ for any $k\in[K]$ and $l\in \mathbb{Z}$.
For $l \neq 0$, the solution to~\eqref{eq:L_1_fourier} with respect to $\boldsymbol{\phi}'_{k,*,l}$ can be expressed as follows:
\begin{align}
\label{eq:hat_phi_prime_kl}
\hat{\boldsymbol{\phi}}'_{k,*,l}
& = 
\left(\overline{\mathbf{E}_{k,l}}^\top \mathbf{B}_k~ \mathbf{E}_{k,l} \right)^{-1} 
\left(\overline{\mathbf{E}_{k,l}}^\top \mathbf{B}_k~ \mathbf{h}_{k,l}\right),
 & \hspace{-0.5in} \text{for } l\neq0,
\\ 
\label{eq:hat_phi_km0}
\hat{\boldsymbol{\phi}}'_{k,*,0} 
& = - \sum_{|l| \leq \ell_0, l \neq 0} \hat{\boldsymbol{\phi}}'_{k,*,l},
 & \hspace{-0.5in} \text{for } l=0.
\end{align}
Here, $\mathbf{E}_{k,l}$ is defined as
\begin{align}
\mathbf{E}_{k,l} \equiv [\exp\{- \operatorname{j} 2 \pi l (\hat{v}_{i,m}+w^*_{r,m}) T^{-1}\} ]_{(i,r) \in \hat{\mathcal{C}}_k \times [R], m \in [M]},
\end{align}
where
$\hat{\mathcal{C}}_k \equiv \{i\in[n]: \hat{z}_i = k\}$.
$\overline{\mathbf{E}_{k,l}}$ denotes the complex conjugate of $\mathbf{E}_{k,l}$,
$\mathbf{B}_k$ is a diagonal matrix of $(\beta_{i,r})_{(i,r) \in \hat{\mathcal{C}}_{k}\times [R]}$,
$\mathbf{h}_{k,l} \equiv (\eta_{i,r,l}N_{i,r}(T)^{-1})_{(i,r) \in \hat{\mathcal{C}}_{k}\times [R]} $,
and $\ell_0 \in \mathbb{N}$ is a truncation parameter introduced to ensure the numerical feasibility of computing $\hat{\boldsymbol{\phi}}'_{k,*,0}$.
For $l \neq 0$, the objective function concerning $\boldsymbol{\phi}'_{k,*,l}$ in~\eqref{eq:L_1_fourier} is a weighted sum of squares, hence the estimate in~\eqref{eq:hat_phi_prime_kl} can be derived using the well-known least squares estimator.
For $l = 0$, the estimate in~\eqref{eq:hat_phi_km0} can be obtained by exploiting the definition of $\mathcal{F}$.
The detailed derivations of~\eqref{eq:hat_phi_prime_kl} and~\eqref{eq:hat_phi_km0}  can be found in Section~\ref*{ssec:detail_centering_setp} of the Supplementary Material.

Upon obtaining $\hat{\boldsymbol{\phi}}'$, the solution to the first  optimization problem in~\eqref{eq:centering_step} can be derived as follows.
For $k\in[K]$ and $m\in[M]$,
\begin{align}
\label{eq:hat_a_k}
\begin{split}
\hat{{a}}_k'
&= T^{-1} - \sum_{m\in[M]} \hat{\phi}'_{k,m,0},
\end{split}
\\
\label{eq:hat_f_km}
 \hat{f}'_{k,m}(t) 
& = \sum_{|l| \leq \ell_0} \hat{\phi}_{k,m,l} \exp( \operatorname{j} 2 \pi l t T^{-1}),
\end{align}
where~\eqref{eq:hat_a_k} is obtained by substituting $\hat{\boldsymbol{\phi}}'$ into~\eqref{eq:L_1_fourier},
and~\eqref{eq:hat_f_km} follows from the inverse Fourier transformation.

The second optimization problem in~\eqref{eq:centering_step} aims to minimize the within-cluster variances of event counts given cluster memberships.
The solution to this optimization problem is straightforward: for any $k \in [K]$,
\begin{align}
\label{eq:hat_Lambda_k}
\hat{\Lambda}_k 
&= \underset{ {\Lambda}_k }{\arg\min} 
\sum_{i\in \hat{\mathcal{C}}_k, r\in [R]} 
\big| N_{i,r}(T) - \Lambda_{k} \big|^2
= \left( {|\hat{\mathcal{C}}_k|~ R} \right)^{-1} \sum_{i \in \hat{\mathcal{C}}_k, r \in [R]} N_{i,r}(T) ,
\end{align}
where $|\hat{\mathcal{C}}_k|$ denotes the cardinality of the set $\hat{\mathcal{C}}_k$.
In summary, the solution  to the centering step is encapsulated by equations~\eqref{eq:hat_a_k},~\eqref{eq:hat_f_km}, and~\eqref{eq:hat_Lambda_k}.

\subsection{The clustering step}
\label{sec:the_clustering_step}
The optimization problem in~\eqref{eq:clustering_step} can be scaled down to the subject level, allowing for the independent estimation of parameters associated with each subject.
For any subject $i\in[n]$, let $\mathbf{v}_i \equiv (v_{i,m})_{m\in[M]}$ denote its associated time shifts.
In~\eqref{eq:clustering_step},
the parameters $z_i$ and $\mathbf{v}_i$ are estimated through the following sub-problem:
\begin{align}
\label{eq:hat_zi_hat_vi_argmin}
\hat{z}_i, \hat{\mathbf{v}}_i 
& = \underset{z_i,\mathbf{v}_i}{\arg\min}
~\{L_{1,i}(z_i, \hat{\mathbf{a}}', \hat{\mathbf{f}}', \mathbf{v}_i )
+ \gamma ~L_{2,i}(z_i, \hat{\boldsymbol{\Lambda}} )\},
\end{align}
where $L_{1,i}$ and $L_{2,i}$ are defined as 
\begin{align}
\label{eq:L1i_zafv}
& L_{1,i}(z_i, \hat{\mathbf{a}}', \hat{\mathbf{f}}', \mathbf{v}_i ) 
 \equiv 
\sum_{r\in [R]} 
\beta_{i,r}~
\frac{1}{T} \Bigg\|
\frac{y_{i,r}(t)}{N_{i,r}(T)} 
- \bigg\{ \hat{a}'_{z_i} 
+ \hspace*{-5pt} \sum_{m\in[M]} S^{v_{i,m}+w^*_{r,m}} \hat{f}'_{z_i,m}(t) \bigg\}  \Bigg\|_t^2,
\\
\label{eq:L2i_zLambda}
& L_{2,i}(z_i, \hat{\boldsymbol{\Lambda}} ) 
 \equiv 
\sum_{r\in [R]} 
\big| N_{i,r}(T) - \hat{\Lambda}_{z_i} \big|^2.
\end{align}
The parameter dimension for the problem in~\eqref{eq:hat_zi_hat_vi_argmin} is significantly reduced compared to the original optimization problem in~\eqref{eq:clustering_step}. 
Consequently, solving the problem in~\eqref{eq:hat_zi_hat_vi_argmin} is computationally more efficient than addressing the original problem stated in~\eqref{eq:clustering_step}.

To solve the optimization problem in~\eqref{eq:hat_zi_hat_vi_argmin}, we employ the following procedure:
\begin{align}
\label{eq:tilde_v_i_k}
\tilde{\mathbf{v}}_{i|k}
& = \underset{\mathbf{v}_{i}}{\arg\min}
~L_{1,i}(k, \hat{\mathbf{a}}', \hat{\mathbf{f}}', \mathbf{v}_i ), 
\quad \text{for } k\in[K],
\\ 
\label{eq:hat_zi_argmin}
\hat{z}_i
& = \underset{z_i \in [K] }{\arg\min}
~\{L_{1,i}(z_i, \hat{\mathbf{a}}', \hat{\mathbf{f}}', \tilde{\mathbf{v}}_{i|z_i} )
+ \gamma ~L_{2,i}(z_i, \hat{\boldsymbol{\Lambda}} )\},
\\ 
\label{eq:hat_v_i_D_i_zi}
\hat{\mathbf{v}}_i
&= \tilde{\mathbf{v}}_{i|\hat{z}_i}.
\end{align}
In the first step~\eqref{eq:tilde_v_i_k}, we determine the optimal time shift for each potential cluster membership $k\in[K]$. 
The optimization problem in this step can be solved in the frequency domain using the Newton's method (see Section~\ref*{ssec:newton_method} of the Supplementary Material).
In the second step~\eqref{eq:hat_zi_argmin}, we evaluate the objective function for each possible cluster membership, leveraging the optimal time shift corresponding to that particular cluster. 
Subsequently, we designate the cluster associated with the minimal objective function value as the estimated cluster membership.
In the last step~\eqref{eq:hat_v_i_D_i_zi}, we choose the optimal time shift for the estimated cluster membership.
In summary, the solution to the clustering step is encapsulated by~\eqref{eq:tilde_v_i_k},~\eqref{eq:hat_zi_argmin}, and~\eqref{eq:hat_v_i_D_i_zi}.

\subsection{Overall estimation procedure}
The objective function in~\eqref{eq:optimization_problem} exhibits a non-convex nature.
This characteristic poses a crucial need for an appropriate initialization scheme and convergence criterion.
Our proposed initialization scheme is described in Remark~\ref{remark:initialization}.
The overall estimation procedure is summarized in Algorithm~\ref{algo}.

\begin{algorithm}[H]{}
\label{algo}
\SetAlgoLined
\KwIn{$\{N_{i,r}(t): i\in[n], r\in[R] \}, K, \gamma, \ell_0$ }
Initialize $ \hat{\mathbf{v}}^{(0)}, \hat{\mathbf{z}}^{(0)}$ via \eqref{eq:init_v},~\eqref{eq:init_z_obj}, let $s=0$, and $L^{(0)} = \infty$\;
\While{not stop}{
    Update $\hat{\mathbf{a}}'^{(s+1)}$, $\hat{\mathbf{f}}'^{(s+1)}$, $\hat{\boldsymbol{\Lambda}}^{(s+1)}$ via~\eqref{eq:hat_a_k} - \eqref{eq:hat_Lambda_k} given 
    $
    (\hat{\mathbf{z}}^{(s)}, \hat{\mathbf{v}}^{(s)})$;
    \\
    Update $\hat{\mathbf{z}}^{(s+1)}$, $\hat{\mathbf{v}}^{(s+1)}$ via~\eqref{eq:tilde_v_i_k} - \eqref{eq:hat_v_i_D_i_zi},  given 
    $
    (\hat{\mathbf{a}}'^{(s+1)}, \hat{\mathbf{f}}'^{(s+1)}, \hat{\boldsymbol{\Lambda}}^{(s+1)} )$;
    \\ 
    Evaluate the loss function: $L^{(s+1)} 
\equiv L_1(\hat{\mathbf{z}}^{(s+1)}, \hat{\mathbf{a}}'^{(s+1)}, \hat{\mathbf{f}}'^{(s+1)}, \mathbf{v}^{(s+1)} )
+ \gamma ~L_2(\hat{\mathbf{z}}^{(s+1)}, \hat{\boldsymbol{\Lambda}}^{(s+1)} )$;
    \\
    Evaluate the stopping criterion: $\{L^{(s)}  - L^{(s+1)}\} /  L^{(s+1)} 
    \leq \epsilon$;
    \\
    $s=s+1$;
 }
\KwOut{$ \hat{\mathbf{z}}^{(s)}, \hat{\mathbf{a}}'^{(s)}, \hat{\mathbf{f}}'^{(s)}, \hat{\mathbf{v}}^{(s)}$, $\hat{\boldsymbol{\Lambda}}^{(s)}$.}
 \caption{ Iterative algorithm for {ASIMM}}
\end{algorithm}

\begin{remark}
\label{remark:initialization}
\textbf{Initialization.}
The subject-specific time shifts are initialized based on the earliest event occurrence following each stimulus.
Specifically, the value of $\hat{\mathbf{v}}^{(0)}$ is defined as follows. 
For $i\in[n]$, $m \in [M]$, 
\begin{align}
\label{eq:init_v}
& \hat{v}^{(0)}_{i,m} \equiv \min\{ t_{i,r,j} - w^*_{r,m}: t_{i,r,j} > w^*_{r,m},  r \in [R], j \in [N_{i,r}(T)]\}.
\end{align}
The cluster memberships are initialized based on the \emph{adjusted} event times that roughly aligned the point processes using the initial subject-specific time shifts in \eqref{eq:init_v}. 
These adjusted event times are calculated by shifting the event times associated with each stimulus to an anchor point for that stimulus.
To be specific, we shift the event times as $\tilde{t}_{i,r,j} \equiv t_{i,r,j} - \hat{u}^{(0)}_{i,r,m} + \min_{r'\in[R]}w^*_{r',m} $ for event $j$ where $t_{i,r,j} \in [\hat{u}^{(0)}_{i,r,m}, \hat{u}^{(0)}_{i,r,m+1}]$. 
Here, $\hat{u}^{(0)}_{i,r,m} \equiv \hat{v}^{(0)}_{i,m}+w^*_{r,m}$ denotes the total time shift associated with stimulus $m$ for $m\in[M]$, $\hat{u}^{(0)}_{i,r,M+1} \equiv T$, and $\min_{r'\in[R]}w^*_{r',m}$ represents the anchor point of stimulus $m$.
Subsequently, cluster memberships are initialized by applying the k-means algorithm on adjusted event times:
\begin{align}
\label{eq:init_z_obj}
\hat{\mathbf{z}}^{(0)} \equiv \underset{\mathbf{z}}{\arg\min} \sum_{k\in[K]} \sum_{i,j: z_i=z_j=k} 
\left\| y'_i(t) - y'_j(t)  \right\|^2,
\end{align}
where $y'_i(t)$ denotes the empirical distribution of $\{\tilde{t}_{i,r,j}: r \in [R], j\in [N_{i,r}(T)]\}$. 
The efficacy of the proposed initialization approach is illustrated in Figure~\ref*{fig:init} of the Supplementary Material, where it is shown to outperform the random initialization with multiple restarts.
\end{remark}


\section{Simulation} \label{sec:simulation}

\subsection{Simulation experiment design}
\label{subsec:generation}

We assess the performance of the proposed method in three synthetic experiments.
In the first experiment, we aim to explore the {intensity decomposition performance}.
To this end, we generate Poisson processes $N_{i,r}$'s whose intensities  follow the additive shape invariant model in~\eqref{model:simplified_2}.
Specifically, we set $T=2.5$, $M=2$, $v_{i,1} \sim \mathrm{Unif}(0,1/64)$, $v_{i,2} \sim \mathrm{Unif}(0,1/16)$, $w^*_{r,1} \sim \mathrm{Unif}(0,\tau)$, $w^*_{r,2} \sim \mathrm{Unif}(0.8,0.8+\tau)$. 
These parameter values remain unchanged across all subsequent experiments. 
In addition, we set $a=20$, and set $\mathbf{f}$ as 
\begin{align}
\label{eq:true_f_Nclus1}
\begin{split}
f_{1}(t) 
& = 70 \times \big[ \{ 2 - 2\cos( 4\pi[t-0.4] ) \} \times \mathbf{1}( t \in [0.4, 0.9]) \big]
\equiv 70 \times {q}_1(t) ,
\\
f_{2}(t) 
& = 70 \times  \big[ \{ 2 - 2\cos( 2\pi{|2t|}^{1/2} ) \} \times \mathbf{1}( t \in [0, 0.5]) \big]
\equiv 70 \times q_2(t).
\end{split}
\end{align}
Notably, $q_1(t)$ and $q_2(t)$ capture the event time distribution of the two components, and both $q_1(t)$ and $q_2(t)$ integrate to unity.
We tune the signal strength in synthetic data by altering the values of $R$, $n$ and $\tau$.
Intuitively, $R$ and $n$ serve as the sample size, hence are positively associated with signal strength. And $\tau$ is associated with the variance of $w^*_{r,m}$'s and thus the identifiability of the intensity components.

In the second experiment, we compare the clustering performance of {the proposed ASIMM~\eqref{model:full}} and two relevant methods: kCFC, introduced in~\cite{Chiou2007}, and k-mean alignment, introduced in~\cite{Sangalli2010a}. Here we do not apply the method by \cite{yin2021row} since it considers a multiplicative model whereas our model assumes additive components. 
We generate Poisson processes using the model in~\eqref{model:full}.
In particular, we set the true cluster memberships by sequentially assigning $n=40$ subjects into \(K=4\) clusters of equal size, that is, \(z_i = \lceil (i/n)K \rceil\), where \(\lceil \cdot \rceil\) denotes the ceiling function.
In addition, we set $a_k = 20$, and set $f_{k,m}$'s as shown in Table~\ref{table:true_f_Nclus4}.
In Table~\ref{table:true_f_Nclus4}, we introduce $\rho\in(0,1)$ to control the distinctiveness of clusters.
Consider the mean intensity associated with the $k$-th cluster.
For $i\in \mathcal{C}_k$ and $r\in[R]$, we can derive from~\eqref{model:full} that $\mathbb{E} \lambda_{i,r}(t) = a_k + \sum_{m\in[M]} (p_m \star f_{k,m}) (t)$, where $p_m$ denotes the probability density function of $v_{i,m}+w^*_{r,m}$, and $\star$ denotes the convolution operator that $(p_m \star f_{k,m}) (t)=\int_0^{T_0} p_m(t-x) f_{k,m}(x) \mathrm{d} x$.
When $\rho=0$, the shapes of $\sum_{m\in[M]} (p_m \star f_{k,m}) (t)$'s are identical across clusters, whereas when $\rho=1$, the shapes of $\sum_{m\in[M]} (p_m \star f_{k,m}) (t)$'s exhibit substantial distinctions across clusters.
In essence, as $\rho$ increases, the clusters become more separable.

\setlength{\tabcolsep}{13pt}
\begin{table}
\caption{ True values of $\{f_{k,m}(t): k\in[K], m\in[M]\}$ in Scenario~2. 
The parameter $\rho$ controls the distinctiveness across clusters, whose value is altered in the experiment.
The functions $q_1(t)$ and $q_2(t)$ are defined in~\eqref{eq:true_f_Nclus1},
$h_1(x) \equiv |\max(x, 0)|^{1/2}$, and $h_2(x) \equiv 1 + \min(x, 0)$.
}
\label{table:true_f_Nclus4} 
\centering
\begin{tabular}{ *{3}{c} }
\hline
$ $ &
${m=1}$ &
${m=2}$ 
\\
\hline
${k=1}$ &
$52.5 \times q_1(t) $ &
$52.5 \times q_2(t) $
\\[10pt]
${k=2}$ &
\begin{tabular}{@{}l@{}}
$60 \times \left[ 1- h_1(2\rho-1) \right] \times q_1(t) $
\\ 
$~ + 48 \times h_2(2\rho-1) \times q_2 \left( 2[t-0.8] \right) $
\end{tabular}
  &
\begin{tabular}{@{}l@{}}
$60 \times \left[ 1 + h_1(2\rho-1) \right] \times q_2(t)$
\\ 
$\quad - 48 \times h_2(2\rho-1)  \times q_2(2t)$ 
\end{tabular}
\vspace*{10pt}
\\
${k=3}$ &
$67.5 \times (1+0.5\rho) \times q_1(t) $ &
$67.5 \times (1-0.5\rho) \times q_2(t) $
\\[10pt]
${k=4}$ &
$75  \times (1+\rho) \times q_1(t)$ &
$75 \times (1-\rho) \times q_2(t)$
\\
\hline
\end{tabular}
\end{table}

The third experiment is a continuation of the second experiment, where our focus is directed towards evaluating the clustering performance of the proposed method. 
We manipulate signal strength by varying the values of variables $R$, $n$, and $\tau$. 
Through this manipulation, we aim to investigate the impact of $R$, $n$, and $\tau$ on the performance of clustering estimation.

\subsection{Intensity estimation performance}

In the first experiment, we investigate the effect of varying $R$, $n$ and $\tau$ on the intensity estimation performance of the proposed method.
When $K=1$, the value of $\gamma$ does not affect the estimation result, hence can be set to zero. 
We let \(\ell_0=10\), and \(\epsilon=0.005\) in all experiments.
A sensitivity analysis concerning \(\ell_0\) demonstrates the robustness of estimation results to changes in \(\ell_0\), as detailed in Section~\ref*{ssec:sensitivity_freqtrun} of the Supplementary Material.

We evaluate the intensity estimation performance via  the mean integrated squared error (MISE).
The MSIE is defined as:
\begin{align}
\text{MISE} \equiv \frac{1}{M}
\sum_{m\in [M]} 
d \left\{\hat{f}'_{m}(t), \frac{f^*_{m}(t)}{\Lambda^*_{}} \right\}.
\end{align}
where $d\{f_1,f_2\} \equiv {\min}_{v \in [-T,T]} \left\| S^v f_1  - f_2  \right\|^2$.
It is worth noting that the definition of MISE considers $\hat{\mathbf{f}}'$ rather than $\hat{\mathbf{f}}$. 
We excludes $\hat{{\Lambda}}$ from the evaluation criterion since its performance, as a sample mean, is well-studied. 

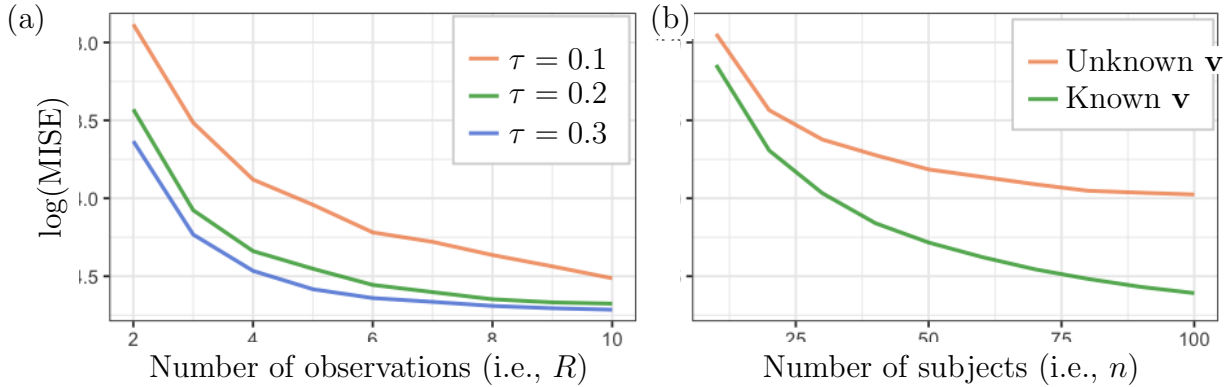
\begin{figure}
\centering
\begin{tikzpicture}
    \node[below right, inner sep=0] (image_12) at (0,0) 
    	{
    		\includegraphics[width=0.5\textwidth]
			{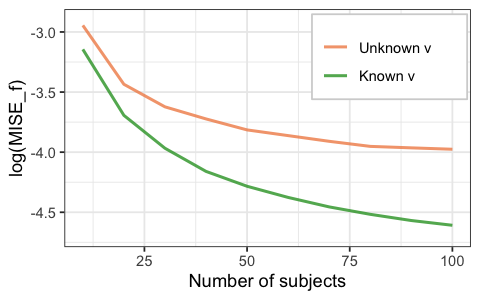}
		};
	\begin{scope}[
		shift={(image_12.south west)},
	    x={($0.1*(image_12.south east)$)},
	    y={($0.1*(image_12.north west)$)}]
		\node[above, fill=white, yshift=-2pt, xshift=12pt, minimum width=4cm] at (5,0){ Number of subjects (i.e., $n$)};
		\node[right, xshift=-3pt, yshift=0.3cm, fill=white, minimum height=3.5cm, minimum width=1.05cm] at (image_12.west){  };
		\node[right, fill=white, yshift=-0pt,  xshift=14pt, inner sep=1] at (6.7,8.5)
		{ Unknown $\mathbf{v}$};
		\node[right, fill=white, yshift=-14pt,  xshift=14pt, inner sep=1] at (6.7,8.5)
		{ {Known $\mathbf{v}$}};
		\node[below, fill=white, inner sep=1] at (1,10){ (b) };
	\end{scope}
	\node[below left, xshift=15pt, inner sep=0] (image_11) at (image_12.north west) 
    	{
    		\includegraphics[width=0.5\textwidth]
			{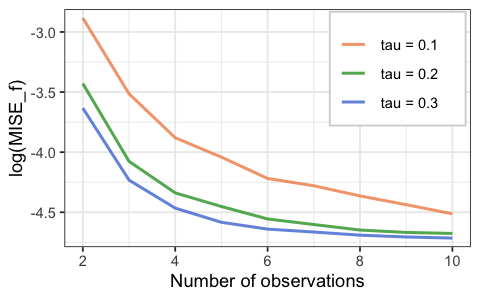}
		};
	\begin{scope}[
		shift={(image_11.south west)},
	    x={($0.1*(image_11.south east)$)},
	    y={($0.1*(image_11.north west)$)}]
	\node[above, fill=white, yshift=-2pt, xshift=12pt, minimum width=4cm] at (5,0){ Number of observations (i.e., $R$)};
	\node[below, fill=white, yshift=-2pt, rotate = 90, xshift=13pt, minimum width=4cm] at (0,5){ $\log$(MISE)};
	\node[below, fill=white, yshift=-2pt,  xshift=13pt, inner sep=1] at (8,9){ $\tau = 0.1$};
	\node[below, fill=white, yshift=-15pt,  xshift=13pt, inner sep=1] at (8,9){ $\tau = 0.2$};
	\node[below, fill=white, yshift=-30pt,  xshift=13pt, inner sep=1] at (8,9){ $\tau = 0.3$};
	\node[below, fill=white, inner sep=1] at (0,10){ (a) };
	\end{scope}
\end{tikzpicture}
\caption{
Intensity estimation performance in Experiment 1 with $5000$ replicates.
Synthetic data is generated with varying $R$, $n$, and $\tau$. 
MISE is shown in log scale for better visualization, where smaller values indicate better performances.
Panel (a) shows the performance of intensity estimation with varying values of $R$ and $\tau$.
Panel (b) demonstrates the performance of intensity estimation with varying $n$. The curve labeled ``Unknown $\mathbf{v}$'' shows results when the algorithm is not provided with the true value of $\mathbf{v}$, while the curve labeled ``Known $\mathbf{v}$'' depicts results when the algorithm is provided with the true value of $\mathbf{v}$.
}
\label{fig:intensity_estimation}
\end{figure}

The intensity estimation performance is shown in Figure~\ref{fig:intensity_estimation}. 
Firstly, the MISE rapidly improves as $R$ increases, because when $R$ increases, each subject is associated with more samples, while the number of unknown parameters remain constant. 
Secondly,  a decrease in MISE is observed as $\tau$ increases.
This is because when $\tau$ is small, there is a potential non-identifiability issue due to limited sample size.
As $\tau$ increases, the variance of $w^*_{r,m}$'s increases, thereby alleviating the non-identifiability issue.
Thirdly, the MISE exhibits a decreasing trend with an increase in $n$, since $n$ serves as the sample size for the estimation of intensity components.
However, it is noteworthy that the MISE decreases slower in response to an increase in $n$ compared to an increase in $R$. 
This is because an increment in $n$ leads to a proportional increase in the quantity of unknown subject-specific time shifts (i.e., $v_{i,m}$'s). 
Consequently, when the algorithm is provided with the true values of $v_{i,m}$'s, the MISE shows a significant reduction.

\subsection{Comparison with relevant methods}
\label{sec:compare_relevant_method}
In the second experiment, we compare the clustering performance of {the proposed ASIMM~\eqref{model:full}} with the kCFC~\citep{Chiou2007} and the k-mean alignment~\citep{Sangalli2010a}.
We apply proposed method with \(K=4\), \(\gamma=0.01\), where the selection of \(\gamma\) follows the procedure outlined in Section~\ref{subsec:optimization_full}.
We apply the kCFC by employing the implementation provided in the \texttt{R} package \emph{fdapace}~\citep{fdapace}, specifically the function named ``kCFC''.
The parameters for this implementation are set as follows: the desired number of clusters is specified as $4$, and the maximum number of principal components is set to $2$. Additionally, we specify the type of design as "dense", and set the maximum number of iterations to $30$.
We employ the k-mean alignment by utilizing the implementation available in the \texttt{R} package \emph{fdasrvf}~\citep{fdasrvf}, specifically the function named ``kmeans\_align''.
In configuring the algorithm parameters, we specify the desired number of clusters specified to \(4\), the maximum number of iterations specified to $50$, and the minimum number of curves per cluster to $2$.
Both the kCFC and the k-mean alignment are applied on the empirical intensities aggregated across observations, which can be expressed as  $y_{i}(t) \equiv \{N_{i}(t+\Delta t) - N_{i}(t)\} \Delta t^{-1}$, where $N_i(t) \equiv  R^{-1} \sum_{r\in[R]}{N_{i,r}(t)}$.

We evaluate the cluster estimation performance via the Adjusted Rand Index (ARI) \citep{Hubert1985}.
Let $\boldsymbol{\mathcal{C}}^* \equiv \{\mathcal{C}^*_{k}: k\in[K]\}$,
$\hat{\boldsymbol{\mathcal{C}}} \equiv \{\hat{\mathcal{C}}_{k'}: {k'}\in[K']\}$ denote the set of true clusters and the set of estimated clusters, where $K$ and $K'$ are the true number of clusters and specified number of clusters.
The ARI is formally defined as 
\begin{align}
\mathrm{ARI}
\equiv
\frac{
	\sum_{k,k'} \binom{d_{k,k'}}{2}
	-\left[ 
	\sum_{k} \binom{b_k}{2}
	\sum_{k'} \binom{c_{k'}}{2}
	\right] 
	\binom{n}{2}^{-1} }
{\frac{1}{2}
	\left[ 
	\sum_{k} \binom{b_k}{2}
	+ \sum_{k'} \binom{c_{k'}}{2}
	\right]
	- \left[ 
	\sum_{k} \binom{b_k}{2}
	\sum_{k'} \binom{c_{k'}}{2}
	\right] 
	\binom{n}{2}^{-1} } ,
\label{eq:ARI}
\end{align}
where $b_k \equiv |\mathcal{C}^*_k|$, $c_{k'} \equiv |\hat{\mathcal{C}}_{k'}|$, and $d_{k,k'} \equiv |\mathcal{C}^*_{k} \cap \hat{\mathcal{C}}_{k'}|$ for $k\in[K]$ and $k'\in[K^\prime]$. 
The ARI quantifies the similarity between $\boldsymbol{\mathcal{C}}^*$ and $\hat{\boldsymbol{\mathcal{C}}}$.
For instance, when $\boldsymbol{\mathcal{C}}^*$ and $\hat{\boldsymbol{\mathcal{C}}}$ are equivalent up to a permutation of cluster labels, the ARI is equal to one.
Conversely, when $\hat{\boldsymbol{\mathcal{C}}}$ is entirely random, the ARI has a mean value of zero.

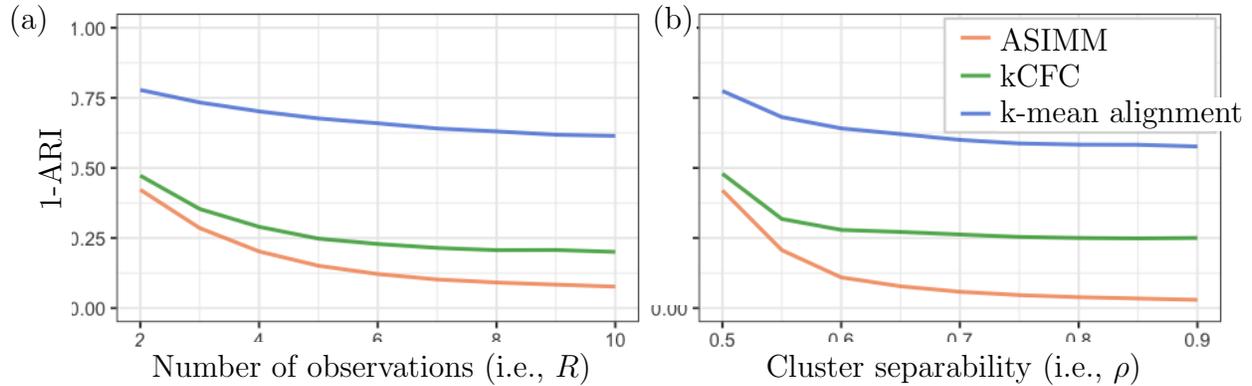
\begin{figure}
\centering
\begin{tikzpicture}
	\node[below right, xshift=0pt, inner sep=0] (image_2) at (0,0) 
    	{
    		\includegraphics[width=0.5\textwidth]
			{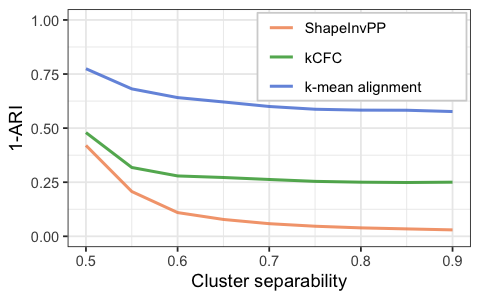}
		};
	\begin{scope}[
		shift={(image_2.south west)},
	    x={($0.1*(image_2.south east)$)},
	    y={($0.1*(image_2.north west)$)}]
		\node[above, fill=white, yshift=-2pt, xshift=12pt, minimum width=4cm] at (5,0){ Cluster separability (i.e., $\rho$) };
		\node[right, xshift=-2.5pt, yshift=0.3cm, fill=white, minimum height=3.5cm, minimum width=1.09cm] at (0,5){  };
		\node[below, fill=white, inner sep=1] at (1,10){ (b) };
	\end{scope}
	\begin{scope}[
		shift={(image_2.south west)},
	    x={($0.1*(image_2.south east)$)},
	    y={($0.1*(image_2.north west)$)}]
		\node[right, fill=white, inner sep=1] at (6.2,9){ {ASIMM~~~} };
		\node[right, fill=white, inner sep=1] at (6.2,8.1){ kCFC };
		\node[right, fill=white, inner sep=1] at (6.2,7.1){ k-mean alignment };
	\end{scope}
	\node[below left, xshift=15pt, inner sep=0] (image_1) at (image_2.north west) 
    	{
    		\includegraphics[width=0.5\textwidth]
			{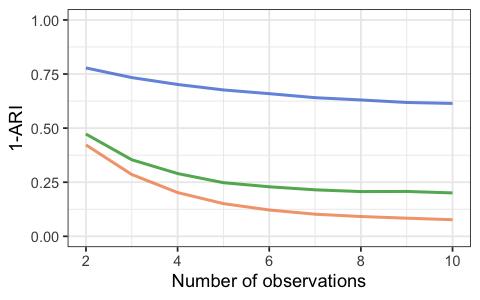}
		};
	\begin{scope}[
		shift={(image_1.south west)},
	    x={($0.1*(image_1.south east)$)},
	    y={($0.1*(image_1.north west)$)}]
		\node[above, fill=white, yshift=-2pt, xshift=12pt, minimum width=4cm] at (5,0){ Number of observations (i.e., $R$)};
		\node[below, fill=white, yshift=-2pt, rotate = 90, xshift=12pt, minimum width=4cm] at (0,5){ 1-ARI};
		\node[below, fill=white, inner sep=1] at (0,10){ (a) };
	\end{scope}
	
\end{tikzpicture}
\caption{
Clustering performance in Experiment 2 with 5000 replicates of our proposal in orange, kCFC by \cite{Chiou2007} in green, and k-mean alignment by \cite{Sangalli2010a} in blue. 
Synthetic data is generated with $n = 40$, $\tau = 0.1$, varying $R$ and $\rho$.
In panel (a), the value of $\rho$ is fixed as $\rho = 0.5$. In panel (b), the value of $R$ is fixed as $R=2$.
}
\label{fig:compare_methods}
\end{figure}

Figure~\ref{fig:compare_methods} shows the clustering performance of the  three considered methods.
Across all depicted scenarios in Figure~\ref{fig:compare_methods}, it is evident that the proposed method consistently outperforms both kCFC and k-mean alignment. 
The superiority of proposed method becomes especially clear when the values of $R$ and $\rho$ are large.
This is expected since the proposed method excels in handling data with additive intensity components and time shifts.
In contrast, kCFC and k-mean alignment were not devised to handle the setting in this experiment.
In essence, the proposed method serves as a valuable complement to existing approaches for functional clustering.

\subsection{Cluster estimation performance}
In the third experiment, we investigate the effect of $R$, $n$ and $\tau$ on the cluster estimation performance of the proposed method.
We apply proposed method with the same set of tuning parameters as in the second experiment.

\begin{figure}
\centering
\begin{tikzpicture}
    \node[below right, inner sep=0] (image_12) at (0,0) 
    	{
    		\includegraphics[width=0.5\textwidth]
			{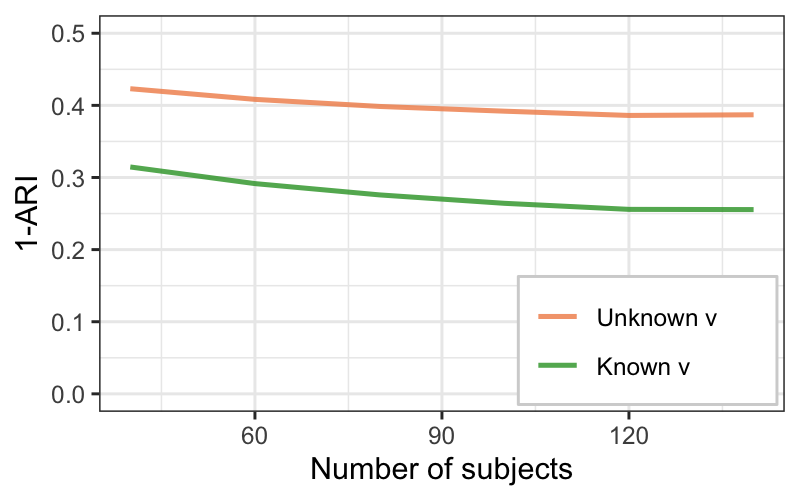}
		};
	\begin{scope}[
		shift={(image_12.south west)},
	    x={($0.1*(image_12.south east)$)},
	    y={($0.1*(image_12.north west)$)}]
		\node[above, fill=white, yshift=-2pt, xshift=12pt, minimum width=4cm] at (5,0){ Number of subjects (i.e., $n$)};
		\node[right, xshift=-3pt, yshift=0.3cm, fill=white, minimum height=3.5cm, minimum width=0.99cm] at (image_12.west){  };
		\node[right, fill=white, yshift=-0pt,  xshift=14pt, inner sep=1] at (6.7,3.8){ Unknown $\mathbf{v}$};
		\node[right, fill=white, yshift=-15pt,  xshift=14pt, inner sep=1] at (6.7,3.8){ {Known $\mathbf{v}$}};
		\node[below, fill=white, inner sep=1] at (0.8,10){ (b) };
	\end{scope}
	\node[below left, xshift=8pt, inner sep=0] (image_11) at (image_12.north west) 
    	{
    		\includegraphics[width=0.5\textwidth]
			{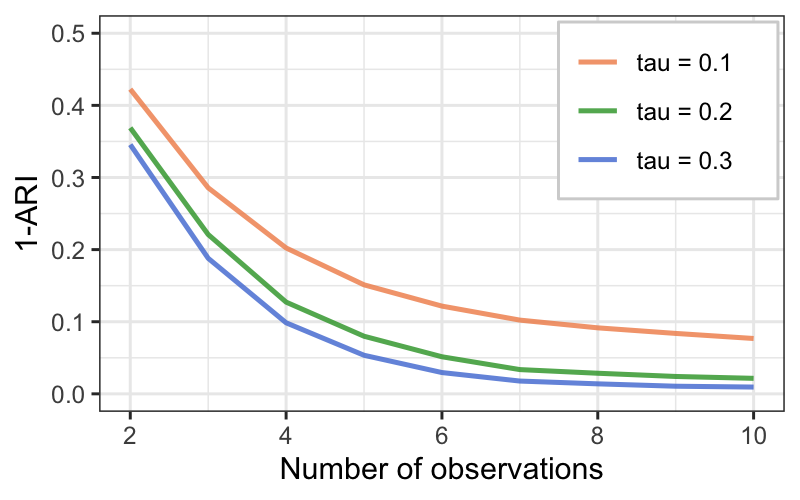}
		};
	\begin{scope}[
		shift={(image_11.south west)},
	    x={($0.1*(image_11.south east)$)},
	    y={($0.1*(image_11.north west)$)}]
	\node[above, fill=white, yshift=-2pt, xshift=12pt, minimum width=4cm] at (5,0){ Number of observations (i.e., $R$)};
	\node[below, fill=white, yshift=-2pt, rotate = 90, xshift=12pt, minimum width=4cm] at (0,5){ 1-ARI};
	\node[below, fill=white, yshift=-2pt,  xshift=14pt, inner sep=1] at (8,9.2){ $\tau = 0.1$};
	\node[below, fill=white, yshift=-16pt,  xshift=14pt, inner sep=1] at (8,9.2){ $\tau = 0.2$};
	\node[below, fill=white, yshift=-29pt,  xshift=14pt, inner sep=1] at (8,9.2){ $\tau = 0.3$};
	\node[below, fill=white, inner sep=1] at (0,10){ (a) };
	\end{scope}
\end{tikzpicture}
\caption{
Clustering performance in Experiment 3 with $5000$ replicates.
Synthetic data is generated under the setting with varying $R$, $n$, and $\tau$.
Panel (a) shows the clustering performance as $R$ and $\tau$ increases, where the value of $n$ is fixed as $n = 40$. 
Panel (b) shows the clustering performance as $n$ increases, where the values of $R$ and $\tau$ are fixed as $R = 2$ and $\tau = 0.1$.
The curve labeled ``Unknown $\mathbf{v}$'' shows results when the algorithm is not provided with the true value of $\mathbf{v}$, while the curve labeled ``Known $\mathbf{v}$'' depicts results when the algorithm is provided with the true value of $\mathbf{v}$.
}
\label{fig:ARI_Nclus4}
\end{figure}
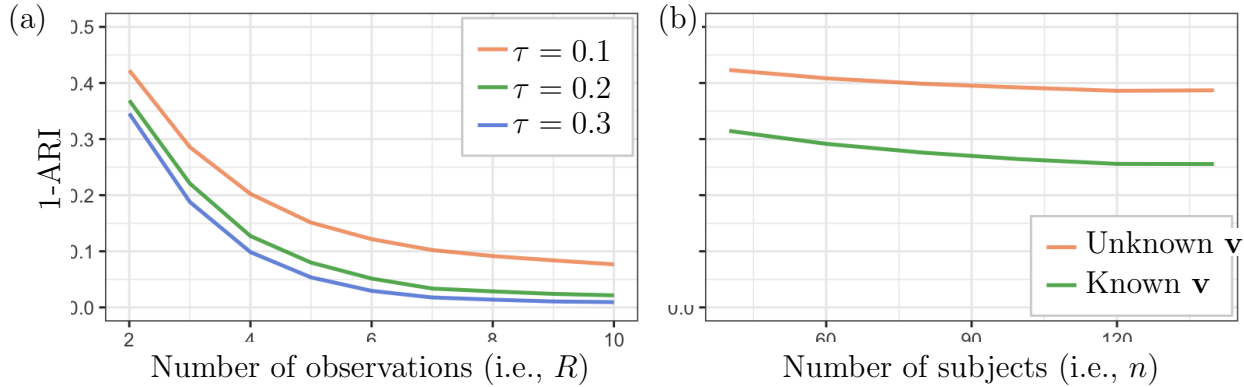

The clustering performance is displayed in Figure~\ref{fig:ARI_Nclus4}.
It is evident that the clustering performance improves as $R$, $\tau$, and $n$ increases.
This is because $R$, $\tau$, and $n$ help in estimating the intensity components, as demonstrated in the first experiment, which in turn improves the cluster estimation.
Furthermore, $R$ serves as sample size for cluster memberships, thereby contributing to improved clustering performance.
However, the impact of increasing $n$ on clustering performance is only marginal. 
This is because as $n$ increases, the number of unknown cluster memberships also increases, meaning that $n$ does not serve as the sample size for cluster memberships.

\section{Real data application} \label{sec:real_data}

We consider the neural data during visual discrimination tasks from \cite{Steinmetz2019}.
In each trial, the mouse encountered a sequential presentation of two stimuli. 
The first stimulus comprised visual gratings of varying contrasts displayed on two screens, one to the left and one to the right of the mouse. 
The second stimulus was an auditory tone cue which was set off after a randomized delay between 0.4 to 0.8 seconds after the onset of the first stimulus.
The mouse could rotate a wheel which, after the auditory cue, would move the visual gratings.
When one contrast is higher than the other, the mouse succeeded and gained rewards if the visual grating of higher contrasts was moved to the center screen. 
The complete criteria for success are detailed in {Table~\ref*{stable:experiment_conditions}} of the Supplementary Material. 
Throughout the experiment, researchers simultaneously recorded firing activities of hundreds of neurons in the \textit{left} hemisphere of the mouse's brain using Neuropixels \citep{jun2017fully,steinmetz2018challenges,steinmetz2021neuropixels}.
We aim to identifying groups of neurons with distinct responses to the two stimuli using our proposed method.

\begin{figure}[ht]
\centering
\begin{tikzpicture}
    \node[below right,inner sep=0] (image_10) at (0,0) 
        {\includegraphics[width=0.33\textwidth, trim=0 0 0 0, clip]
            {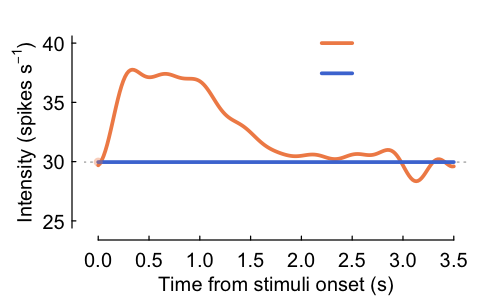}
        }; 
    \begin{scope}[
        shift={($(image_10.south west)$)},
        x={($0.1*(image_10.south east)$)},
        y={($0.1*(image_10.north west)$)}]
        \draw[fill=white, draw=white] (0,2) rectangle (0.8, 9);
        \node[below, rotate=90] at ($(0,5.5)$) {\scriptsize Intensity (spikes/s)};
        \draw[fill=white, draw=white] (2,0) rectangle (9, 1);
        \node[above] at ($(5.7,-0.3)$) {\scriptsize Time from visual stimulus onset (s)};
        \node[above right] at ($(0,9)$) {\footnotesize (1a)};
        \node[right] at ($(7.2,8.6)$) {\tiny $\hat{a}_{1}+\hat{f}_{1,1}$};
        \node[right] at ($(7.2,7.6)$) {\tiny  $\hat{a}_{1}+\hat{f}_{1,2}$};
    \end{scope}  
    \node[below right,inner sep=0] (image_11) at (image_10.north east) 
    {\includegraphics[width=0.33\textwidth, trim=0 0 0 0, clip]
        {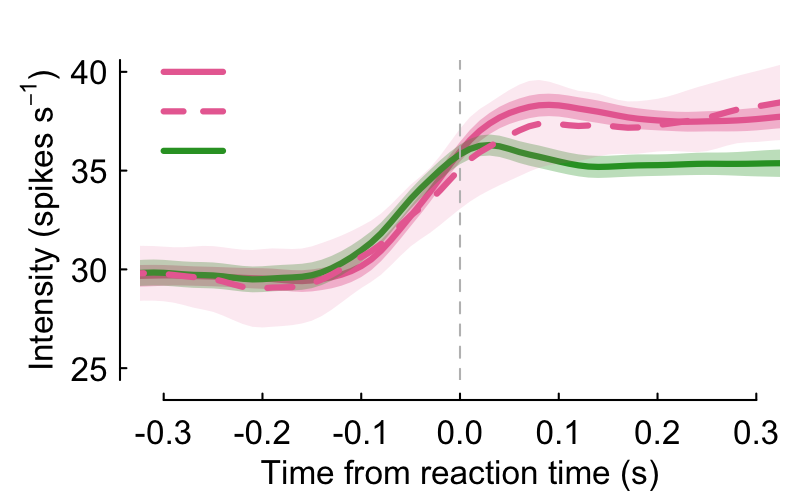}
    }; 
    \begin{scope}[
        shift={($(image_11.south west)$)},
        x={($0.1*(image_11.south east)$)},
        y={($0.1*(image_11.north west)$)}]
        \draw[fill=white, draw=white] (0,2) rectangle (0.8, 9);
        \draw[fill=white, draw=white] (2,0) rectangle (9, 1);
        \node[above] at ($(5.7,-0.3)$) {\scriptsize Time from movement onset (s)};
        \node[right] at ($(2.8,8.5)$) {\tiny L,L};
        \node[right] at ($(2.8,7.8)$) {\tiny R,L};
        \node[right] at ($(2.8,7.0)$) {\tiny R,R};
        \node[above right] at ($(0,9)$) {\footnotesize (1b)};
    \end{scope}  
    \node[below right,inner sep=0] (image_12) at (image_11.north east) 
    {\includegraphics[width=0.33\textwidth, trim=0 0 0 0, clip]
        {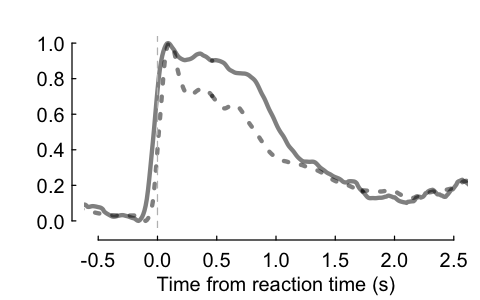}
    };  
    \begin{scope}[
        shift={($(image_12.south west)$)},
        x={($0.1*(image_12.south east)$)},
        y={($0.1*(image_12.north west)$)}]
        \node[below, rotate=90] at ($(0,5.5)$) {\scriptsize Standardized value};
        \draw[fill=white, draw=white] (2,0) rectangle (9, 1);
        \node[above] at ($(5.7,-0.3)$) {\scriptsize Time from movement onset (s)};
        \draw[very thick, black, opacity=0.8] ($(6.0,8.3)$) -- ($(6.6,8.3)$) ;
        \node[right] at ($(6.5,8.3)$) {\tiny Intensity};
        \draw[very thick, black, dotted, opacity=0.8] ($(6.0,7.5)$) -- ($(6.6,7.5)$) ;
        \node[right] at ($(6.5,7.5)$) {\tiny Wheel velocity};
        \node[above right] at ($(0,9)$) {\footnotesize (1c)};
    \end{scope}   
    \node[below right = 0.3 and 0 of image_10.south west,inner sep=0] (image_20) at (image_10.south west) 
        {\includegraphics[width=0.33\textwidth, trim=0 0 0 0, clip]
            {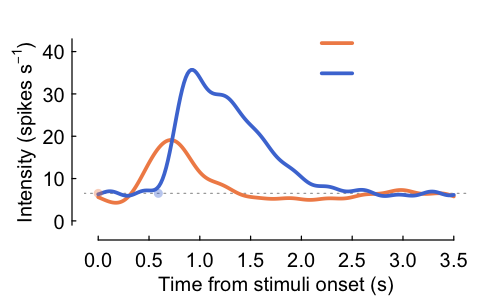}
        };  
    \begin{scope}[
        shift={($(image_20.south west)$)},
        x={($0.1*(image_20.south east)$)},
        y={($0.1*(image_20.north west)$)}]
        \draw[fill=white, draw=white] (0,2) rectangle (0.8, 9);
        \node[below, rotate=90] at ($(0,5.5)$) {\scriptsize Intensity (spikes/s)};
        \draw[fill=white, draw=white] (2,0) rectangle (9, 1);
        \node[above] at ($(5.7,-0.3)$) {\scriptsize Time from visual stimulus onset (s)};
        \node[right] at ($(7.2,8.6)$) {\tiny $\hat{a}_{2}+\hat{f}_{2,1}$};
        \node[right] at ($(7.2,7.6)$) {\tiny  $\hat{a}_{2}+\hat{f}_{2,2}$};
        \node[above right] at ($(0,9)$) {\footnotesize (2a)};
    \end{scope}   
    \node[below right,inner sep=0] (image_21) at (image_20.north east) 
    {\includegraphics[width=0.33\textwidth, trim=0 0 0 0, clip]
        {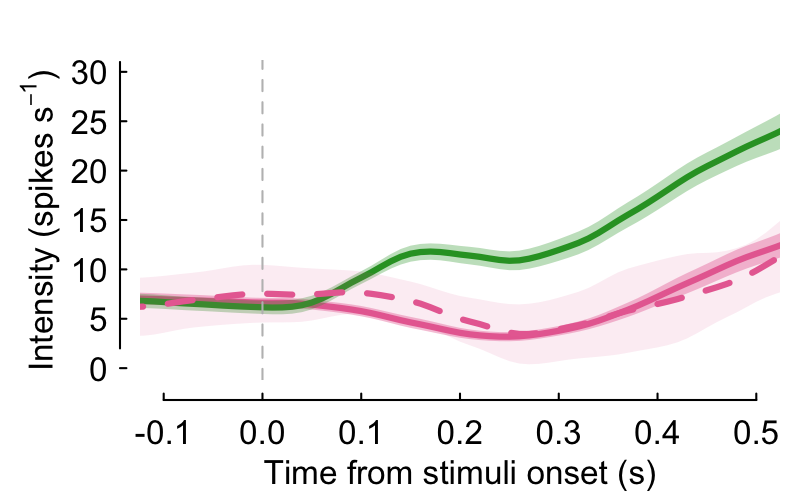}
    }; 
    \begin{scope}[
        shift={($(image_21.south west)$)},
        x={($0.1*(image_21.south east)$)},
        y={($0.1*(image_21.north west)$)}]
        \draw[fill=white, draw=white] (0,2) rectangle (0.8, 9);
        \draw[fill=white, draw=white] (2,0) rectangle (9, 1);
        \node[above] at ($(5.7,-0.3)$) {\scriptsize Time from visual stimulus onset (s)};
        \node[above right] at ($(0,9)$) {\footnotesize (2b)};
    \end{scope}
    \node[below right,inner sep=0] (image_22) at (image_21.north east) 
    {\includegraphics[width=0.33\textwidth, trim=0 0 0 0, clip]
        {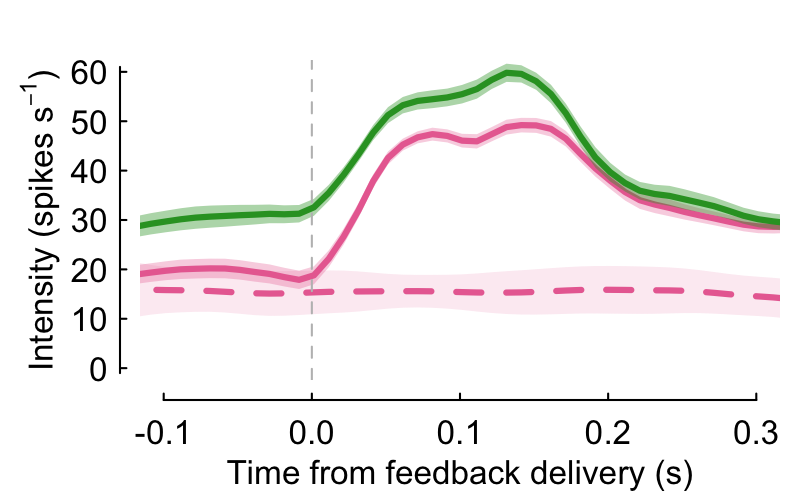}
    }; 
    \begin{scope}[
        shift={($(image_22.south west)$)},
        x={($0.1*(image_22.south east)$)},
        y={($0.1*(image_22.north west)$)}]
        \draw[fill=white, draw=white] (0,2) rectangle (0.8, 9);
        \draw[fill=white, draw=white] (2,0) rectangle (9, 1);
        \node[above] at ($(5.7,-0.3)$) {\scriptsize Time from feedback delivery (s)};
        \node[above right] at ($(0,9)$) {\footnotesize (2c)};
    \end{scope}
    \node[below right = 0.3 and 0 of image_20.south west,inner sep=0] (image_30) at (image_20.south west) 
        {\includegraphics[width=0.33\textwidth, trim=0 0 0 0, clip]
            {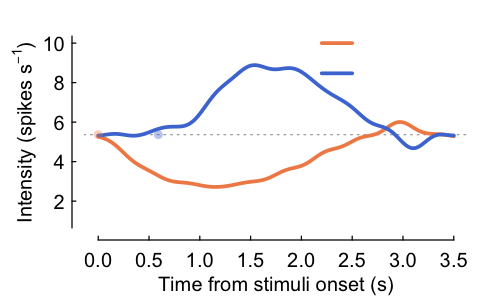}
        };  
    \begin{scope}[
        shift={($(image_30.south west)$)},
        x={($0.1*(image_30.south east)$)},
        y={($0.1*(image_30.north west)$)}]
        \draw[fill=white, draw=white] (0,2) rectangle (0.8, 9);
        \node[below, rotate=90] at ($(0,5.5)$) {\scriptsize Intensity (spikes/s)};
        \draw[fill=white, draw=white] (2,0) rectangle (9, 1);
        \node[above] at ($(5.7,-0.3)$) {\scriptsize Time from visual stimulus onset (s)};
        \node[right] at ($(7.2,8.6)$) {\tiny $\hat{a}_{3}+\hat{f}_{3,1}$};
        \node[right] at ($(7.2,7.6)$) {\tiny  $\hat{a}_{3}+\hat{f}_{3,2}$};
        \node[above right] at ($(0,9)$) {\footnotesize (3a)};
    \end{scope}   
    \node[below right,inner sep=0] (image_31) at (image_30.north east) 
    {\includegraphics[width=0.33\textwidth, trim=0 0 0 0, clip]
        {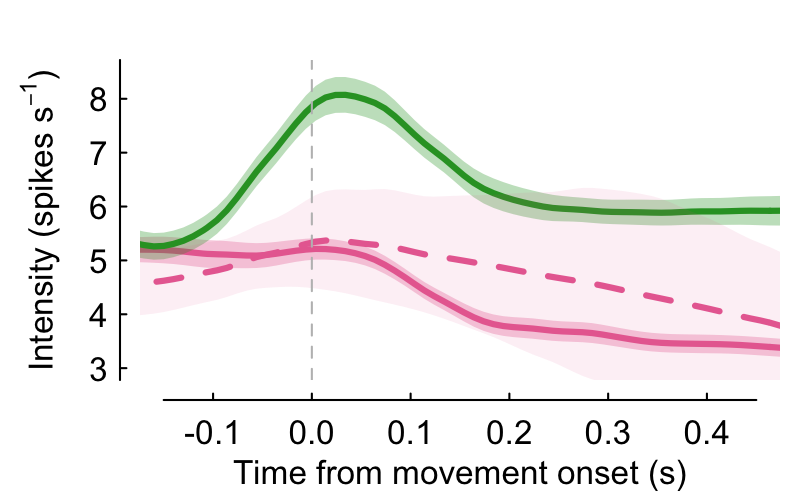}
    }; 
    \begin{scope}[
        shift={($(image_31.south west)$)},
        x={($0.1*(image_31.south east)$)},
        y={($0.1*(image_31.north west)$)}]
        \draw[fill=white, draw=white] (0,2) rectangle (0.8, 9);
        \draw[fill=white, draw=white] (2,0) rectangle (9, 1);
        \node[above] at ($(5.7,-0.3)$) {\scriptsize Time from movement onset (s)};
        \node[above right] at ($(0,9)$) {\footnotesize (3b)};
    \end{scope}
    \node[below right,inner sep=0] (image_32) at (image_31.north east) 
    {\includegraphics[width=0.33\textwidth, trim=0 0 0 0, clip]
        {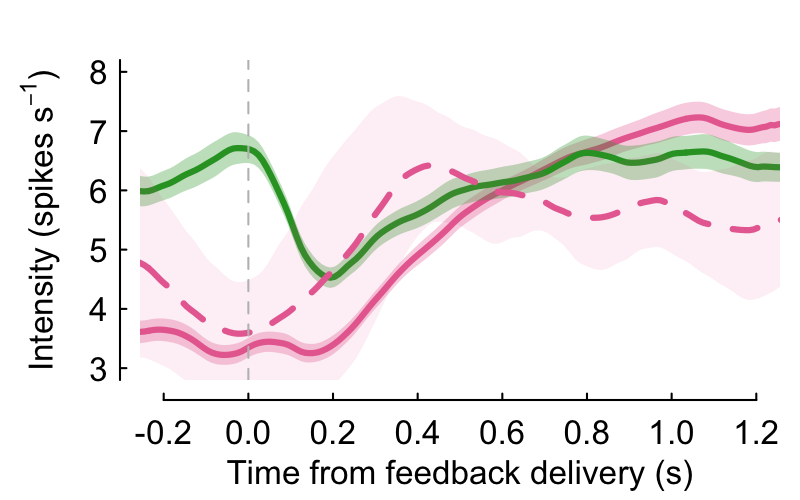}
    };
    \begin{scope}[
        shift={($(image_32.south west)$)},
        x={($0.1*(image_32.south east)$)},
        y={($0.1*(image_32.north west)$)}]
        \draw[fill=white, draw=white] (0,2) rectangle (0.8, 9);
        \draw[fill=white, draw=white] (2,0) rectangle (9, 1);
        \node[above] at ($(5.7,-0.3)$) {\scriptsize Time from feedback delivery (s)};
        \node[above right] at ($(0,9)$) {\footnotesize (3c)};
    \end{scope}    
\end{tikzpicture}
\caption{ 
Estimation of intensity components and average firing intensities in various conditions.
Each row corresponds to one estimated cluster. 
The first column presents the estimated intensity components.
The second and third columns display the average neural firing intensities from both the training trials and other  trials that are not used to fit the model.
The shaded area represents  the mean firing intensities plus or minus two standard errors of the mean.
The legend in panel~(1b) represents ``scenario, choice'', for instance, ``R,L'' represents the trials where the right grating was of a higher contrast, and the mouse chose to move the left grating to the center screen resulting in a failure in that trial.
Panel~(1c) illustrates the average firing intensity and wheel velocity, where both the firing intensity and the wheel velocity are standardized to range from~0 to~1 for alignment.
The condition ``L,R'' is omitted in this figure since there are only three trials, but its firing intensity can be found in Figure~\ref*{sfig:firing_intensity_LR} of the Supplementary Material.
}
\label{fig:RDA}
\end{figure}
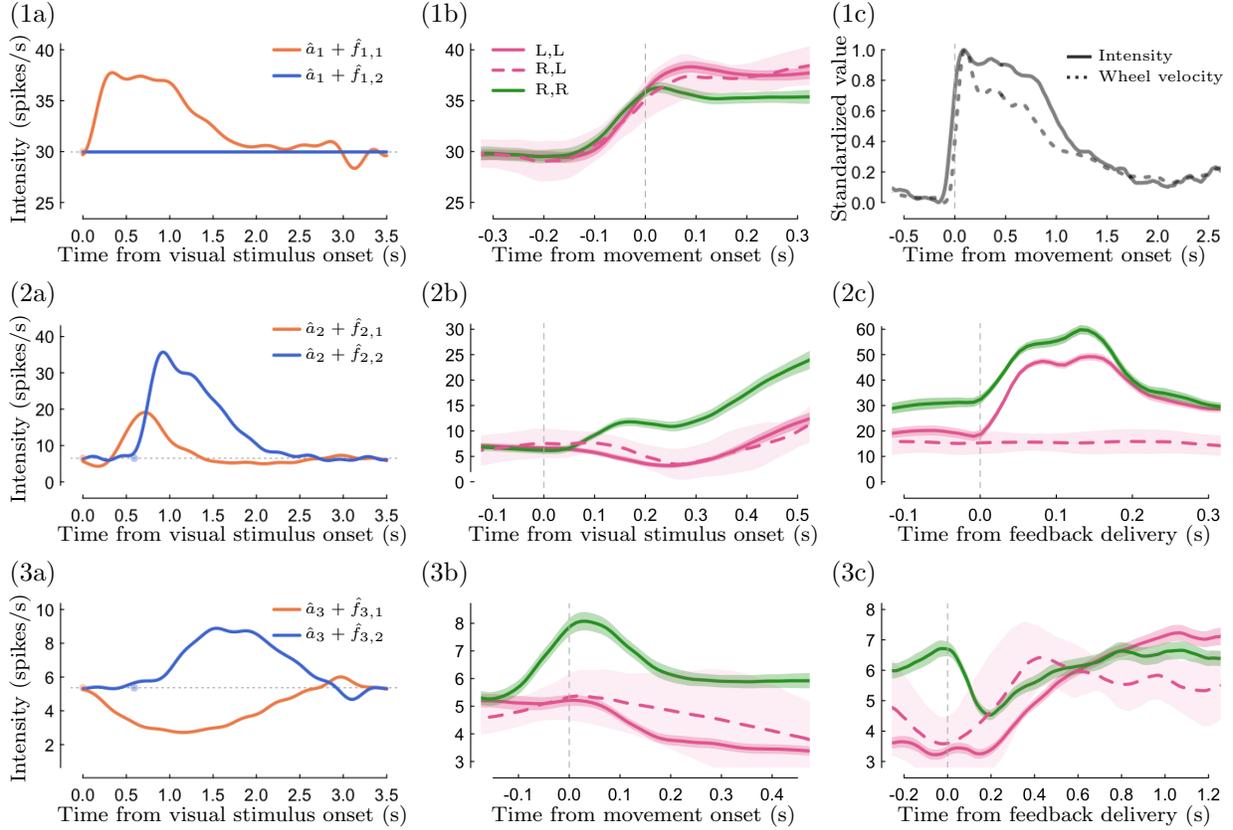

Following the notation in \eqref{model:full}, we index trials using $r \in \{1,\ldots,R\}$ and neurons using  $i \in \{1,\ldots,n\}$. 
Consequently, the firing activities of neuron $r$ in trial $i$ is $N_{i,r}(t), t \in [0,T]$ where $T=3.5$.
We set $m=1$ for the visual grating and $m=2$ for the auditory tone cue.
We further denote the onset time of two stimuli in trial $r$ as $w^*_{r,1}$ and $w^*_{r,2}$. 
Neurons might exhibit firing latencies in response to each stimulus denoted as $v_{i,m}$ for $m\in \{1,2\}$ and $i \in \{1,\ldots,n\}$. 
To demonstrate the usage of proposed method, we focus on $R=102$ experimental trials where the left visual grating was of higher contrast and the mouse successfully gained rewards. 
We study $n = 225$ neurons in the midbrain region, where we remove neurons with fewer than one spike per trial on average.
We apply the proposed algorithm with $K = 3$, $\gamma = 10^{-4}$,  $\ell_0 = 10$ and $\epsilon = 0.005$.
Using the proposed method, we identify three clusters of neurons with distinct responses to the two stimuli shown in Fig~\ref{fig:RDA}. 
The first column of Figure~\ref{fig:RDA} contains the refined intensity components.
To better understand the roles of each clusters, we illustrate the average firing rates from the training trials and the trials that are not used to fit the model in the second and third columns of Figure~\ref{fig:RDA}. 
More details of this analysis can be found in Section~\ref*{ssec:supp_data_analysis} in Supplementary Material.

Recalling that the first stimulus is the visual gratings and the second stimulus is the auditory cue, we have the following observations. 
\begin{enumerate}
    \item[Cluster 1.] There seems to be only one non-zero component in Cluster 1. 
    It is immediately clear from Figure~\ref{fig:RDA}(1c) that the firing rates of neurons in Cluster 1 are highly in sync with the wheel velocity. 
    When aligning the firing rates of neurons in Cluster 1 by movement onset in Figure~\ref{fig:RDA}(1b), we can see that the firing rates share almost the same trajectory across conditions, but their peaks depend on the choice the mouse made. 
    In particular, the firing rate has a higher peak when the mouse chose the left visual grating. 
    This preference, also known as laterality, is likely due to the fact that these neurons are from the left hemisphere.
    We hypothesize that neurons in Cluster 1 are responsible for executing the turning of the wheel. 
    The crucial role of midbrain in coordinating movement has been identified in prior studies \citep[see, e.g., ][]{boecker2008role,coddington2019learning,inagaki2022midbrain}.
    
    \item[Cluster 2.]  The first intensity component exhibits a decrease-then-increase pattern post stimuli onset, and the second component shows a sharp increase after the auditory cue, as shown in  Figure~\ref{fig:RDA}(2a). 
    Figure~\ref{fig:RDA}(2b) shows that the decrease-then-increase pattern seems common in trials when the mouse chose the left visual grating.
    There is, however, no period of suppression in trials when the mouse chose the right visual grating. 
    When aligning the firing rates by feedback delivery time, it is clear that the firing rates peaked almost immediately after rewards delivery regardless of the choice, and, in the absence of rewards, the firing rates remain stationary. 
    We hypothesize that neurons in Cluster 2 might respond to perceptions of stimuli (e.g., visual gratings, rewards), and their activities are further regulated by laterality.
    Similar firing patterns of neurons in midbrain have been identified in prior studies \citep[see, e.g., ][]{coddington2018timing,Steinmetz2019}.

    \item[Cluster 3.] The estimated first component exhibits suppressed activities throughout the trial, and the second component shows increased activities after the auditory cue in Figure~\ref{fig:RDA}(3a). 
    Neural responses in Cluster 3 bear a resemblance to those in Cluster 2 that they both peaked after reward delivery, and they both show preference to contralateral choices (i.e., choosing the right visual grating), as shown in Figures~\ref{fig:RDA}(3b)~and~\ref{fig:RDA}(3c). 
    Unlike Cluster 2 where rewards trigger immediate response,  there is a one-second delay from reward delivery till the firing rates peaked in Cluster~3 in Figure~\ref{fig:RDA}(3c). 
    It might be possible that neurons in Cluster 3 are involved in the initialization of the next trial, when the mouse needs to hold the wheel stationary after reward delivery. 
    Furthermore, Figure~\ref{fig:RDA}(3b) shows that the suppression of activities started after the movement onset in trials with ipsilateral choices, but activities increased before the movement onset in trials with contralateral choices.    
\end{enumerate}

\definecolor{darkgreen}{rgb}{0,0.5,0}
\definecolor{myorange}{RGB}{243, 160, 98}
\definecolor{myblue}{RGB}{60, 86, 240}

\section{Discussion}
\label{sec:discussion}

In this article, we tackle the problem of simultaneously decomposing, aligning, and clustering recurrent event data.
We note that there are a few directions that can be explored in future works.  First, the proposed method assumes that all subjects exhibit consistent intensity across observations, which may not hold true in cases where the subjects might respond randomly. 
For instance, neurons may exhibit periods of reduced responsiveness during tasks known as 
\emph{local sleep}~\citep[see, e.g.,][]{krueger2008sleep,vyazovskiy2011local,vyazovskiy2013sleep}. 
To address this limitation, we can generalize the proposed model by incorporating a hidden Markov model for the responsive status \citep{tokdar2010detection,escola2011hidden,mews2023markov}.
Second, the proposed method only considers observations from the same experimental condition.
A natural extension is to allow observations from different experimental conditions.
For instance, it would be beneficial to learn neural functions based on their firing patterns in various visual conditions and feedback types. 
To achieve this, we can incorporate a cluster structure of observations by employing bi-clustering models \citep{madeira2004biclustering,slimen2018model,galvani2021funcc}.
Third, theoretical properties of the proposed estimators have not been investigated. 
Previous studies have established the consistency of the k-means clustering~\citep{pollard1981strong,sun2012regularized} and clustering based on FPCA \citep{yin2021row}.
In addition, theoretical properties of simultaneous registration and clustering models have been established in~\cite{tang2023model}. 
Building on previously established framework \citep{pollard1981strong,yin2021row, tang2023model}, it may be possible to incorporate the theory of shape-invariant models \citep{Bigot2013,Bigot2013a} to analyze the theoretical properties of the proposed method.

\bibliographystyle{plainnat} 
\bibliography{reference.bib}       

\appendix

\begin{center}
	\huge
	Supplementary Materials
\end{center}


\section{Proof of Proposition~\ref*{prop:identifiability_simplified}}
\label{ssec:proof_prop1}

Suppose $({a}, \mathbf{f}, \mathbf{v}) \in \Theta_0$ are such that for all $t \in [0,T], \mathbf{w}^* \in [0,W]^M, i\in[n]$ the following equality holds:
\begin{align}
\label{seq:tilde_a_sum_over_m}
\begin{split}
& {a}^* + \sum_{m\in[M]} S^{{v}^*_{i,m}+w^*_{m}} {f}^*_{m} (t)
= a + \sum_{m\in[M]} S^{v_{i,m}+w^*_{m}} f_{m} (t).
\end{split}
\end{align}
Noting that  $T_0+V+W \leq T$ by Assumption~\ref*{assump:no_censorship_s}, the equation in~\eqref{seq:tilde_a_sum_over_m} can be formulated in the frequency domain as follows: for $\xi \neq 0$, $\mathbf{w}^* \in [0,W]^M$, $i\in[n]$,
\begin{align}
\label{seq:sum_m_exp_phi}
\sum_{m\in[M]} \exp \left\{-\operatorname{j} 2 \pi \xi ({v}^*_{i,m}+w^*_{m}) \right\} {\phi}^*_{m}(\xi) 
= \sum_{m\in[M]} \exp \left\{-\operatorname{j} 2 \pi \xi (v_{i,m}+w^*_{m}) \right\} \phi_{m}(\xi),
\end{align}
where ${\phi}^*_{m}(\xi)$ and $\phi_{m}(\xi)$ are the Fourier coefficients of ${f}^*_{m}(t)$ and $f_{m}(t)$ at frequency $\xi$.

We first show an intermediate result that $\exp \left\{-\operatorname{j} 2 \pi \xi ({v}^*_{i,m}-v_{i,m}) \right\}
 {\phi}^*_{m}(\xi) 
=  \phi_{m}(\xi)$.
By employing matrix notations,~\eqref{seq:sum_m_exp_phi} can be written as, for $\xi \neq 0$, $\mathbf{w}^* \in [0,W]^M$, $i\in[n]$,
\begin{align}
\label{seq:eta_star_t_phi_star_sq_phi_nostar}
\boldsymbol{\eta^*}(\xi)^\top {\boldsymbol{\psi}}^*_i(\xi) 
= \boldsymbol{\eta^*}(\xi)^\top \boldsymbol{\psi}_i(\xi),
\end{align}
where
\begin{align}
\begin{split}
{\boldsymbol{\eta}}^*(\xi) \equiv
\left[ \exp \left\{-\operatorname{j} 2 \pi \xi w^*_{1} \right\}
~~ \cdots ~~ \exp \left\{-\operatorname{j} 2 \pi \xi w^*_{M} \right\}
\right]^\top
\end{split},
\\
\label{seq:def_phi_star_xi_1}
\begin{split}
{{\boldsymbol{\psi}}^*_i}(\xi) \equiv
\left[\exp \left\{-\operatorname{j} 2 \pi \xi {v}^*_{i,1} \right\} {\phi}^*_{1}(\xi) 
~~ \cdots ~~ 
\exp \left\{-\operatorname{j} 2 \pi \xi {v}^*_{i,M} \right\} {\phi}^*_{M}(\xi) 
\right]^\top
\end{split},
\\ 
\label{seq:def_phi_nostar_xi_1}
\begin{split}
{\boldsymbol{\psi}_i}(\xi) \equiv
\left[\exp \left\{-\operatorname{j} 2 \pi \xi v_{i,1} \right\} \phi_{1}(\xi) 
~~ \cdots ~~ 
\exp \left\{-\operatorname{j} 2 \pi \xi v_{i,M} \right\} \phi_{M}(\xi) 
\right]^\top
\end{split}.
\end{align}
From Assumption~\ref*{assump:time_shifts_different_s}, we know that $\mathbb{E}[\overline{\boldsymbol{\eta}^*(\xi)} {\boldsymbol{\eta}^*(\xi)}^\top]$ is invertible for $\xi \in \mathbb{R} \setminus \{0\}$.
Therefore, we know from~\eqref{seq:eta_star_t_phi_star_sq_phi_nostar} that, for $\xi \neq 0$, $i\in[n]$,
\begin{align}
\label{seq:psi_star_eq_psi_nostar_1}
{\boldsymbol{\psi}}^*_i(\xi) = \boldsymbol{\psi}_i(\xi).
\end{align}
Substituting the definitions of ${\boldsymbol{\psi}}^*_i(\xi)$ and $\boldsymbol{\psi}_i(\xi)$ in~\eqref{seq:def_phi_star_xi_1} and~\eqref{seq:def_phi_nostar_xi_1} into~\eqref{seq:psi_star_eq_psi_nostar_1}, we obtain that, for $\xi \neq 0$, $i\in[n]$, $m\in[M]$,
\begin{align}
\label{seq:exp_phi_star}
\exp \left\{-\operatorname{j} 2 \pi \xi ({v}^*_{i,m}-v_{i,m}) \right\}
 {\phi}^*_{m}(\xi) 
=  \phi_{m}(\xi) .
\end{align}

Now we show that, for $m\in[M]$ such that the set $\{t: f^*_{m}(t) \neq 0\}$ is of positive measure,
the subject-specific time shifts are identifiable up to a constant, i.e., ${v}^*_{i, m} - v_{i,m} = c_{m}$ for $i \in [n]$, where $c_m$ is a constant independent of $i$. 
Based on~\eqref{seq:exp_phi_star}, it can be derived that for any $\xi \neq 0$, $m\in[M]$, $i,i'\in[n]$,
\begin{align}
\label{seq:exp_v_diff_phi}
\exp \left\{-\operatorname{j} 2 \pi \xi ({v}^*_{i,m}-v_{i,m}) \right\}
 {\phi}^*_{m}(\xi) 
=  \exp \left\{-\operatorname{j} 2 \pi \xi ({v}^*_{i',m}-v_{i',m}) \right\}
 {\phi}^*_{m}(\xi).
\end{align}
Consider function $g(t) \equiv S^{{v}^*_{i, m} - v_{i,m}} {f}^*_{m} (t)- S^{{v}^*_{i', m} - v_{i',m}} {f}^*_{m}(t)$.
From~\eqref{seq:exp_v_diff_phi} we know that the Fourier coefficient of $g(t)$ is zero at frequency $\xi \neq 0$.
Moreover, from the definition of the set $\mathcal{F}$, we know that the function $g(t)$ has a bounded support.
By applying Lemma~\ref{lemma:mA_mB} to the function $g(t)$, we deduce that $g(t)=0$ almost everywhere.
Consequently, for any $i, i' \in [n]$ and $m \in [M]$, 
\begin{align}
\label{seq:shifted_f_star_diff_0_ae}
S^{{v}^*_{i, m} - v_{i,m}} {f}^*_{m}(t)
- S^{{v}^*_{i', m} - v_{i',m}} {f}^*_{m}(t) = 0\quad \text { a.e. }
\end{align}
or equivalently,
\begin{align}
\label{seq:S_v_star_minus_v_nostar_f_star_eq_f_star}
S^{({v}^*_{i, m} - v_{i,m})-({v}^*_{i', m} - v_{i',m})} {f}^*_{m}(t)
= {f}^*_{m}(t), \quad \text { a.e. }
\end{align}
The equation in~\eqref{seq:S_v_star_minus_v_nostar_f_star_eq_f_star} implies that, for $m\in[M]$ such that the set $\{t: f^*_{m}(t) \neq 0\}$ is of positive measure, $({v}^*_{i, m} - v_{i,m})-({v}^*_{i', m} - v_{i',m}) = 0$ for any $i,i'\in[n]$.
Consequently, for $i\in[n]$,
\begin{align}
\label{seq:v_start_diff_eq_cm}
{v}^*_{i, m}-v_{i,m}  = c_{m},
\end{align}
where $c_m$ is a constant independent of $i$.
In other words, Statement~\ref*{prop_res:v_s} is proved.

Next we show that, for $m\in[M]$, $S^{c_{m}}{f}^*_{m}(t) = f_{m}(t)$ for almost every $t \in \mathbb{R}$.
Plugging~\eqref{seq:v_start_diff_eq_cm} into~\eqref{seq:exp_phi_star}, we obtain that, for $m\in[M]$ such that the set $\{t: f^*_{m}(t) \neq 0\}$ is of positive measure and $\xi \neq 0$,
\begin{align}
\label{seq:exp_j_xi_cm_phi_star}
\exp \left\{-\operatorname{j} 2 \pi \xi c_{m} \right\}
 \phi^*_{m}(\xi) 
=  \phi_{m}(\xi) . 
\end{align}
Applying the inverse Fourier transformation to~\eqref{seq:exp_j_xi_cm_phi_star} yields
\begin{align}
\label{seq:S_cm_f_star}
S^{c_{m}}f^*_{m}(t) = f_{m}(t) + c \quad \text {a.e.},
\end{align}
where $c \in \mathbb{R}$ is a constant.
From the definition of $\mathcal{F}$, we know that 
\begin{align}
\label{seq:f_star_m_t_eq_0}
f^*_{m}(t) = f_{m}(t) = 0, \text{ for } t \in \mathbb{R} \setminus [0,T_0].
\end{align}
Combining~\eqref{seq:S_cm_f_star} and~\eqref{seq:f_star_m_t_eq_0}, we can derive that $c=0$.
Inserting the value of $c$ to~\eqref{seq:S_cm_f_star} yields that, for  $m\in[M]$ such that the set $\{t: f^*_{m}(t) \neq 0\}$ is of positive measure,
\begin{align}
\label{seq:S_cm_f_star_eq_f}
   S^{c_{m}}f^*_{m}(t) = f_{m}(t) \quad \text {a.e.}
\end{align}
Notably, for $m\in[M]$ such that $f^*_{m} = 0$ almost everywhere,~\eqref{seq:S_cm_f_star_eq_f} always holds. 
Thus~\eqref{seq:S_cm_f_star_eq_f} holds for all $m\in[M]$, in other words, Statement~\ref*{prop_res:f_s} is proved.

Finally, substituting~\eqref{seq:S_cm_f_star_eq_f} and~\eqref{seq:v_start_diff_eq_cm}  into~\eqref{seq:tilde_a_sum_over_m}, we derive that, 
\begin{align}
\label{seq:alpha_star_alpha}
a^* = a.
\end{align}
In other words, Statement~\ref*{prop_res:a_s} is proved. \hfill $\square$

\section{Proof of Proposition~\ref*{prop:identifiability}}
\label{ssec:proof_prop2}
Suppose there exist $(\mathbf{z}, \boldsymbol{a}, \mathbf{f}, \mathbf{v}) \in \Theta_1$ such that for all  $t \in [0,T]$, $\mathbf{w}^* \in [0,W]^M$, $i\in[n]$, the following equation holds:
\begin{align}
\label{seq:a_star_z_star_i}
\begin{split}
a^*_{z^*_i} + \sum_{m\in[M]} S^{v^*_{i,m}+w^*_{m}} f^*_{z^*_i,m} (t)
= a_{z_i} + \sum_{m\in[M]} S^{v_{i,m}+w^*_{m}} f_{z_i,m} (t). 
\end{split}
\end{align}
Using $T_0+V+W \leq T$ by Assumption~\ref*{assump:no_censorship_s},~\eqref{seq:a_star_z_star_i} can be formulated in the frequency domain as follows: for $\xi \neq 0$, $\mathbf{w}^* \in [0,W]^M$, $i\in[n]$,
\begin{align}
\label{seq:sum_m_exp_phi_zi}
\sum_{m\in[M]} \exp \left\{-\operatorname{j} 2 \pi \xi (v^*_{i,m}+w^*_{m}) \right\} \phi^*_{z^*_i,m}(\xi) 
= \sum_{m\in[M]} \exp \left\{-\operatorname{j} 2 \pi \xi (v_{i,m}+w^*_{m}) \right\} \phi_{z_i,m}(\xi),
\end{align}
where $\phi^*_{k,m}(\xi)$ and $\phi_{k,m}(\xi)$ are the Fourier coefficients of $f^*_{k,m}$ and $f_{k,m}$ at frequency $\xi$.

We first show an intermediate result that $\exp \left\{-\operatorname{j} 2 \pi \xi ({v}^*_{i,m}-v_{i,m}) \right\}
 {\phi}^*_{z^*_i,m}(\xi) 
=  \phi_{z_i,m}(\xi)$.
By employing matrix notations,~\eqref{seq:sum_m_exp_phi_zi} can be written as
\begin{align}
\boldsymbol{\eta^*}(\xi)^\top \boldsymbol{\psi}_i^*(\xi) = \boldsymbol{\eta^*}(\xi)^\top \boldsymbol{\psi}_i(\xi),
\end{align}
where
\begin{align}
\begin{split}
{\boldsymbol{\eta}}^*(\xi) \equiv
\left[ \exp \left\{-\operatorname{j} 2 \pi \xi w^*_{1} \right\}
~~ \cdots ~~ \exp \left\{-\operatorname{j} 2 \pi \xi w^*_{M} \right\}
\right]^\top
\end{split},
\\
\label{seq:def_phi_star_xi}
\begin{split}
{\boldsymbol{\psi}_i}^*(\xi) \equiv
\left[\exp \left\{-\operatorname{j} 2 \pi \xi v^*_{i,1} \right\} \phi^*_{z^*_i,1}(\xi) 
~~ \cdots ~~ 
\exp \left\{-\operatorname{j} 2 \pi \xi v^*_{i,M} \right\} \phi^*_{z^*_i,M}(\xi) 
\right]^\top
\end{split},
\\ 
\label{seq:def_phi_nostar_xi}
\begin{split}
{\boldsymbol{\psi}_i}(\xi) \equiv
\left[\exp \left\{-\operatorname{j} 2 \pi \xi v_{i,1} \right\} \phi_{z_i,1}(\xi) 
~~ \cdots ~~ 
\exp \left\{-\operatorname{j} 2 \pi \xi v_{i,M} \right\} \phi_{z_i,M}(\xi) 
\right]^\top
\end{split}.
\end{align}
From Assumption~\ref*{assump:time_shifts_different_s} we know that $\mathbb{E}[\overline{\boldsymbol{\eta}^*(\xi)} {\boldsymbol{\eta}^*(\xi)}^\top]$ is invertible for $\xi \in \mathbb{R} \setminus \{0\}$.
Therefore, we have that, for $\xi \neq 0$, 
\begin{align}
\label{seq:psi_star_eq_psi_nostar}
\boldsymbol{\psi}^*_i(\xi) = \boldsymbol{\psi}_i(\xi), 
\end{align}
Substituting the definitions of ${\boldsymbol{\psi}}^*_i(\xi)$ and $\boldsymbol{\psi}_i(\xi)$ in~\eqref{seq:def_phi_star_xi} and~\eqref{seq:def_phi_nostar_xi} into~\eqref{seq:psi_star_eq_psi_nostar}, we obtain that, for $\xi \neq 0$, $i\in[n]$, $m\in[M]$,
\begin{align}
\label{seq:exp_phi_star_zi_star}
\exp \left\{-\operatorname{j} 2 \pi \xi (v^*_{i,m}-v_{i,m}) \right\}
 \phi^*_{z^*_i,m}(\xi) 
=  \phi_{z_i,m}(\xi) .
\end{align}

Now we show that cluster memberships are identifiable up to a permutation of cluster labels, i.e., $z_i = \sigma(z^*_i)$ where $\sigma: [K] \to [K]$ is a permutation of $[K]$.
To achieve this, we prove the following two statements: (i) $z_i = z_{i'} \Rightarrow z^*_i = z^*_{i'}$; and (ii) $z_i \neq z_{i'} \Rightarrow z^*_i \neq z^*_{i'}$.
First, based on~\eqref{seq:exp_phi_star_zi_star}, we can derive that for $i,i'\in[n]$ such that $z_i=z_{i'}$, $\xi \neq 0$, and $m\in[M]$,
\begin{align}
\label{seq:exp_v_diff_phi_zi}
\exp \left\{-\operatorname{j} 2 \pi \xi (v^*_{i,m}-v_{i,m}) \right\}
 \phi^*_{z^*_i,m}(\xi) 
=  \exp \left\{-\operatorname{j} 2 \pi \xi (v^*_{i',m}-v_{i',m}) \right\}
 \phi^*_{z^*_{i'},m}(\xi) . 
\end{align}
Consider a function $g(t)$ defined as $g(t) \equiv S^{v^*_{i, m} - v_{i,m}} {f}^*_{z^*_i,m} (t)- S^{v^*_{i', m} - v_{i',m}} {f}^*_{z^*_{i'},m}(t)$.
From~\eqref{seq:exp_v_diff_phi_zi} we know that the Fourier transform of $g(t)$ is zero for $\xi \neq 0$.
Moreover, from the definition of $\mathcal{F}$ we know that the function $g(t)$ has a bounded support.
Applying Lemma~\ref{lemma:mA_mB} to $g(t)$, we deduce that $g(t)=0$ almost everywhere. 
Consequently, for any $i, i' \in [n]$ that $z_i = z_{i'}$ and $m \in [M]$, the following equation holds:
\begin{align}
\label{seq:shifted_f_star_diff_0_ae}
S^{v^*_{i, m} - v_{i,m}} {f}^*_{z^*_i,m}(t)
- S^{v^*_{i', m} - v_{i',m}} {f}^*_{z^*_{i'},m}(t) = 0,\quad \text { a.e. }
\end{align}
According to Assumption~\ref*{assump}, if $z^*_i \neq z^*_{i'}$, then there exists $m_0\in[M]$ such that  for any $x \in \mathbb{R}$, $\{t\in \mathbb{R}: S^x f^*_{z^*_i,m_0}(t) \neq f^*_{z^*_{i'},m_0}(t) \}$ has a positive measure.
Therefore, in order for~\eqref{seq:shifted_f_star_diff_0_ae} to hold, we must have $z^*_i = z^*_{i'}$. In other words,
\begin{align}
\label{seq:z_eq_to_z_star_eq_ii}
z_i = z_{i'} \Rightarrow z^*_i = z^*_{i'} .
\end{align}

Second, suppose there exist $i_0,i_1$ such that $z_{i_0} \neq z_{i_1}$ and $z^*_{i_0} = z^*_{i_1}$.
Given that there are $K$ non-empty clusters by definition, it is always possible to find indices $i_2,\ldots,i_{K}$ such that $z^*_{i_1},\ldots,z^*_{i_{K}}$ are pairwise distinct.
Based on~\eqref{seq:z_eq_to_z_star_eq_ii}, it follows that $z_{i_1},\ldots,z_{i_{K}}$ are also pairwise distinct. 
Since $z_{i_0} \neq z_{i_1}$ by definitions of $i_0$ and $i_1$, there must exist $k \in [K]\setminus\{1\}$ such that $z_{i_0} = z_{i_k}$, which, according to~\eqref{seq:z_eq_to_z_star_eq_ii}, implies $z^*_{i_0} = z^*_{i_k}$.
Thus, by definitions of $i_0$ and $i_1$, we have $z^*_{i_0} = z^*_{i_1} = z^*_{i_k}$.
This contradicts with the definition of $i_2,\ldots,i_K$, which asserts that $z^*_{i_1}$ is different from $z^*_{i_2},\ldots,z^*_{i_K}$.
Consequently, such $i_0$ and $i_1$ cannot exist. Hence we know that, for all $i,i'\in[n]$,
\begin{align}
\label{seq:z_star_neq_to_z_eq_ii}
z_i \neq z_{i'} \Rightarrow z^*_i \neq z^*_{i'} .
\end{align}
Combining~\eqref{seq:z_star_neq_to_z_eq_ii} and~\eqref{seq:z_eq_to_z_star_eq_ii}, we obtain that $z^*_i = z^*_{i'}$ if and only if $z_i = z_{i'}$.
This further implies that there exists a permutation of $[K]$, denoted by $\sigma: [K] \to [K]$, such that for $i\in[n]$,
\begin{align}
\label{seq:z_i_eq_z_i_star}
 z_i = \sigma(z^*_i).
\end{align}
In other words, Statement~\ref*{prop_res:z} is proved.

Now we show that, for $k\in[K],m\in[M]$ such that the set $\{t: f^*_{k,m}(t) \neq 0\}$ is of positive measure, the subject-specific time shifts are identifiable up to a constant, i.e., $v^*_{i, m} - v_{i,m} = c_{k,m}$ for $i \in \{i:z^*_i = k\}$, where $c_{k,m}$ is a constant independent of $i$.
Plugging~\eqref{seq:z_i_eq_z_i_star} into~\eqref{seq:shifted_f_star_diff_0_ae}, we have that, for $i,i'\in[n]$ such that $z^*_i = z^*_{i'} = k$ and $m\in[M]$,
\begin{align}
\label{seq:S_v_star_diff_f_star_km}
S^{v^*_{i, m} - v_{i,m}} {f}^*_{k,m}(t)
= S^{v^*_{i', m} - v_{i',m}} {f}^*_{k,m}(t), \quad \text { a.e. }
\end{align}
From the definition of $\mathcal{F}$ we know that $f^*_{k,m}(t)$ has a bounded support for $k\in[K]$ and $m\in[M]$. 
Therefore,~\eqref{seq:S_v_star_diff_f_star_km} implies that, for $k\in[K]$ and $m\in[M]$ such that the set $\{t: f^*_{k,m}(t) \neq 0\}$ is of positive measure, and $i,i'\in[n]$ such that $z^*_i = z^*_{i'}$,
\begin{align}
\label{seq:v_start_diff_eq_v_diff_im}
v^*_{i, m}-v^*_{i', m} = v_{i,m}-v_{i',m}. 
\end{align}
It follows from~\eqref{seq:v_start_diff_eq_v_diff_im} that
\begin{align}
\label{seq:v_star_diff_const}
v^*_{i, m} - v_{i,m} = c_{k,m},
\end{align}
where $c_{k,m} \in \mathbb{R}$ is a constant independent of $i$. 
In other words, Statement~\ref*{prop_res:v} is proved.

Next we show that $S^{c_{k,m}}f^*_{k,m}(t) = f_{\sigma(k),m}(t)$ for almost every $t \in \mathbb{R}$.
Plugging~\eqref{seq:v_star_diff_const} and~\eqref{seq:z_i_eq_z_i_star} into~\eqref{seq:exp_phi_star_zi_star}, we obtain that, for $k\in[K]$ and $m\in[M]$ such that the set $\{t: f^*_{k,m}(t) \neq 0\}$ is of positive measure and $\xi \neq 0$,
\begin{align}
\label{seq:exp_j_xi_cm_phi_star_2}
\exp \left\{-\operatorname{j} 2 \pi \xi c_{k,m} \right\}
 \phi^*_{k,m}(\xi) 
=  \phi_{\sigma(k),m}(\xi) , 
\end{align}
Applying the inverse Fourier transformation to~\eqref{seq:exp_j_xi_cm_phi_star_2} yields
\begin{align}
\label{seq:S_ckm_f_star}
S^{c_{k,m}}f^*_{k,m}(t) = f_{\sigma(k),m}(t) + c \quad \text {a.e.},
\end{align}
where $c \in \mathbb{R}$ is a constant.
From the definition of $\mathcal{F}$, we know that 
\begin{align}
\label{seq:f_star_km_t_eq_0}
f^*_{k,m}(t) = f_{k,m}(t) = 0, \text{ for } t \in \mathbb{R} \setminus [0,T_0].
\end{align}
Combining~\eqref{seq:S_ckm_f_star} and~\eqref{seq:f_star_km_t_eq_0}, we can derive that $c=0$.
Inserting the value of $c$ to~\eqref{seq:S_ckm_f_star} yields that, for $k\in[K]$ and $m\in[M]$ such that  the set $\{t: f^*_{k,m}(t) \neq 0\}$ is of positive measure, 
\begin{align}
\label{seq:S_ckm_f_star_eq_f_sigma_k}
   S^{c_{k,m}}f^*_{k,m}(t) = f_{\sigma(k),m}(t) \quad \text {a.e.}
\end{align}
Notably, for $k\in[K]$ and $m\in[M]$ such that $f^*_{k,m} = 0$ almost everywhere,~\eqref{seq:S_ckm_f_star_eq_f_sigma_k} always holds. 
In other words,~\eqref{seq:S_ckm_f_star_eq_f_sigma_k} holds for all $k\in[K]$ and $m\in[M]$.
Therefore, Statement~\ref*{prop_res:f} is proved.

Finally, substituting~\eqref{seq:S_ckm_f_star_eq_f_sigma_k},~\eqref{seq:v_star_diff_const} and~\eqref{seq:z_i_eq_z_i_star} into~\eqref{seq:a_star_z_star_i}, we derive that, for $k\in[K]$,
\begin{align}
\label{seq:alpha_star_alpha}
a^*_{k} = a_{\sigma(k)}.
\end{align}
In other words, Statement~\ref*{prop_res:alpha}  is proved. \hfill $\square$

\section{Connection between the additive shape invariant model and the functional principal component analysis (FPCA)}
\label{ssec:connection_FPCA}
When the variances of $v_{i,m}$'s and $w^*_{r,m}$'s are both close to zero, the proposed model in~(\ref*{model:simplified_2}) can be approximated using the Taylor expansion:
\begin{align}
\label{seq:lambda_approx_fpca}
\begin{split}
\lambda_{i, r}(t) 
&= a + 
\sum_{m \in[M]} S^{u_{i,r,m}} f_{m}(t)  
\\
& = a + 
\sum_{m \in[M]} f_{m} \left( \{t-\mathbb{E}u_{i,r,m}\}-\{u_{i,r,m}-\mathbb{E}u_{i,r,m}\} \right)
\\
&\approx 
a + \sum_{m \in[M]} \left\{ f_{m}(t-\mathbb{E}u_{i,r,m}) - (u_{i,r,m}-\mathbb{E}u_{i,r,m}) Df_{m}(t-\mathbb{E}u_{i,r,m}) \right\}
\\
&= 
a + \sum_{m \in[M]} f_{m}(t-\mathbb{E}u_{i,r,m}) 
- \sum_{m \in[M]} (u_{i,r,m}-\mathbb{E}u_{i,r,m}) Df_{m}(t-\mathbb{E}u_{i,r,m}) 
,
\end{split}
\end{align}
where $u_{i,r,m} \equiv v_{i,m} + w^*_{r,m}$, 
and $Df_m$ denotes the first order derivative of $f_m$.
In~\eqref{seq:lambda_approx_fpca}, the first equality follows from the definition of $u_{i,r,m}$, the second equality follows from the definition of the shift operator, the approximation in the third line follows from the Taylor expansion.
To elucidate the connection between the proposed additive shape invariant model and the FPCA, we define a new set of parameters:
\begin{align}
\label{seq:def_mu}
\mu(t) 
& \equiv a + \sum_{m \in[M]} f_{m}(t-\mathbb{E}u_{i,r,m}) ,
\\ 
\label{seq:def_zeta_irm}
\zeta_{i,r,m} 
& \equiv -(u_{i,r,m}-\mathbb{E}u_{i,r,m}) \left\| Df_{m}(t-\mathbb{E}u_{i,r,m}) \right\|_t, 
\\ 
\label{seq:def_psi_m}
\psi_m(t) 
& \equiv Df_{m}(t-\mathbb{E}u_{i,r,m}) \left\| Df_{m}(t-\mathbb{E}u_{i,r,m}) \right\|_t^{-1}.
\end{align}
By definitions of $\zeta_{i,r,m}$ and $\psi_m(t)$ in~\eqref{seq:def_zeta_irm} and~\eqref{seq:def_psi_m}, we know that $\mathbb{E} \zeta_{i,r,m}  = 0$, and $\|\psi_m(t)\|_t=1$.
Using the new set of parameters in~\eqref{seq:def_mu},~\eqref{seq:def_zeta_irm} and~\eqref{seq:def_psi_m}, the model approximation in~\eqref{seq:lambda_approx_fpca} can be expressed as follows:
\begin{align}
\label{seq:lambda_mu_sum_zeta_psi}
\lambda_{i, r}(t) 
& \approx 
\mu(t) + \sum_{m \in[M]} \zeta_{i,r,m} ~\psi_m(t)
.
\end{align}
When $\{f_m(t - \mathbb{E}u_{i,r,m}): m\in[M]\}$ have non-overlapping supports, it follows that $\psi_m(t)$'s are mutually orthogonal.
Consequently, the approximate model in~\eqref{seq:lambda_mu_sum_zeta_psi} is an FPCA model, where $\psi_1(t),\cdots,\psi_M(t)$ correspond to the first $M$ eigenfunctions, and $\zeta_{i,r,1},\ldots,\zeta_{i,r,M}$ correspond to the principal components associated with the eigenfunctions.

\section{Detailed derivation of solutions in~(\ref*{eq:hat_phi_prime_kl}) and~(\ref*{eq:hat_phi_km0}) }
\label{ssec:detail_centering_setp}
The first optimization problem in the centering step~(\ref*{eq:centering_step}) can be elaborated as follows: 
\begin{align}
\label{seq:subproblem_a_f}
\begin{split}
\hat{\mathbf{a}}', \hat{\mathbf{f}}'
& =  \underset{\mathbf{a}', \mathbf{f}' }{\arg\min} 
~L_1(\hat{\mathbf{z}}, \mathbf{a}', \mathbf{f}', \hat{\mathbf{v}} )
\\ 
&= \underset{\mathbf{a}', \mathbf{f}' }{\arg\min} 
\sum_{i\in[n], r\in [R]} 
\hspace*{-10pt} \beta_{i,r}~
\frac{1}{T} \Bigg\|
\frac{y_{i,r}(t)}{N_{i,r}(T)} 
- \bigg\{ a'_{\hat{z}_i} + \sum_{m\in[M]} S^{\hat{v}_{i,m}+w^*_{r,m}} f'_{\hat{z}_i,m}(t) \bigg\}  \Bigg\|_t^2
.
\end{split}
\end{align}
Utilizing the renowned Parseval’s theorem, the optimization problem in~\eqref{seq:subproblem_a_f} can be formulated as follows:
\begin{align}
\label{seq:L_1_fourier}
\begin{split}
\hat{\mathbf{a}}',\hat{\boldsymbol{\phi}}'
& =  \underset{\mathbf{a}', \boldsymbol{\phi}' }{\arg\min} 
\sum_{i\in [n], r\in [R]} 
\hspace*{-10pt} \beta_{i,r}~
\sum_{l\in \mathbb{Z}} 
\Bigg| \frac{\eta_{i,r,l}}{N_{i,r}(T)}
- \bigg\{ a'_{\hat{z}_i} \mathbf{1}(l=0) 
\hspace*{5pt} + 
\\ & \hspace*{1.5in} 
\sum_{m\in[M]} \exp\big\{- \operatorname{j} 2 \pi l (\hat{v}_{i,m} +w^*_{r,m}) T^{-1} \big\} \phi'_{\hat{z}_i,m,l}
\bigg\}
 \Bigg|^2 ,
\end{split}
\end{align}
where $\boldsymbol{\phi}' \equiv (\phi'_{k,m,l})_{k\in[K],m\in[M],l\in\mathbb{Z}}$, $\{\phi'_{\hat{z}_i,m,l}: l\in \mathbb{Z}\}$ denotes the Fourier coefficients of $f'_{z_i,m}(t)$,
$\{\eta_{i,r,l}: l \in \mathbb{Z}\}$ denotes the Fourier coefficients of $y_{i,r}(t)$, and $\operatorname{j}$ denotes the imaginary unit.

The optimization problem in~\eqref{seq:L_1_fourier} can be solved by breaking it down to smaller independent problems.
For $k\in[K]$ and $l\in \mathbb{Z}$, let $\boldsymbol{\phi}'_{k,*,l} \equiv ({\phi}'_{k,m,l})_{m\in[M]}$.
For $l \neq 0$, the objective function associated with $\boldsymbol{\phi}'_{k,*,l}$ is essentially a weighted sum of squares:
\begin{align}
\label{seq:phi_prime_hat_argmin}
\hat{\boldsymbol{\phi}}'_{k,*,l}
&= \underset{ \boldsymbol{\phi}'_{k,*,l} }{\arg\min} 
~(\mathbf{h}_{k,l}-\mathbf{E}_{k,l} \boldsymbol{\phi}'_{k,*,l} )^\top \mathbf{B}_k~\overline{(\mathbf{h}_{k,l}-\mathbf{E}_{k,l} \boldsymbol{\phi}'_{k,*,l}) }
, 
\end{align}
where
$\mathbf{h}_{k,l} \equiv (\eta_{i,r,l}N_{i,r}(T)^{-1})_{(i,r) \in \hat{\mathcal{C}}_{k}\times [R]} $,
$\hat{\mathcal{C}}_k \equiv \{i\in[n]: \hat{z}_i = k\}$,
$\mathbf{E}_{k,l} \equiv [\exp\{- \operatorname{j} 2 \pi l (\hat{v}_{i,m}+w^*_{r,m}) T^{-1}\} ]_{(i,r) \in \hat{\mathcal{C}}_k \times [R], m \in [M]} $,
$\mathbf{B}_k$ is a diagonal matrix of $(\beta_{i,r})_{(i,r) \in \hat{\mathcal{C}}_{k}\times [R]}$,
and $\overline{z}$ denotes the complex conjugate of $z$ for any $z \in \mathbb{C}$.
As a result, the solution to~\eqref{seq:phi_prime_hat_argmin} is:
\begin{align}
\label{seq:hat_phi_prime_kl}
\hat{\boldsymbol{\phi}}'_{k,*,l}
= 
\left(\overline{\mathbf{E}_{k,l}}^\top \mathbf{B}_k~ \mathbf{E}_{k,l} \right)^{-1} 
\left(\overline{\mathbf{E}_{k,l}}^\top \mathbf{B}_k~ \mathbf{h}_{k,l}\right),
\quad \text{for } l\neq0.
\end{align}
For $l= 0$, the parameter $\boldsymbol{\phi}'_{k,*,0}$ can be estimated by exploiting the definition of $\mathcal{F}$.
From the definition of $\mathcal{F}$ we know that ${f}_{k,m}(0) = 0$ for $k\in[K], m\in[M]$.
Consequently, ${f}'_{k,m}(0)= {f}_{k,m}(0) \Lambda_{k}^{-1} = 0$.
In addition, using the inverse Fourier transformation, we can derive that $\sum_{l \in \mathbb{Z}} {\phi}'_{k,m,l}  = {f}'_{k,m}(0)$. Hence we know that $\sum_{l \in \mathbb{Z}} {\phi}'_{k,m,l}  = 0$.
This leads to the estimate of $\boldsymbol{\phi}'_{k,*,0}$ as
\begin{align}
\label{seq:hat_phi_km0}
 \hat{\boldsymbol{\phi}}'_{k,*,0} 
 & = - \sum_{|l| \leq \ell_0, l \neq 0} \hat{\boldsymbol{\phi}}'_{k,*,l},
\end{align}
where $\ell_0$ is the truncation frequency to facilitate the numerical feasibility of computing $\hat{\boldsymbol{\phi}}'_{k,*,0}$.

Based on $\hat{\boldsymbol{\phi}}'$, it is straightforward to obtain estimation for $\mathbf{a}'$ as follows:
\begin{align}
\label{eq:hat_a_k}
\begin{split}
\hat{{a}}_k'
& =  \underset{{a}_k' }{\arg\min} 
\sum_{i\in \hat{\mathcal{C}}_k, r\in [R]} 
\hspace*{-10pt} \beta_{i,r}~
\Bigg| \frac{\eta_{i,r,0}}{N_{i,r}(T)}
- \bigg\{ a'_{k} 
+ \sum_{m\in[M]} \hat{\phi}'_{k,m,0}
\bigg\}
\Bigg|^2 
\\ 
&= \Bigg[\sum_{i\in \hat{\mathcal{C}}_k, r\in [R]} \beta_{i,r} \bigg\{\frac{\eta_{i,r,0}}{N_{i,r}(T)}  - \sum_{m\in[M]} \hat{\phi}'_{k,m,0} \bigg\} \Bigg] 
{\Bigg[ \sum_{i\in \hat{\mathcal{C}}_k, r\in [R]} \beta_{i,r}\Bigg]}^{-1}
\\ 
&= T^{-1} - \sum_{m\in[M]} \hat{\phi}'_{k,m,0},
\end{split}
\end{align}
where in the third equality we use $\eta_{i,r,0} = T^{-1} \int_0^T y_{i,r}(t) \mathrm{d}t = T^{-1} N_{i,r}(T)$.
Moreover, using the inverse Fourier transform, the estimation for ${\mathbf{f}}'$ can be derived as follows:
\begin{align}
\label{seq:hat_f_km}
 \hat{f}'_{k,m}(t) 
& = \sum_{|l| \leq \ell_0} \hat{\phi}_{k,m,l} \exp( ~\operatorname{j} 2 \pi l t T^{-1}).
\end{align}

\section{Implementation of Newton's method in the clustering step}
\label{ssec:newton_method}
We employ the Newton's method to solve the optimization problem in~(\ref*{eq:tilde_v_i_k}) of the main text.
The objective function in~(\ref*{eq:tilde_v_i_k}) can be formulated in the frequency domain:
\begin{align}
\label{seq:L1i_zafv_2}
L_{1,i}(k, \hat{\mathbf{a}}', \hat{\mathbf{f}}', \mathbf{v}_i ) 
= \sum_{r\in [R]} 
\beta_{i,r}~
\sum_{l \in \mathbb{Z}} L_{1,i,l} (\mathbf{v}_i),
\end{align}
where $L_{1,i,l}(\mathbf{v}_i)$ is defined as
\begin{align}
\label{seq:L_1il}
L_{1,i,l}(\mathbf{v}_i) \equiv \Bigg| \frac{\eta_{i,r,l}}{N_{i,r}(T)}
- \bigg\{ \hat{a}'_{k} \mathbf{1}(l=0) 
+
\sum_{m\in[M]} \exp\big\{- \operatorname{j} 2 \pi l ({v}_{i,m} +w^*_{r,m}) T^{-1} \big\} \hat{\phi}'_{k,m,l}
\bigg\}
 \Bigg|^2.
\end{align}
The definition in~\eqref{seq:L_1il} suggests that $L_{1,i,l}(\mathbf{v}_i)$ remains constant with respect to $\mathbf{v}_i$ when $l=0$.
Moreover, since $\hat{f}_{k,m}$ is calculated using the truncated Fourier series (see~(\ref*{eq:hat_f_km}) of the main text), it follows that $\hat{\phi}'_{k,m,l} = 0$ for $|l| > \ell_0$.  
As a result, $L_{1,i,l}(\mathbf{v}_i)$ remains constant with respect to $\mathbf{v}_i$ for $|l| > \ell_0$.
Therefore, the optimization problem in~(\ref*{eq:tilde_v_i_k}) can be formulated in the frequency domain as
\begin{align}
\label{seq:tilde_v_Qi}
\begin{split}
\tilde{\mathbf{v}}_{i|k}
& = \underset{\mathbf{v}_{i}}{\arg\min}
\sum_{r\in [R]} 
\beta_{i,r}~
\sum_{|l| \leq \ell_0, l \neq 0} L_{1,i,l}(\mathbf{v}_i)
\\ 
&= \underset{\mathbf{v}_{i}}{\arg\min}
\sum_{r\in [R]} 
\beta_{i,r}
\sum_{|l| \leq \ell_0, l \neq 0}
\Bigg| \frac{\eta_{i,r,l}}{N_{i,r}(T)}
- \sum_{m\in[M]} \exp\big\{- \operatorname{j} 2 \pi l ({v}_{i,m} +w^*_{r,m}) T^{-1} \big\} \hat{\phi}'_{k,m,l}
 \Bigg|^2 
\\ 
& \equiv 
\underset{\mathbf{v}_{i}}{\arg\min}~
Q_{i}(\mathbf{v}_i).
\end{split}
\end{align}
where the first equality follows from the fact that $L_{1,i,l}(\mathbf{v}_i)$ remains constant with respect to~$\mathbf{v}_i$ for $l=0$ or $|l|>\ell_0$, and the second equality follows from the definition of $L_{1,i,l}(\mathbf{v}_i)$ in~\eqref{seq:L_1il}.

We solve the optimization problem in~\eqref{seq:tilde_v_Qi} using the Newton's method.
Specifically, the estimate of $\mathbf{v}_i$ is updated iteratively via
\begin{align}
\label{seq:hat_v_i_update}
\hat{\mathbf{v}}_i \leftarrow \hat{\mathbf{v}}_i - \mathrm{trunc}\{[\nabla^2 Q_{i}(\hat{\mathbf{v}}_i)]^{-1} \nabla Q_{i}(\hat{\mathbf{v}}_i)\},
\end{align}
where $\mathrm{trunc}\{x\}$ is a function defined as
\begin{align}
\mathrm{trunc}\{x\} \equiv 
\begin{cases}
-T/10, &\mathrm{if} ~x< -T/10,
\\
x, &\mathrm{if} ~x\in[-T/10,T/10],
\\ 
T/10, &\mathrm{if} ~x> T/10,
\end{cases}
\end{align}
$\nabla^2 Q_{i}(\hat{\mathbf{v}}_i) \equiv ({\partial^2 Q_{i}(\mathbf{v}_i)}/{\partial v_{i,m_1} \partial v_{i,m_2}})_{(m_1,m_2)\in[M]^2}$ denotes the Hessian matrix,
and $\nabla Q_{i}(\hat{\mathbf{v}}_i) \equiv ({\partial Q_{i}(\mathbf{v}_i)}/{\partial v_{i,m}})_{m\in[M]}$ denotes the gradient.
The gradient can be calculated through the following partial derivatives: for $m\in[M]$,
\begin{align}
\label{seq:first_order_derivative}
\begin{split}
\frac{\partial Q_{i}(\mathbf{v}_i)}{\partial v_{i,m}} 
&= 
- 2 \sum_{r\in [R]} 
\beta_{i,r} \hspace*{-5pt}
\sum_{|l| \leq \ell_0, l \neq 0}
\hspace*{-5pt}
\Re{e} 
\Bigg( 
\bigg[ 
(- \operatorname{j} 2 \pi l T^{-1})
\exp\big\{- \operatorname{j} 2 \pi l ({v}_{i,m} +w^*_{r,m}) T^{-1} \big\} \hat{\phi}'_{k,m,l} \bigg]
\\ & \hspace*{1.03in}
\times 
\overline{\bigg[ \frac{\eta_{i,r,l}}{N_{i,r}(T)}
- \hspace*{-10pt} \sum_{m'\in[M]\setminus\{m\}} \hspace*{-10pt} \exp\big\{- \operatorname{j} 2 \pi l ({v}_{i,m'} +w^*_{r,m'}) T^{-1} \big\} \hat{\phi}'_{k,m',l} \bigg]}
\Bigg).
\end{split}
\end{align}
The Hessian matrix can be calculated through the following second order partial derivatives: for $m\in[M]$, 
\begin{align}
\label{seq:second_order_derivative}
\begin{split}
\frac{\partial^2 Q_{i}(\mathbf{v}_i)}{\partial v_{i,m}^2}
&= 
- 2 \sum_{r\in [R]} 
\beta_{i,r}
\sum_{|l| \leq \ell_0, l \neq 0}
\Re{e} 
\Bigg( 
\bigg[ 
(- \operatorname{j} 2 \pi l T^{-1})^2
\exp\big\{- \operatorname{j} 2 \pi l ({v}_{i,m} +w^*_{r,m}) T^{-1} \big\} \hat{\phi}'_{k,m,l} \bigg]
\\ & \hspace*{1.03in}
\times 
\overline{\bigg[ \frac{\eta_{i,r,l}}{N_{i,r}(T)}
- \hspace*{-10pt} \sum_{m'\in[M]\setminus\{m\}} \hspace*{-10pt} \exp\big\{- \operatorname{j} 2 \pi l ({v}_{i,m'} +w^*_{r,m'}) T^{-1} \big\} \hat{\phi}'_{k,m',l} \bigg]}
 \Bigg),
\end{split}
\end{align}
while for $m_1,m_2\in[M]$ such that $m_1 \neq m_2$,
\begin{align}
\label{seq:second_order_derivative_2}
\begin{split}
\frac{\partial^2 Q_{i}(\mathbf{v}_i)}{\partial v_{i,m_1} \partial v_{i,m_2}} 
&= 
2 \sum_{r\in [R]} 
\beta_{i,r}
\sum_{|l| \leq \ell_0, l \neq 0}
\Re{e} 
\Bigg( 
\bigg[ 
(- \operatorname{j} 2 \pi l T^{-1})
\exp\big\{- \operatorname{j} 2 \pi l ({v}_{i,m_1} +w^*_{r,m_1}) T^{-1} \big\} \hat{\phi}'_{k,m_1,l} \bigg]
\\ & \hspace*{1.5in}
\times 
\overline{\bigg[ 
(- \operatorname{j} 2 \pi l T^{-1})
\exp\big\{- \operatorname{j} 2 \pi l ({v}_{i,m_2} +w^*_{r,m_2}) T^{-1} \big\} \hat{\phi}'_{k,m_2,l} \bigg]}
 \Bigg).
\end{split}
\end{align}

\section{Additional simulation results}

\subsection{Heuristic selection method for $\gamma$}
\label{ssec:heuristic_select_gamma}
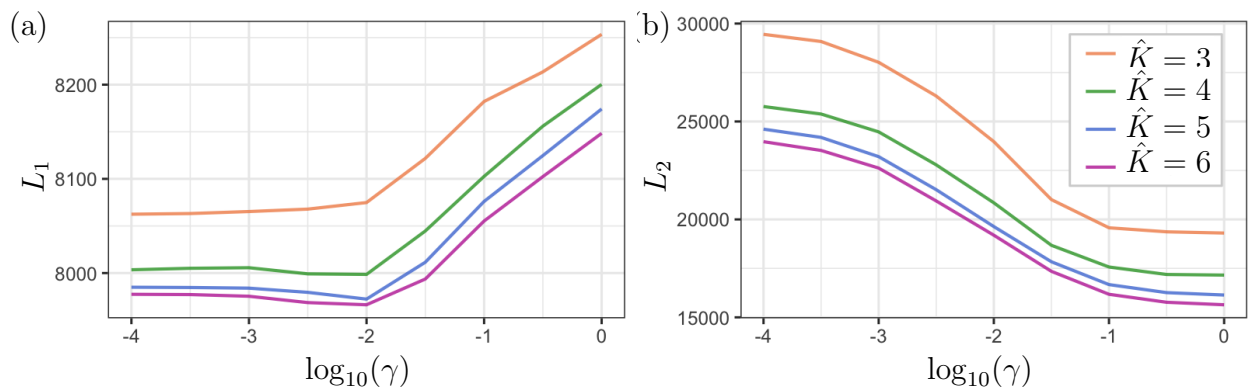
\begin{figure}
\centering
\begin{tikzpicture}
    \node[below right, inner sep=0] (image_12) at (0,0) 
      {
         \includegraphics[width=0.5\textwidth]
         {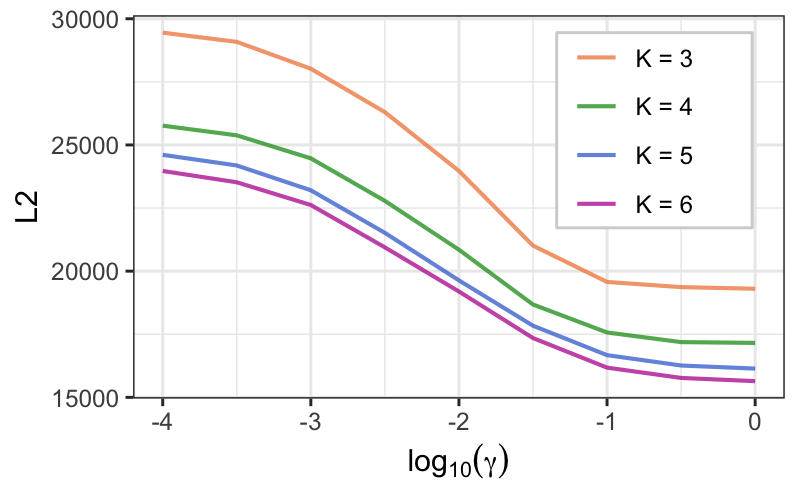}
      };
   \begin{scope}[
      shift={(image_12.south west)},
       x={($0.1*(image_12.south east)$)},
       y={($0.1*(image_12.north west)$)}]
      \node[above, fill=white, yshift=-0pt, xshift=12pt, minimum width=4cm] at (5,0){ $\log_{10}(\gamma)$};
      \node[below, fill=white, yshift=-2pt, rotate = 90, xshift=12pt, minimum width=3cm, inner sep=0] at (0.1,5){ $L_2$ };
      \node[right, fill=white, yshift=5pt,  xshift=14pt, inner sep=1] at (7.2,8.5){ $\hat{K}=3$};
      \node[right, fill=white, yshift=-7pt,  xshift=14pt, inner sep=1] at (7.2,8.5){ $\hat{K}=4$};
      \node[right, fill=white, yshift=-21pt,  xshift=14pt, inner sep=1] at (7.2,8.5){ $\hat{K}=5$};
      \node[right, fill=white, yshift=-34pt,  xshift=14pt, inner sep=1] at (7.2,8.5){ $\hat{K}=6$};
      \node[below, xshift=5pt, fill=white, inner sep=1] at (0,10){ (b) };
   \end{scope}
   \node[below left, xshift=0pt, inner sep=0] (image_11) at (image_12.north west) 
      {
         \includegraphics[width=0.5\textwidth]
         {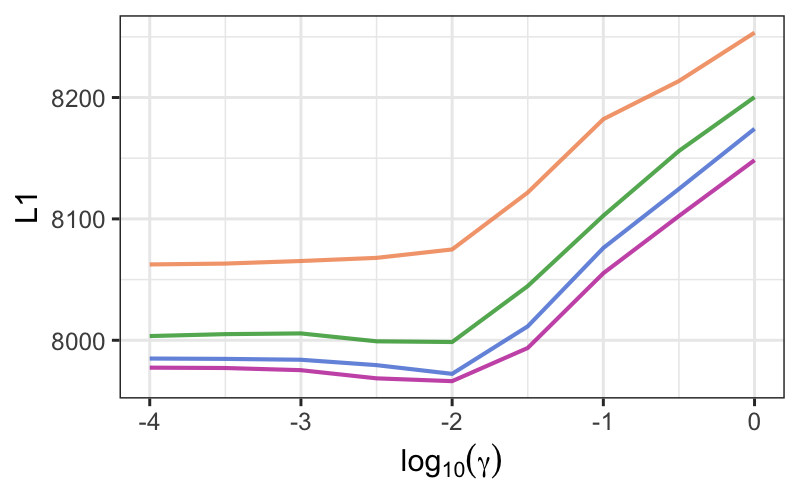}
      };
   \begin{scope}[
      shift={(image_11.south west)},
       x={($0.1*(image_11.south east)$)},
       y={($0.1*(image_11.north west)$)}]
   \node[above, fill=white, yshift=-0pt, xshift=12pt, minimum width=4cm] at (5,0){ $\log_{10}(\gamma)$};
   \node[below, fill=white, yshift=-2pt, rotate = 90, xshift=12pt, minimum width=3cm, inner sep=0] at (0.1,5){ $L_1$ };
    \node[below, xshift=5pt, fill=white, inner sep=1] at (0,10){ (a) };
   \end{scope}
\end{tikzpicture}
\caption{
Trends of $L_1$ and $L_2$ as $\gamma$ changes averaged across $5000$ replicates.
Synthetic data is generated with  $K=4$, $n=40$, $R=3$, $\tau=0.3$, $\rho=0.5$.
In the legend, $\hat{K}$ represents the designated number of clusters as input of the algorithm.
It is evident that the trend of $L_1$ and $L_2$ is consistent across the designated number of clusters.
 }
\label{Sfig:L1_L2_vs_gamma}
\end{figure}

Figure~\ref{Sfig:L1_L2_vs_gamma} shows the trend of $L_1$ and $L_2$ as $\gamma$ changes, where the candidate range of $\gamma$ is discussed in Section~\ref{sec:range_of_gamma}.
From Figure~\ref{Sfig:L1_L2_vs_gamma}, we can see consistent trends of $L_1$ and $L_2$ across the designated number of clusters $\hat{K}$, indicating that the choice of $\gamma$ using the proposed heuristic method is insensitive to the change in~$\hat{K}$.

\begin{figure}
\centering
\begin{tikzpicture}
    \node[below right, inner sep=0] (image_11) at (0,0) 
      {
         \includegraphics[width=0.65\textwidth]
         {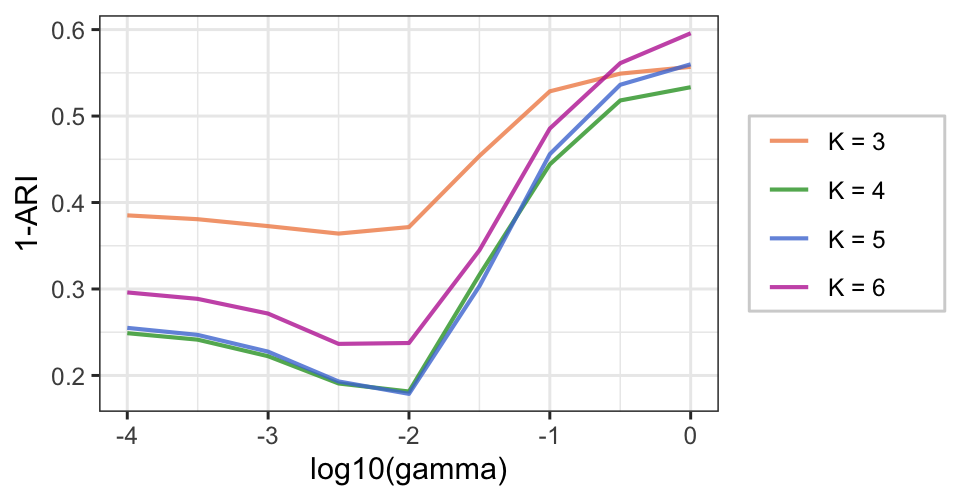}
      };
      \begin{scope}[
      shift={(image_11.south west)},
       x={($0.1*(image_11.south east)$)},
       y={($0.1*(image_11.north west)$)}]
      \node[above, fill=white, yshift=-0pt, xshift=12pt, minimum width=4cm] at (3.7,0){ $\log_{10}(\gamma)$};
      \node[below, fill=white, yshift=-2pt, rotate = 90, xshift=12pt, minimum width=3cm, inner sep=2] at (0.1,5){ 1-ARI };
      \node[right, fill=white, yshift=7pt,  xshift=14pt, inner sep=1] at (8.0,6.7){ $\hat{K}=3$};
      \node[right, fill=white, yshift=-6pt,  xshift=14pt, inner sep=1] at (8.0,6.7){ $\hat{K}=4$};
      \node[right, fill=white, yshift=-20pt,  xshift=14pt, inner sep=1] at (8.0,6.7){ $\hat{K}=5$};
      \node[right, fill=white, yshift=-34pt,  xshift=14pt, inner sep=1] at (8.0,6.7){ $\hat{K}=6$};
   \end{scope}
\end{tikzpicture}
\caption{
Clustering performance as $\gamma$ changes averaged across $5000$ replicates.
Synthetic data is generated with $K=4$, $n=40$, $R=3$, $\tau=0.3$, $\rho=0.5$.
In the legend, $\hat{K}$ represents the designated number of clusters as input of the algorithm.
 }
\label{Sfig:ARI_vs_gamma}
\end{figure}
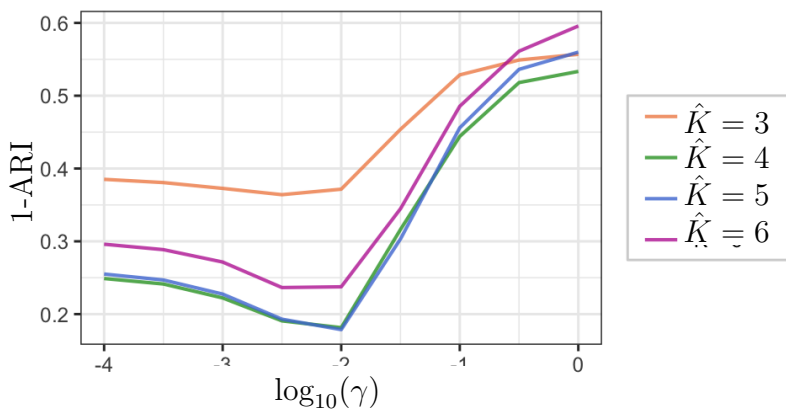

Figure~\ref{Sfig:ARI_vs_gamma} shows the clustering performance as $\gamma$ changes.
We can see that the clustering performance slightly improves when$\gamma$ increases from $10^{-4}$ to $10^{-2}$. 
This is because the within-cluster heterogeneity of the distribution of event times remains unchanged (see Figure~\ref{Sfig:L1_L2_vs_gamma}(1a)), while the within-cluster heterogeneity of event counts decreases (see Figure~\ref{Sfig:L1_L2_vs_gamma}(1b)).
Moreover, we can see from Figure~\ref{Sfig:ARI_vs_gamma} that once $\gamma$ exceeds $0.01$, the clustering performance rapidly deteriorates. 
This is because the estimated clusters have increasing within-cluster heterogeneity of the distribution of event times when $\gamma$ exceeds $0.01$ (see Figure~\ref{Sfig:L1_L2_vs_gamma}(1a)).
Therefore, by choosing the largest $\gamma$ before observing a significant increase in $L_1$, we are able to achieve optimal clustering performance.

\subsection{Sensitivity analysis with respect to $\ell_0$}
\label{ssec:sensitivity_freqtrun}
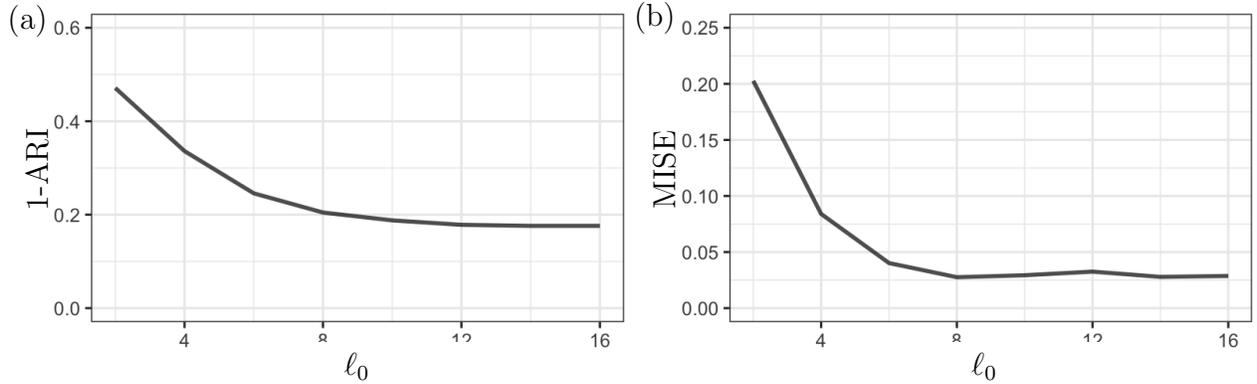
\begin{figure}
\centering
\begin{tikzpicture}
   \node[below right, inner sep=0] (image_1) at (0,0) 
      {
         \includegraphics[width=0.5\textwidth]
         {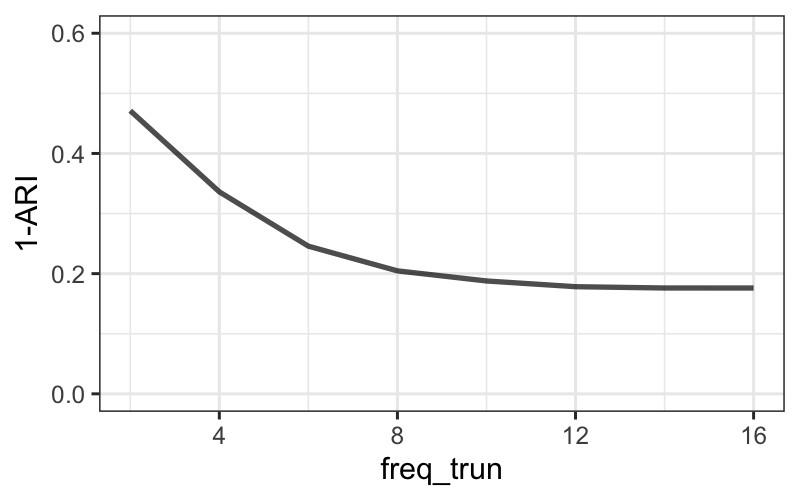}
      };
   \begin{scope}[
      shift={(image_1.south west)},
       x={($0.1*(image_1.south east)$)},
       y={($0.1*(image_1.north west)$)}]
      \node[above, fill=white, yshift=-0pt, xshift=12pt, minimum width=4cm] at (5,0){ $\ell_0$ };
      \node[below, fill=white, yshift=-2pt, rotate = 90, xshift=12pt, minimum width=3cm, inner sep=2] at (0.1,5){ 1-ARI };
      \node[below, xshift=5pt, fill=white, inner sep=1] at (0,10){ (a) };
   \end{scope}
   
   \node[below right, xshift=2pt, inner sep=0] (image_3) at (image_1.north east) 
      {
         \includegraphics[width=0.5\textwidth]
         {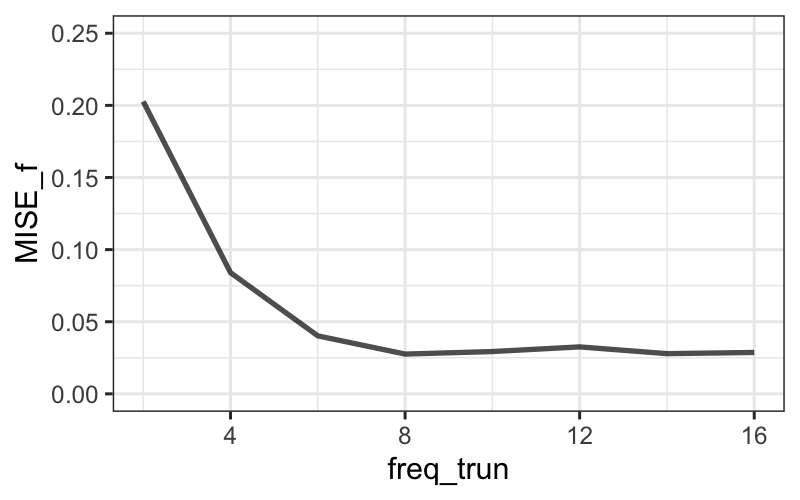}
      };
   \begin{scope}[
      shift={(image_3.south west)},
       x={($0.1*(image_3.south east)$)},
       y={($0.1*(image_3.north west)$)}]
      \node[above, fill=white, yshift=-0pt, xshift=12pt, minimum width=4cm] at (5,0){ $\ell_0$ };
      \node[below, fill=white, yshift=-2pt, rotate = 90, xshift=12pt, minimum width=3cm, inner sep=2] at (0.1,5){ MISE };
      \node[below, xshift=5pt, fill=white, inner sep=0] at (0,10){ (b) };
   \end{scope}
\end{tikzpicture}
\caption{
Sensitivity analysis with respect to $\ell_0$ averaged over 5000 replicates. 
Synthetic data is generated with $K=4$, $n=40$, $R=3$, $\tau=0.3$, $\rho=0.5$. 
Panel (a) and (b) show that the performance of clustering  and intensity component estimation  become stable when $\ell_0 \geq 8$.
 }
\label{Sfig:sensitivity_freqtrun}
\end{figure}

Figure~\ref{Sfig:sensitivity_freqtrun} shows how the performance of clustering and intensity estimation changes with different values of $\ell_0$.
Both criteria decrease when $\ell_0$ increases from 2 to 8, because higher values of $\ell_0$ allow the estimated intensities to capture more information in the distribution of event times. 
As $\ell_0$ further increases, both 1-ARI and MISE stabilize, indicating that the estimation results are not significantly affected by further changes in $\ell_0$ as long as $\ell_0$ is sufficiently large to capture the signal in the distribution of event times, for instance, for the synthetic data in Figure~\ref{Sfig:sensitivity_freqtrun}, $\ell_0 = 8$ is large enough. 

\subsection{Effectiveness of proposed initialization scheme}
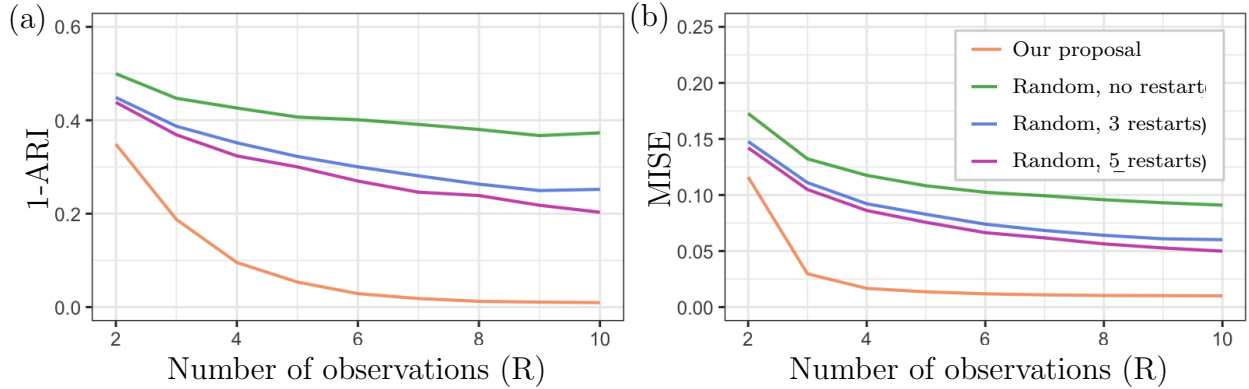
\begin{figure}
\centering
\begin{tikzpicture}
   \node[below right, inner sep=0] (image_12) at (0,0) 
      {
         \includegraphics[width=0.5\textwidth]
         {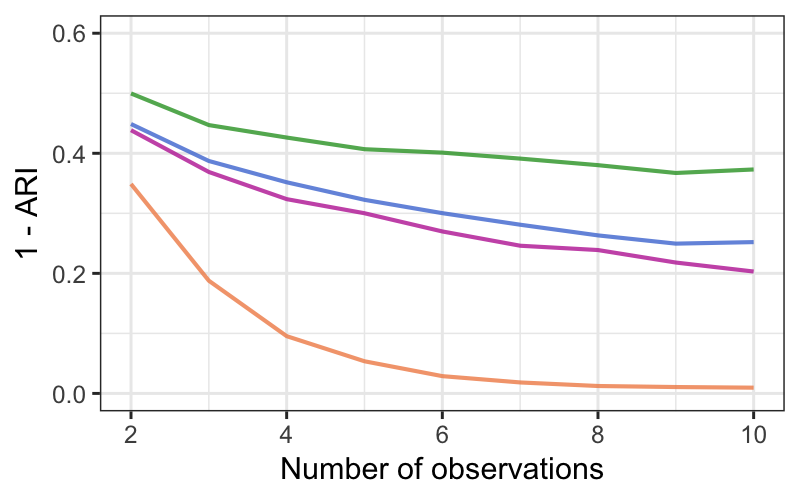}
      };
   \begin{scope}[
      shift={(image_12.south west)},
       x={($0.1*(image_12.south east)$)},
       y={($0.1*(image_12.north west)$)}]
      \node[above, fill=white, yshift=-3pt, xshift=12pt, minimum width=4cm] at (5,0){ Number of observations (R)};
      \node[below, fill=white, yshift=-2pt, rotate = 90, xshift=12pt, minimum width=3cm, inner sep=2] at (0.1,5){ 1-ARI };
      \node[below, xshift=5pt, fill=white, inner sep=1] at (0,10){ (a) };
   \end{scope}
   \node[below right, xshift=0pt, inner sep=0] (image_11) at (image_12.north east) 
      {
         \includegraphics[width=0.5\textwidth]
         {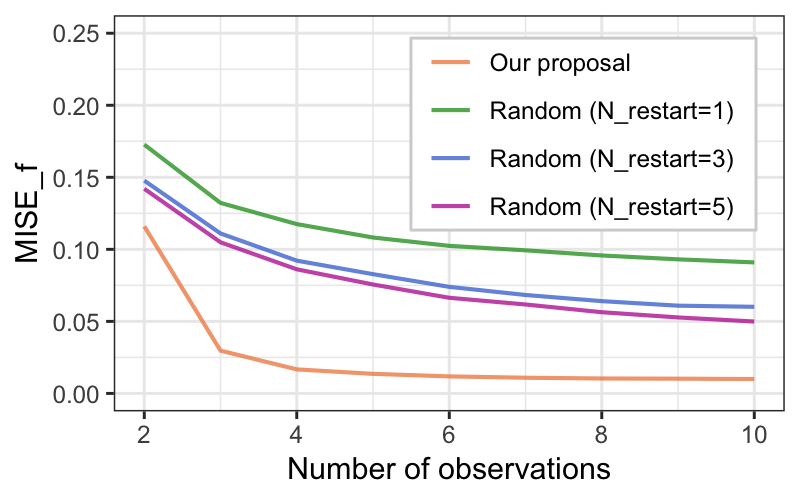}
      };
   \begin{scope}[
      shift={(image_11.south west)},
       x={($0.1*(image_11.south east)$)},
       y={($0.1*(image_11.north west)$)}]
   \node[above, fill=white, yshift=-3pt, xshift=12pt, minimum width=4cm] at (5,0){ Number of observations (R)};
   \node[below, fill=white, yshift=-2pt, rotate = 90, xshift=12pt, minimum width=3cm, inner sep=2] at (0.1,5){ MISE };
   \node[right, fill=white, yshift=2pt,  xshift=14pt, inner sep=1, minimum width=2.5cm, minimum height=1.5cm] at (5.2,7.0){ };
   \node[right, fill=white, yshift=5pt,  xshift=15pt, inner sep=0] at (5.4,8.4){\scriptsize Our proposal};
   \node[right, fill=white, yshift=-9pt,  xshift=15pt, inner sep=0] at (5.4,8.4){\scriptsize Random, no restart};
   \node[right, fill=white, yshift=-23pt,  xshift=15pt, inner sep=0] at (5.4,8.4){\scriptsize Random, 3 restarts};
   \node[right, fill=white, yshift=-37pt,  xshift=15pt, inner sep=0] at (5.4,8.4){\scriptsize Random, 5 restarts};
   \node[below, xshift=5pt, fill=white, inner sep=0] at (0,10){ (b) };
   \end{scope}
\end{tikzpicture}
\caption{
Performance of proposed initialization scheme averaged over $5000$ replicates.
Synthetic data is generated with $K=4$, $n=40$, $\tau=0.3$, $\rho=0.5$, and varying $R$.
In the legend, ``our proposal'' stands for the proposed initialization scheme, ``random'' stands for the random initialization scheme with restarts.
For the random initialization scheme, time shifts are initialized using $\text{Unif}(0, \hat{v}^{(0)}_{i,m})$, and cluster memberships are initialized randomly.
The best result among restarts is selected according to the smallest objective function.
Panel (a) and (b) show that the proposed initialization scheme leads to better estimation than the random initialization scheme with 5 restarts.
 }
\label{fig:init}
\end{figure}

Figure~\ref{fig:init} shows the efficacy of proposed initialization scheme compared to a random initialization scheme.
In the random initialization scheme, initial time shifts are set using $\text{Unif}(0, \hat{v}^{(0)}_{i,m})$, and initial cluster memberships randomly assigned.
The results in Figure~\ref{fig:init} demonstrate that the proposed initialization method yields  better estimation than the random initialization with multiple restarts.
Moreover, keep increasing the number of restarts only results in diminishing improvement in performance.   
Overall, using proposed initialization scheme is computationally efficient and achieve promising estimation performance.

\section{Supplement for real data analysis}
\label{ssec:supp_data_analysis}
\subsection{Data preprocessing}
\subsubsection*{Experimental trials}

Table~\ref{stable:experiment_conditions} presents the feedback types under different experimental conditions, along with the number of trials for each condition.
The trials where the left visual grating was of higher contrast and the feedback was reward are analyzed using our proposed method.
We refer to these trials as \emph{training trials}.
The rest of trials, which are not used to fit the proposed model, are employed to explore the roles of identified clusters under different tasks.

\begin{table}
\caption{ 
Experimental conditions and total numbers of trials. 
In the ``scenario'' column, ``L'' and ``R'' denote higher contrast in the left and right gratings, respectively.
The ``choice'' column shows ``L'' for moving the left grating towards the center and ``R'' for moving the right grating
The ``feedback'' column shows "1" for reward and "-1" for penalty.
}
\label{stable:experiment_conditions} 
\centering
\begin{tabular}{ *{4}{c} }
\hline
Scenario & Choice & Feedback & \# trials
\\ 
\hline
L & L & 1 & 102
\\ 
R & R & 1 & 81
\\ 
R & L & -1 & 8
\\ 
L & R & -1 & 3
\\ 
\hline
\end{tabular}
\end{table}

\subsubsection*{Selection of time window}
Due to the design of the experiment, the duration of time between stimulus onset and feedback delivery across trials. 
In order to create comparable samples, we analyze a 3.5-second time window whose center is positioned at the midpoint between 0.1 seconds before the visual stimuli onset and 2 seconds after the feedback delivery.
To be more specific, for observation $r\in\{1,\cdots,R\}$, the start and end time of observation $r$ is set as
\begin{align}
\mathrm{ObsStartTime}_r \equiv  \{(\mathrm{VisTime}_r - 0.1) + (\mathrm{FeedTime}_r+2)\}/2 -  3.5/2,
\\ 
\mathrm{ObsEndTime}_r \equiv  \{(\mathrm{VisTime}_r - 0.1) + (\mathrm{FeedTime}_r+2)\}/2 +  3.5/2,
\end{align}
where $\mathrm{ObsStartTime}_r$ denotes the start time of the time window for trial $r$, 
$\mathrm{VisTime}_r$ denotes the visual stimulus onset time of trial $r$, 
$\mathrm{FeedTime}_r$ denotes the feedback delivery time of trial $r$,
and 
$\mathrm{ObsEndTime}_r$ denotes the end time of the time window for trial $r$.  
In cases where the time window started within 1 second after the feedback delivery time of the previous trial, or ended after the visual stimulus onset time of the next trial, we augment the spike trains as detailed in the next section.

\begin{figure}[ht]
	\centering
	\begin{tikzpicture}
		\node[below right,inner sep=0] (image2) at (0,0) 
		{\includegraphics[width=0.5\textwidth, trim=0 0 0 0, clip]
			{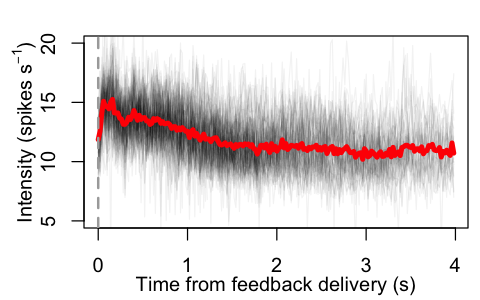}
		};    
	\end{tikzpicture}
	\caption{ 
		Average neural firing intensities over neurons and trials. 
		Each gray curve represents an average firing intensity of the midbrain region in a single training trial. 
		All trials are aligned based on the feedback delivery time.
		The red curve represents the average firing intensity across all training trials.
		It is observed that the average firing intensity stabilizes approximately 2 seconds after feedback delivery.
	}
	\label{sfig:intensity_post_feedback}
\end{figure}

Here we analyze the firing activity within a $T=3.5$ seconds time window due to the following reasons.
Figure~\ref{sfig:intensity_post_feedback} displays the average neural firing rate post feedback delivery over all training trials.
It is evident that the average firing rate stabilizes at around 2 seconds post feedback delivery.
Moreover, among the 102 training trials, the maximum duration between visual stimulus onset and feedback delivery is 1.4 seconds.
Therefore, by setting $T=3.5$, we can ensure that each observation includes at least 0.1 second before visual stimulus onset and 2 seconds after feedback delivery.

\subsubsection*{Augmentation of spike trains}
\begin{figure}[ht]
   \centering
   \begin{tikzpicture}
      \node[below right,inner sep=0] (image2) at (0,0) 
      {\includegraphics[width=0.5\textwidth, trim=0 0 0 0, clip]
         {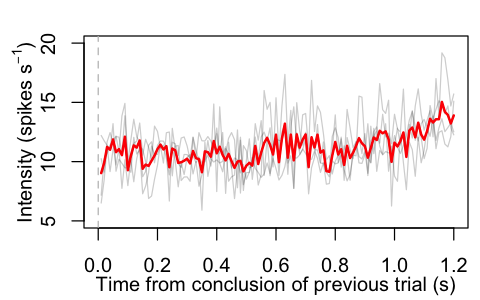}
      };
      \begin{scope}[
        shift={($(image2.south west)$)},
        x={($0.1*(image2.south east)$)},
        y={($0.1*(image2.north west)$)}]
        \node[above right] at ($(0,9)$) { (a)};
      \end{scope}       
      \node[below right,inner sep=0] (image3) at (image2.north east) 
      {\includegraphics[width=0.5\textwidth, trim=0 0 0 0, clip]
         {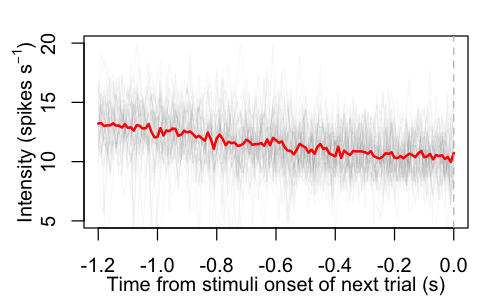}
      };    
      \begin{scope}[
        shift={($(image3.south west)$)},
        x={($0.1*(image3.south east)$)},
        y={($0.1*(image3.north west)$)}]
        \node[above right] at ($(0,9)$) { (b)};
      \end{scope}   
   \end{tikzpicture}
   \caption{ 
      Neural firing patterns after the previous trial's conclusion or before the next trial's visual stimulus onset.
      Each gray curve represents an average firing intensity of the midbrain region in one trial. 
      The red curves represent the average of gray curves.
      Panel (a) shows the average neural firing intensities post conclusion of the previous trials for the trial set $\{r: \mathrm{ObsStartTime}_r < \mathrm{ConcluTime}_{r-1} \}$. 
      The trials are aligned by the conclusion time of the previous trials.
      Panel (b) shows the average neural firing intensities prior visual stimuli onset of the next trials for the trial set $\{r: \mathrm{ObsEndTime}_r > \mathrm{VisTime}_{r+1}\}$. 
      The trials are aligned by the visual stimulus onset time of the next trials.
   }
   \label{sfig:intensity_post_prev_trial_conclusion}
\end{figure}

For trials with $\mathrm{ObsStartTime}_r < \mathrm{ConcluTime}_{r-1}$, where $\mathrm{ConcluTime}_{r-1}$ denotes the trial conclusion time (i.e., 1 second post feedback delivery time) of trial $r-1$, we augment the spike trains to impute the missing data. 
Figure~\ref{sfig:intensity_post_prev_trial_conclusion}(a) shows that, for trials with $\mathrm{ObsStartTime}_r < \mathrm{ConcluTime}_{r-1}$, the average intensity remains stable within 0.4 second post the conclusion time of previous trial.
Therefore, we augment spikes by making shifted copies of spikes within 0.4 second post $\mathrm{ConcluTime}_{r-1}$.
We append the following artificial spikes to the firing activity of neuron $i \in [n]$ in observation $r \in \{r: \mathrm{ObsStartTime}_r < \mathrm{ConcluTime}_{r-1}\}$:
\begin{align}
\begin{split}
\bigg\{t_{i,r,j}-0.4 \times k: 
& ~t_{i,r,j}\in[\mathrm{ConcluTime}_{r-1}, \mathrm{ConcluTime}_{r-1}+0.4], 
\\ 
& ~t_{i,r,j}-0.4 \times k \in [\mathrm{ObsStartTime}_r, \mathrm{ConcluTime}_{r-1}], k\in \mathbb{N} \bigg\}.
\end{split}
\end{align}

Additionally, we augment the spike trains for trials with $\mathrm{ObsEndTime}_r > \mathrm{VisTime}_{r+1}$.
Figure~\ref{sfig:intensity_post_prev_trial_conclusion}(b) shows that, for trials with $\mathrm{ObsEndTime}_r > \mathrm{VisTime}_{r+1}$, the average intensity remains stable within 0.4 seconds before the visual stimulus onset of the next trials.
Therefore, we append the following spikes to the firing activity of neuron $i\in[n]$ in observation $r \in \{r: \mathrm{ObsEndTime}_r > \mathrm{VisTime}_{r+1} \}$:
\begin{align}
\begin{split}
\bigg\{t_{i,r,j}+0.4 \times k: 
& ~t_{i,r,j}\in[\mathrm{VisTime}_{r+1}-0.4, \mathrm{VisTime}_{r+1}], 
\\ 
& ~t_{i,r,j}+0.4 \times k \in [\mathrm{VisTime}_{r+1}, \mathrm{ObsEndTime}_r], k\in \mathbb{N} \bigg\}
\end{split}
\end{align}

\subsection{Tuning parameter selection}

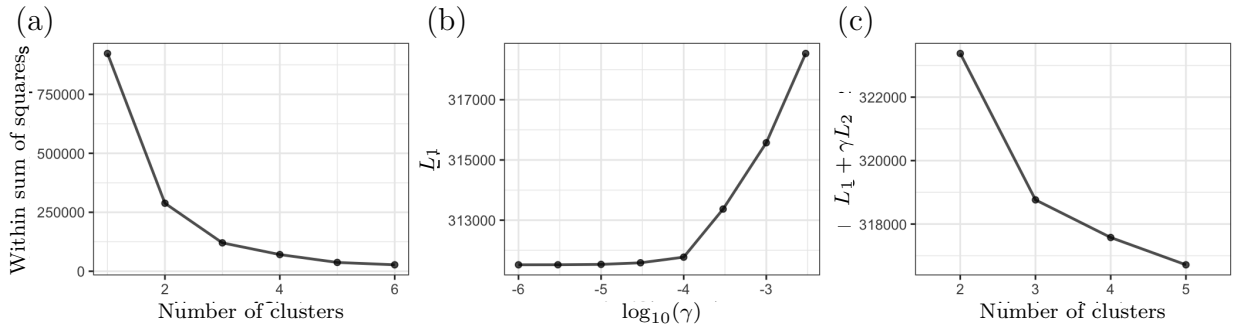
\begin{figure}[ht]
   \centering
   \begin{tikzpicture}
      \node[below right,inner sep=0] (image2) at (0,0) 
      {\includegraphics[width=0.33\textwidth, trim=0 0 0 0, clip]
         {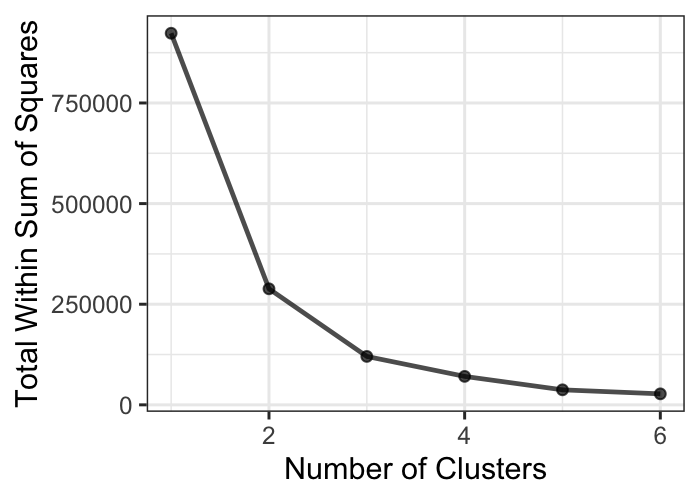}
      };
      \begin{scope}[
        shift={($(image2.south west)$)},
        x={($0.1*(image2.south east)$)},
        y={($0.1*(image2.north west)$)}]
        \node[above right, xshift=-1pt] at ($(0,9.5)$) { (a)};
        \node[above, xshift=-1pt, fill=white, inner sep=0] at ($(6,0.3)$) {\scriptsize Number of clusters };
        \node[below, rotate=90, minimum width=78pt, xshift=6pt, yshift=-1pt, fill=white, inner sep=0] at ($(0,5)$) {\scriptsize Within sum of squares };
      \end{scope}       
      \node[below right = 0 and 0 of image2.north east,inner sep=0] (image3) at (image2.north east) 
      {\includegraphics[width=0.33\textwidth, trim=0 0 0 0, clip]
         {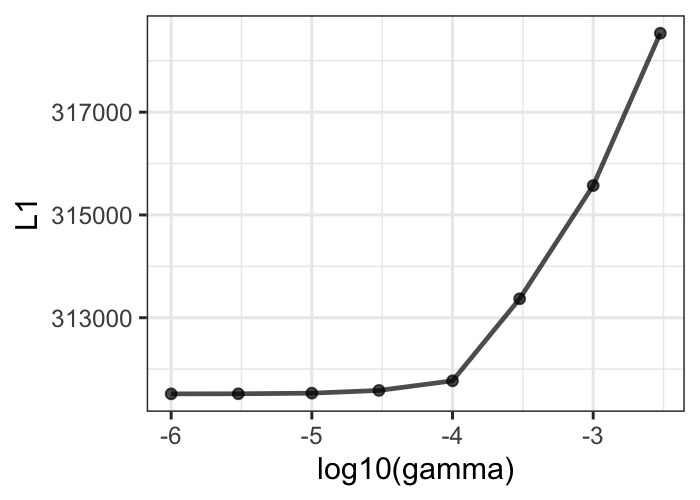}
      };    
      \begin{scope}[
        shift={($(image3.south west)$)},
        x={($0.1*(image3.south east)$)},
        y={($0.1*(image3.north west)$)}]
        \node[above right, xshift=-1pt] at ($(0,9.5)$) { (b)};
        \node[above, xshift=-1pt, yshift=-2pt, fill=white, minimum width=2cm, inner sep=0] at ($(6,0.3)$) {\scriptsize $\log_{10}(\gamma)$ };
        \node[below, rotate=90, xshift=8pt, yshift=-0.5pt,
        minimum width=50pt,  fill=white, inner sep=0] at ($(0,5)$) {\scriptsize $L_1$ };
      \end{scope}
      \node[below right,inner sep=0] (image4) at (image3.north east) 
      {\includegraphics[width=0.33\textwidth, trim=0 0 0 0, clip]
         {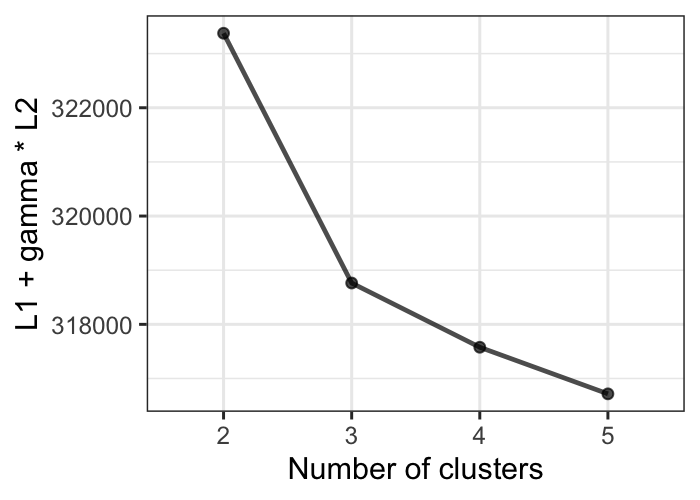}
      };   
      \begin{scope}[
        shift={($(image4.south west)$)},
        x={($0.1*(image4.south east)$)},
        y={($0.1*(image4.north west)$)}]
        \node[above right, xshift=-1pt] at ($(0,9.5)$) { (c)};
        \node[above, xshift=-1pt, yshift=0pt, fill=white, minimum width=2cm, inner sep=0] at ($(6,0.3)$) {\scriptsize Number of clusters };
        \node[below, rotate=90, xshift=8pt, yshift=-2pt, minimum width=50pt, fill=white, inner sep=0] at ($(0,5)$) {\scriptsize $L_1+\gamma L_2$ };
      \end{scope} 
   \end{tikzpicture}
   \caption{ 
   Results of the heuristic method for selection of $\gamma$ and $K$. 
   Panel (a) shows the within cluster variance of $N_{i}(T)$'s obtained from the k-means algorithm against the number of clusters, where $N_i(T) \equiv  R^{-1} \sum_{r\in[R]}{N_{i,r}(T)}$.
   Using the elbow method, we set $K=3$ as a preliminary estimation of $K$.
   Panel (b) shows the values of $L_1$ as $\gamma$ increases.
   The upward trend of $L_1$ when $\gamma$ exceeds $10^{-4}$ suggests the choice of $\gamma = 10^{-4}$.
   Panel (c) shows the overall objective function value obtained from various numbers of clusters. 
   The curve suggests a diminishing decrease in the objective function value when $K$ exceeds 3.
   }
   \label{sfig:gamma_selection}
\end{figure}

We choose the values of $\gamma$ and $K$ using the heuristic method proposed in Section~\ref*{subsec:optimization_full}.
Figure~\ref{sfig:gamma_selection} presents the results of the heuristic method that informs our choice of $\gamma$ and $K$.
Firstly, we establish a preliminary estimate of $K$ by applying the k-means algorithm on $N_{i}(T)$'s where $N_i(T) \equiv  R^{-1} \sum_{r\in[R]}{N_{i,r}(T)}$, and selecting $K$ using the elbow method.
Figure~\ref{sfig:gamma_selection}(a) illustrates that the within-cluster variance has only marginal reduction as the number of clusters exceeds 3. 
Therefore, we set $K=3$ as a preliminary estimation of $K$. 
Secondly, given the preliminary estimation of $K$, we choose the largest $\gamma$ before observing a significant increase in $L_1$.
Figure~\ref{sfig:gamma_selection}(b) shows that the value of $L_1$ a significant upward trend once $\gamma$ exceeds $10^{-4}$.
Therefore, we set $\gamma = 10^{-4}$.
Finally, with $\gamma$ fixed at $10^{-4}$, we refine the value of $K$ by identifying the elbow point on the curve of overall objective function (i.e., $L_1 + \gamma L_2$) against $K$. 
Figure~\ref{sfig:gamma_selection}(c) suggests that $K=3$ is a feasible choice.

\subsection{Intensity component refinement}
We refine the estimated intensity components by employing the proposed additive shape invariant model in~(\ref*{model:simplified_2}) on each estimated cluster.
For Cluster 1, we apply Algorithm~\ref*{algo} on spike trains of neurons in Cluster 1, denoted as $\mathcal{N}_1 \equiv \{N_{i,r}(t):i\in \hat{\mathcal{C}}_1, r\in[R]\}$.
We set $M=1$ and let the observation-specific time shifts $w^*_{r,1}$'s to be the visual stimulus onset time. 
The rest of the parameters are set as follows: $K=1$, $\gamma=0$, $\ell_0=10$, $\epsilon=0.005$.
The algorithm is applied with 20 restarts, where each restart involves distinct initial values for subject-specific time shifts.
These initial values are obtained by jittering the proposed initial subject-specific time shifts  as follows: for $i \in \hat{\mathcal{C}}_1$, $m=1$ and $x\in\{1,\cdots,20\}$,
\begin{align}
\tilde{v}_{i,m,x}^{(0)} \equiv \hat{v}_{i,m}^{(0)} + \varepsilon_{i,m,x},
\quad \varepsilon_{i,m,x} \sim \mathrm{unif}(-T/50,T/50),
\end{align}
where $\tilde{v}_{i,m,x}^{(0)}$ represents the initial subject-specific time shifts associated subject $i$ and stimulus $m$ in restart $x$,
and $\hat{v}_{i,m}^{(0)}$ represents the proposed initial subject-specific time shifts defined in~(\ref*{eq:init_v}).
The best result of intensity components among the 20 restarts is selected based on the smallest value of the objective function.

For Cluster 2 and 3, Algorithm~\ref*{algo} is applied to $\mathcal{N}_k \equiv \{N_{i,r}(t):i\in \hat{\mathcal{C}}_k, r\in[R]\}$, $k\in\{2,3\}$, in a similar manner as for $\mathcal{N}_1$. 
The only difference lies in the specification of $M=2$ for Clusters 2 and 3, where $m=1$ corresponds to the visual stimulus and $m=2$ corresponds to the auditory tone cue.

\subsection{Supplementary results of neural data analysis}
\subsubsection*{Neural firing intensity in condition ``L,R''}
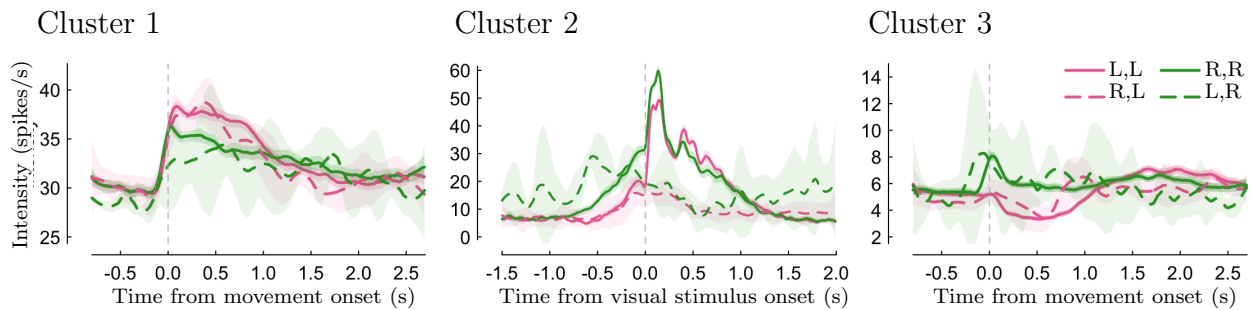
\begin{figure}[ht]
   \centering
   \begin{tikzpicture}
      \node[below right,inner sep=0] (image2) at (0,0) 
      {\includegraphics[width=0.361\textwidth, trim=0 0 0 0, clip]
         {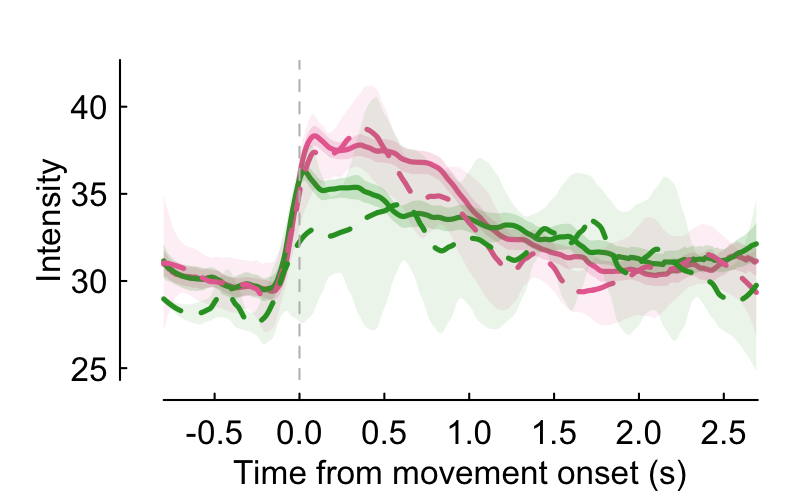}
      };
      \begin{scope}[
        shift={($(image2.south west)$)},
        x={($0.1*(image2.south east)$)},
        y={($0.1*(image2.north west)$)}]
        \node[above right, xshift=10pt] at ($(0,9.5)$) { Cluster 1};
        \node[above, xshift=-2pt, yshift=1pt,fill=white, inner sep=0] at ($(6,0)$) {\scriptsize Time from movement onset (s) };
        \node[below, rotate=90, minimum width=68pt, minimum height=8.5pt, xshift=6pt, yshift=-4pt, fill=white, inner sep=0] at ($(0,5)$) {\scriptsize Intensity (spikes/s) };
      \end{scope}       
      \node[below right = 0 and 0 of image2.north east,inner sep=0] (image3) at (image2.north east) 
      {\includegraphics[width=0.33\textwidth, trim=70 0 0 0, clip]
         {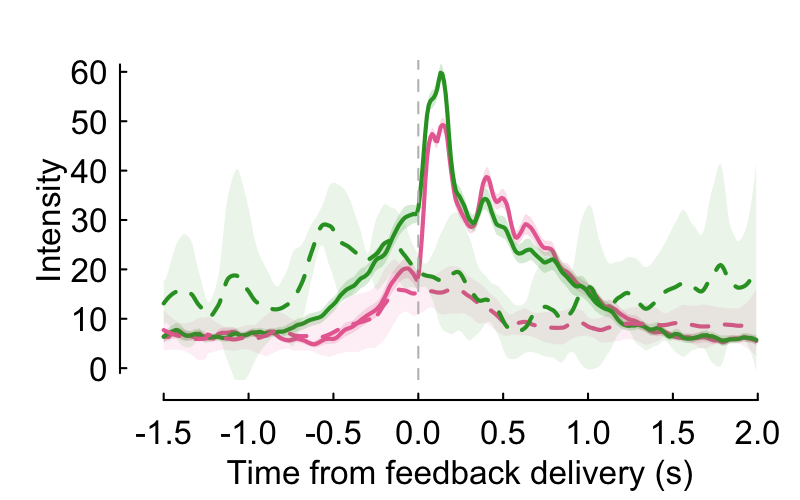}
      };    
      \begin{scope}[
        shift={($(image3.south west)$)},
        x={($0.1*(image3.south east)$)},
        y={($0.1*(image3.north west)$)}]
        \node[above right, xshift=-2pt] at ($(0,9.5)$) {Cluster 2 };
        \node[above, xshift=-8pt, yshift=1pt, fill=white, inner sep=0] at ($(6,0)$) {\scriptsize Time from visual stimulus onset (s) };        
     \end{scope}
      \node[below right,inner sep=0] (image4) at (image3.north east) 
      {\includegraphics[width=0.33\textwidth, trim=70 0 0 0, clip]
         {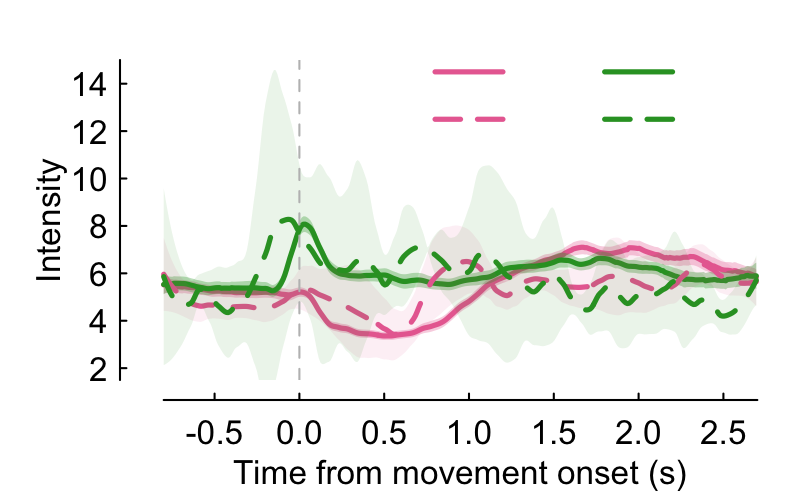}
      };   
      \begin{scope}[
        shift={($(image4.south west)$)},
        x={($0.1*(image4.south east)$)},
        y={($0.1*(image4.north west)$)}]
        \node[above right, xshift=-1pt] at ($(0,9.5)$) { Cluster 3};
        \node[above, xshift=-8pt, yshift=1pt, fill=white, inner sep=0] at ($(6,0)$) {\scriptsize Time from movement onset (s) };        
        \node[right, xshift=3pt, yshift=-5pt] at ($(5.6,9.0)$) {\scriptsize L,L};
        \node[right, xshift=3pt, yshift=-13pt] at ($(5.6,9.0)$) {\scriptsize R,L};
        \node[right, xshift=2pt, yshift=-5pt] at ($(8.0,9.0)$) {\scriptsize R,R};
        \node[right, xshift=2pt, yshift=-13pt] at ($(8.0,9.0)$) {\scriptsize L,R};
      \end{scope} 
   \end{tikzpicture}
   \caption{ 
   Firing patterns of the three clusters in four different experimental conditions.
   The lines represent the mean firing intensities.
   The shaded regions represent  the mean firing intensities plus or minus two standard errors of the mean.
   The legends represent ``scenario, choice'', for instance, ``L,R'' represents the trials where the left grating was of a higher contrast, and the mouse chose to move the right grating.
   In the ``L,R'' condition, there is high uncertainty due to the limited number of trials. 
   }
   \label{sfig:firing_intensity_LR}
\end{figure}

Figure~\ref{sfig:firing_intensity_LR} displays the average firing patterns of the three clusters in four different experimental conditions. 
From the figure we can see that, there is high uncertainty in condition ``L,R'' since there are only 3 trials. 
However, the firing patterns in the ``L,R'' condition generally support our hypothesis regarding the roles of the clusters.
For instance, in ``L,R'' trials, the firing rates of neurons in Cluster~1 show an upward trend before movement onset, which is consistent with our hypothesis that Cluster~1 neurons are responsible for executing the turning of the wheel.   
Moreover, the firing rates of neurons in Cluster 2 in ``L,R'' trials remain stationary after feedback delivery, aligning with our hypothesis that Cluster 2 neurons might respond to perceptions of stimuli such as rewards.

\subsection{Results obtained using kCFC}

For comparison, we apply the kCFC~\citep{Chiou2007} to the neural data. 
Specifically, we apply the kCFC to the aggregated spike trains across the training trials with $K=3$, and set the remaining parameter values consistent with those specified in Section~\ref*{sec:compare_relevant_method} of the main text.
Table~\ref{stable:membership_compare} shows the association between clusters obtained from the kCFC and clusters obtained from the proposed ASIMM.
From the table we see that the clusters from the kCFC are mixtures of the clusters from ASIMM.
Particularly, for Cluster 1 and 2 identified by kCFC, approximately half of the neurons in these clusters are not grouped together in ASIMM's results.

\begin{figure}[ht]
\centering
\begin{tikzpicture}
    \node[below right,inner sep=0] (image_10) at (0,0) 
        {\includegraphics[width=0.33\textwidth, trim=0 0 0 0, clip]
            {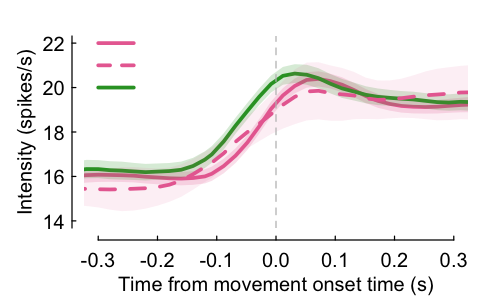}
        }; 
    \begin{scope}[
        shift={($(image_10.south west)$)},
        x={($0.1*(image_10.south east)$)},
        y={($0.1*(image_10.north west)$)}]
        \draw[fill=white, draw=white] (0,2) rectangle (0.8, 9);
        \node[below, rotate=90] at ($(0,5.5)$) {\scriptsize Intensity (spikes/s)};
        \node[right] at ($(2.8,8.5)$) {\tiny L,L};
        \node[right] at ($(2.8,7.8)$) {\tiny R,L};
        \node[right] at ($(2.8,7.0)$) {\tiny R,R};
        \draw[fill=white, draw=white] (2,0) rectangle (9, 1);
        \node[above] at ($(5.7,-0.3)$) {\scriptsize Time from movement onset (s)};
        \node[above right] at ($(0,9)$) {\footnotesize (1a)};
        \node[above, xshift=5pt] at ($(5,10)$) { Cluster 1};
    \end{scope}  
    \node[below right,inner sep=0] (image_11) at (image_10.north east) 
    {\includegraphics[width=0.33\textwidth, trim=0 0 0 0, clip]
        {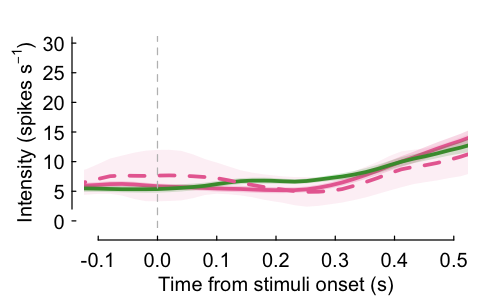}
    }; 
    \begin{scope}[
        shift={($(image_11.south west)$)},
        x={($0.1*(image_11.south east)$)},
        y={($0.1*(image_11.north west)$)}]
        \draw[fill=white, draw=white] (0,2) rectangle (0.8, 9);
        \draw[fill=white, draw=white] (2,0) rectangle (9, 1);
        \node[above] at ($(5.7,-0.3)$) {\scriptsize Time from visual stimulus onset (s)};
        \node[above right] at ($(0,9)$) {\footnotesize (2a)};
        \node[above, xshift=5pt] at ($(5,10)$) { Cluster 2};
    \end{scope}  
    \node[below right,inner sep=0] (image_12) at (image_11.north east) 
    {\includegraphics[width=0.33\textwidth, trim=0 0 0 0, clip]
        {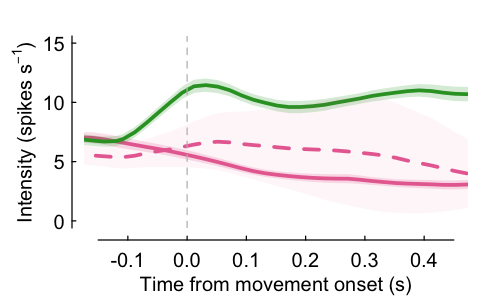}
    };  
    \begin{scope}[
        shift={($(image_12.south west)$)},
        x={($0.1*(image_12.south east)$)},
        y={($0.1*(image_12.north west)$)}]
        \draw[fill=white, draw=white] (0,2) rectangle (0.8, 9);
        \draw[fill=white, draw=white] (2,0) rectangle (9, 1);
        \node[above] at ($(5.7,-0.3)$) {\scriptsize Time from movement onset (s)};
        \node[above right] at ($(0,9)$) {\footnotesize (3a)};
        \node[above, xshift=5pt] at ($(5,10)$) { Cluster 3};
    \end{scope}   
    \node[below right = 0.3 and 0 of image_10.south west,inner sep=0] (image_20) at (image_10.south west) 
        {\includegraphics[width=0.33\textwidth, trim=0 0 0 0, clip]
            {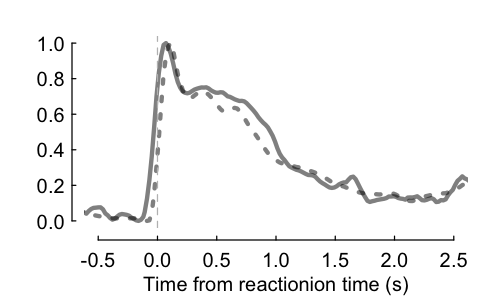}
        };  
    \begin{scope}[
        shift={($(image_20.south west)$)},
        x={($0.1*(image_20.south east)$)},
        y={($0.1*(image_20.north west)$)}]
        \draw[fill=white, draw=white] (0,2) rectangle (0.7, 9);
        \node[below, rotate=90] at ($(0,5.5)$) {\scriptsize Standardized value};
        \draw[fill=white, draw=white] (2,0) rectangle (9, 1);
        \node[above] at ($(5.7,-0.3)$) {\scriptsize Time from movement onset (s)};
        \node[above right] at ($(0,9)$) {\footnotesize (1b)};
        \draw[very thick, black, opacity=0.8] ($(6.0,8.3)$) -- ($(6.6,8.3)$) ;
        \node[right] at ($(6.5,8.3)$) {\tiny Intensity};
        \draw[very thick, black, dotted, opacity=0.8] ($(6.0,7.5)$) -- ($(6.6,7.5)$) ;
        \node[right] at ($(6.5,7.5)$) {\tiny Wheel velocity};

    \end{scope}   
    \node[below right,inner sep=0] (image_21) at (image_20.north east) 
    {\includegraphics[width=0.33\textwidth, trim=0 0 0 0, clip]
        {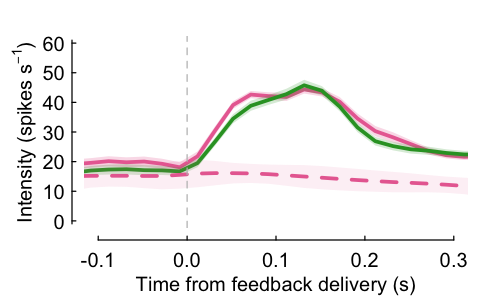}
    }; 
    \begin{scope}[
        shift={($(image_21.south west)$)},
        x={($0.1*(image_21.south east)$)},
        y={($0.1*(image_21.north west)$)}]
        \draw[fill=white, draw=white] (0,2) rectangle (0.8, 9);
        \node[below, rotate=90] at ($(0,5.5)$) {\scriptsize Intensity (spikes/s)};
        \draw[fill=white, draw=white] (2,0) rectangle (9, 1);
        \node[above] at ($(5.7,-0.3)$) {\scriptsize Time from feedback delivery (s)};
        \node[above right] at ($(0,9)$) {\footnotesize (2b)};
    \end{scope}
    \node[below right,inner sep=0] (image_22) at (image_21.north east) 
    {\includegraphics[width=0.33\textwidth, trim=0 0 0 0, clip]
        {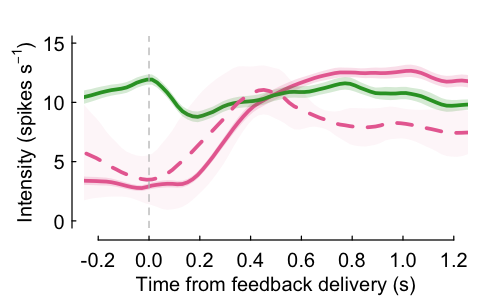}
    }; 
    \begin{scope}[
        shift={($(image_22.south west)$)},
        x={($0.1*(image_22.south east)$)},
        y={($0.1*(image_22.north west)$)}]
        \draw[fill=white, draw=white] (0,2) rectangle (0.8, 9);
        \draw[fill=white, draw=white] (2,0) rectangle (9, 1);
        \node[above] at ($(5.7,-0.3)$) {\scriptsize Time from feedback delivery (s)};
        \node[above right] at ($(0,9)$) {\footnotesize (3b)};
    \end{scope}
   
\end{tikzpicture}
\caption{ 
Firing patterns of clusters from kCFC.
The three columns correspond to the three estimated clusters.
The shaded area represents the mean firing intensities plus or minus two standard errors of the mean.
The legend in panel~(1a) represents ``scenario, choice'', for instance, ``R,L'' represents the trials where the right grating was of a higher contrast, and the mouse chose to move the left grating.
Panel~(1b) illustrates the average firing intensity and wheel velocity, where both firing intensity and wheel velocity are standardized to range from~0 to~1.
The firing intensities of Cluster 1 and 2 do not exhibit laterality.
}
\label{sfig:RDA_kcfc}
\end{figure}
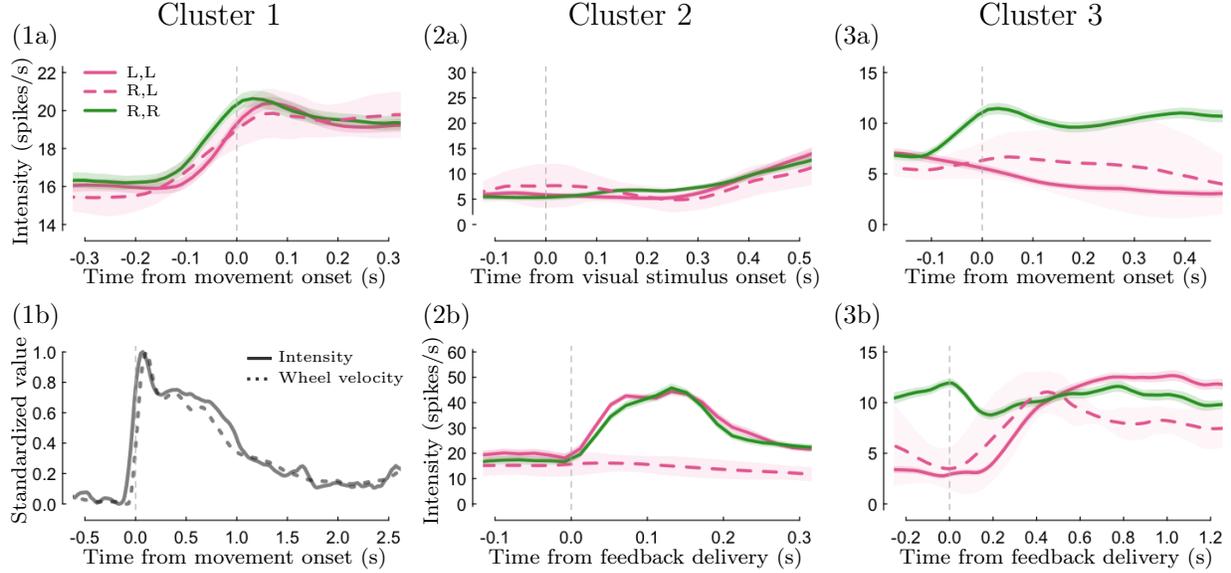

Figure~\ref{sfig:RDA_kcfc} shows the firing patterns of clusters from the kCFC.
We observe that firing patterns bear some resemblance to those obtained from ASIMM.
However, Cluster~1 and~2 from the kCFC do not exhibit laterality. 
This lack of laterality is likely because Cluster~1 and~2 are mixtures of clusters from the ASIMM.
For instance, the firing intensity of Cluster~1 from the ASIMM has a higher peak when the mouse chose the \emph{left} visual grating (see Figure~\ref*{fig:RDA}(1b)), whereas the firing intensity of Cluster~3 from the ASIMM has a higher peak when the mouse chose the \emph{right} grating (see Figure~\ref*{fig:RDA}(3b)).
Consequently, when Cluster~1 and~3 from our proposed method are mixed, their individual lateral biases are counteracted, resulting in the absence of clear laterality.

\begin{table}
\caption{ 
Contingency table of the clusters from the proposed ASIMM and the clusters from the kCFC.
}
\label{stable:membership_compare} 
\centering
\begin{tabular}{cccc}
\hline
 &
\multicolumn{3}{c}{ASIMM} \\
\cline{2-4}
kCFC & Cluster 1 & Cluster 2 & Cluster 3 \\
\hline
Cluster 1 & 60 & 2 & 71 \\
\hline
Cluster 2 & 3 & 14 & 10 \\
\hline
Cluster 3 & 0 & 7 & 58 \\
\hline
\end{tabular}
\end{table}

\section{Auxiliary Lemmas}
\label{ssec:proof_of_lemma}
\begin{lemma}
\label{slemma:independent_invertible}
Let $\eta_m \equiv \exp\{-\operatorname{j} 2 \pi \xi w^*_m\}$ and $ \boldsymbol{\eta} \equiv (\eta_m)_{m\in[M]}$, where $\xi \in \mathbb{R} \setminus \{0\}$.
If $\{w^*_m \in \mathbb{C}: m\in[M]\}$  are independent random variables, and among them, at least $M-1$ variables have non-zero variance, i.e., $|\{m\in[M]: \mathrm{var}(w^*_m) > 0\}| \geq M-1$,
then the matrix $\mathbb{E}[\bar{\boldsymbol{\eta}} \boldsymbol{\eta}^\top]$ is invertible.
\end{lemma}
\begin{proof}
Without loss of generality, we assume that $\mathrm{var}(w^*_m) > 0$ for $m=2,\ldots,M$.
Suppose $\mathbb{E}[\bar{\boldsymbol{\eta}} \boldsymbol{\eta}^\top]$ is not invertible. 
By the definition of invertible matrix, there exists $\mathbf{x} \in \mathbb{C}^M \setminus\{\mathbf{0}\}$ such that $\mathbb{E}[\bar{\boldsymbol{\eta}} \boldsymbol{\eta}^\top] \mathbf{x} = \mathbf{0}$.
Thus we can derive that
\begin{align}
\label{seq:E_xT_eta_etaT}
\mathbb{E}[\bar{\mathbf{x}}^\top \bar{\boldsymbol{\eta}} \boldsymbol{\eta}^\top \mathbf{x}] = 0.
\end{align}
Moreover, we have
\begin{align}
\label{seq:E_xT_eta_var_eta}
\mathbb{E}[\bar{\mathbf{x}}^\top \bar{\boldsymbol{\eta}} \boldsymbol{\eta}^\top \mathbf{x}] 
& = \mathbb{E}\left[ \bigg|\sum_{m\in[M]} \eta_m x_m \bigg|^2 \right] 
\geq \mathrm{var}\bigg( \sum_{m\in[M]} \eta_m x_m \bigg)
= \sum_{m\in[M]} |x_m|^2 ~\mathrm{var}(\eta_m) \geq 0,
\end{align}
where the second equality follows from the assumption that $\{w^*_m: m\in[M]\}$ are independent.
In order for~\eqref{seq:E_xT_eta_etaT} and~\eqref{seq:E_xT_eta_var_eta} to hold simultaneously, we must have $|x_m|^2 ~\mathrm{var}(\eta_m) = 0$ for $m\in[M]$.
From the assumption that $\mathrm{var}(w^*_m)>0$ for $m=2,\ldots,M$, we obtain that  $\mathrm{var}(\eta_m) > 0$ for $m=2,\ldots,M$. Thus, we know that $x_m=0$ for $m=2,\ldots,M$.
Since $\mathbf{x} \neq 0$ by definition, we deduce that $x_1 \neq 0$.
Hence, in order for $|x_1|^2 ~\mathrm{var}(\eta_1) = 0$ to hold we must have
\begin{align}
\label{seq:var_eta_1_eq_0}
\textrm{var}(\eta_1) = 0.
\end{align}

Additionally, in order for~\eqref{seq:E_xT_eta_etaT} and~\eqref{seq:E_xT_eta_var_eta} to hold simultaneously, we must have \[\mathbb{E}\left[ \bigg|\sum_{m\in[M]} \eta_m x_m \bigg|^2 \right] 
= \mathrm{var}\bigg( \sum_{m\in[M]} \eta_m x_m \bigg),\]
or equivalently,
\begin{align}
\label{seq:E_sum_m_eta_m}
   \left| \mathbb{E}\bigg( \sum_{m\in[M]} \eta_m x_m  \bigg) \right|^2 = 0.
\end{align}
Plugging $x_m = 0$ for $m=2,\ldots,M$ into~\eqref{seq:E_sum_m_eta_m}, we obtain that
\begin{align}
\label{seq:E_sum_eta_x_1}
\mathbb{E}\left( \sum_{m\in[M]} \eta_m x_m  \right) = \mathbb{E}(\eta_1) x_1.
\end{align}
Combining~\eqref{seq:E_sum_m_eta_m},~\eqref{seq:E_sum_eta_x_1}, and that $x_1 \neq 0$, we deduce that
\begin{align}
\label{seq:E_eta_1_eq_0}
   \mathbb{E}[\eta_1] = 0.
\end{align}
Combining~\eqref{seq:var_eta_1_eq_0} and~\eqref{seq:E_eta_1_eq_0}, we can derive that $\eta_1 = 0$.
However, we know that $|\eta_1| = 1$ by definition, implying that $\eta_1 = 0$ is impossible.
Therefore, $\mathbb{E}[\bar{\boldsymbol{\eta}} \boldsymbol{\eta}^\top]$ must be invertible.
\end{proof}

\begin{lemma}
\label{lemma:independent_increment_between_components}
Let $\eta_m \equiv \exp\{-\operatorname{j} 2 \pi \xi w^*_m\}$ and $ \boldsymbol{\eta} \equiv (\eta_m)_{m\in[M]}$, where $\xi \in \mathbb{R} \setminus \{0\}$.
If $w^*_m = w^*_{m-1} + \delta_{m-1}$ for $m=2,\ldots,M$, where $\{\delta_m \in \mathbb{C}: m\in[M-1]\}$  are independent random variables with non-zero variance,
then the matrix $\mathbb{E}[\bar{\boldsymbol{\eta}} \boldsymbol{\eta}^\top]$ is invertible.
\end{lemma}
\begin{proof}
Suppose $\mathbb{E}[\bar{\boldsymbol{\eta}} \boldsymbol{\eta}^\top]$ is not invertible. 
By the definition of invertible matrix, there exists $\mathbf{x} \in \mathbb{C}^M \setminus\{\mathbf{0}\}$ such that $\mathbb{E}[\bar{\boldsymbol{\eta}} \boldsymbol{\eta}^\top] \mathbf{x} = \mathbf{0}$.
Thus we can derive that
\begin{align}
\label{seq:E_xT_eta_etaT_2}
\mathbb{E}[\bar{\mathbf{x}}^\top \bar{\boldsymbol{\eta}} \boldsymbol{\eta}^\top \mathbf{x}] = 0.
\end{align}
On the other hand, we have
\begin{align}
\label{seq:E_x1_sum_exp_j_delta}
\begin{split}
\mathbb{E}[\bar{\mathbf{x}}^\top \bar{\boldsymbol{\eta}} \boldsymbol{\eta}^\top \mathbf{x}]
&= \mathbb{E} \left| \sum_{m\in[M]} \eta_m x_m \right|^2
\\ 
&= \mathbb{E} \left| \sum_{m\in[M]} \exp\{-\operatorname{j} 2 \pi \xi w^*_m\} x_m \right|^2
\\ 
&= \mathbb{E} \left| \exp\{-\operatorname{j} 2 \pi \xi w^*_1\} \left[  x_1 + \sum_{m=2}^M \exp\{-\operatorname{j} 2 \pi \xi (\delta_1 + \cdots + \delta_{m-1} )\} x_m \right] \right|^2
\\ 
&= \mathbb{E} \left| x_1 + \sum_{m=2}^M \exp\{-\operatorname{j} 2 \pi \xi (\delta_1 + \cdots + \delta_{m-1} )\} x_m \right|^2
\end{split}
\end{align}
where the second equation follows from the definition of ${\eta}_m$, the third equation follows from the assumption that $w^*_{m} = w^*_{m-1}+\delta_{m-1}$, and the last equation follows from the fact that $|\exp\{-\operatorname{j} \theta\}| = 1$ for any $\theta \in \mathbb{R}$.
Define $A_m \equiv x_m + \sum_{m'=m+1}^M \exp\{-\operatorname{j} 2 \pi \xi (\delta_{m} + \cdots + \delta_{m'-1} )\} x_{m'}$ for $m\in[M-1]$, and $A_M \equiv x_M$. 
Combining~\eqref{seq:E_xT_eta_etaT_2} and~\eqref{seq:E_x1_sum_exp_j_delta} we have $\mathbb{E}|A_1|^2 = 0$, which further implies that
\begin{align}
\label{seq:E_A1}
A_1 = 0.
\end{align}
From the definition of $A_m$'s we know that $A_m = x_m + \exp\{-\operatorname{j} 2 \pi \xi \delta_{m}\} A_{m+1}$. Thus~\eqref{seq:E_A1} implies that $x_1 + \exp\{-\operatorname{j} 2 \pi \xi \delta_{1}\} A_{2} = 0$, or equivalently,
\begin{align}
\label{seq:A_mplus1}
A_{2} = -x_1  \exp\{~\operatorname{j} 2 \pi \xi \delta_{1}\}.
\end{align}
From the definition of $A_{2}$, we know that $A_{2}$ is a function of $\{\delta_{2}, \ldots, \delta_{M}\}$, which are independent to $\delta_{1}$ by our assumption.
Furthermore, $\delta_{1}$ has a positive variance based on our assumption.
Therefore, in order for~\eqref{seq:A_mplus1} to hold, it must holds that $x_1 = 0$, or equivalently,
\begin{align}
A_{2} = 0.
\end{align}
Similarly, we can derive that $A_3 = ... = A_M = 0$. 
Therefore, using the definition of $A_m$'s, we know that, 
\begin{align}
\label{seq:xm_Am_0}
x_m & = A_m - \exp\{-\operatorname{j} 2 \pi \xi \delta_{m}\} A_{m+1} = 0,
& \quad  & m = 1, \cdots, M-1,
\\ 
\label{seq:xM_0}
x_M & = A_M = 0,
& \quad & m = M.
\end{align}
Equation~\eqref{seq:xm_Am_0} and~\eqref{seq:xM_0} contradict the definition of $\mathbf{x}$ that asserts $\mathbf{x} \neq \mathbf{0}$. Therefore, $\mathbb{E}[\bar{\boldsymbol{\eta}} \boldsymbol{\eta}^\top]$ must be invertible.

\end{proof}

\begin{lemma}
\label{lemma:mA_mB}
Let $f \subseteq L^1\left(\mathbb{R}\right)$ and $\hat{f}$ be its Fourier transform, and let $A \equiv \left\{x \in \mathbb{R} : f(x) \neq 0\right\}$ and
$B \equiv \{\xi \in \mathbb{R} : \hat{f}(\xi) \neq 0\}$. Then
\begin{align}
m(A)<\infty \text { and } {m}(B)<\infty \Rightarrow f=0 \quad \text { a.e. },
\end{align}
where $m$ denotes the Lebesgue measure.
\end{lemma}
The proof of Lemma~\ref{lemma:mA_mB} can be found in~\cite{BENEDICKS1985}.

\section{Reasonable range of $\gamma$}
\label{sec:range_of_gamma}
To find a reasonable range for $\gamma$, we explore the magnitude of $\mathbb{E} [L_1(\mathbf{z}^*, {\mathbf{a}^*}', {\mathbf{f}^*}', \mathbf{v}^* )]$  and  $\mathbb{E} [L_2 (\mathbf{z}^*, \boldsymbol{\Lambda}^* )]$, where $\mathbf{z}^*, {\boldsymbol{a}^*}', {\mathbf{f}^*}', \mathbf{v}^*, \boldsymbol{\Lambda}^*$ denote the true parameters.
The magnitude of $\mathbb{E} [L_1(\mathbf{z}^*, {\mathbf{a}^*}', {\mathbf{f}^*}', \mathbf{v}^* )]$  and  $\mathbb{E} [L_2 (\mathbf{z}^*, \boldsymbol{\Lambda}^* )]$ can be approximated as follows:
\begin{gather}
\label{seq:E_shape_error}
\mathbb{E}L_1 (\mathbf{z}^*, {\mathbf{a}^*}', {\mathbf{f}^*}', \mathbf{v}^* )
\lessapprox (nR) {(T \Delta t)}^{-1},
\\ 
\label{seq:E_scale_error}
\mathbb{E}L_2 (\mathbf{z}^*, \boldsymbol{\Lambda}^* )
\approx \sum_{i\in[n],r\in[R]} N_{i,r}(T) .
\end{gather}
The derivation of~\eqref{seq:E_shape_error} and~\eqref{seq:E_scale_error} is provided later in this section.
Combining~\eqref{seq:E_shape_error} and~\eqref{seq:E_scale_error}, we obtain that
\begin{align}
\begin{split}
\frac{\mathbb{E}L_1 (\mathbf{z}^*, {\mathbf{a}^*}', {\mathbf{f}^*}', \mathbf{v}^* )}
{\mathbb{E}L_2 (\mathbf{z}^*, \boldsymbol{\Lambda}^* )}
& \lessapprox  (nR) {(T \Delta t)}^{-1} \bigg\{\sum_{i\in[n],r\in[R]} N_{i,r}(T) \bigg\}^{-1}
\equiv \gamma_0.
\end{split}
\end{align}
Consequently, we suggested to explore $\gamma$ in the range $[10^{-5} \times \gamma_0, 10 \times \gamma_0]$.

\subsubsection*{Derivation of~\eqref{seq:E_shape_error}}
From the definition of $L_1$ in~(\ref*{eq:L1_zafv}) of the main text, we know that
\begin{align}
\mathbb{E}L_1 (\mathbf{z}^*, {\mathbf{a}^*}', {\mathbf{f}^*}', \mathbf{v}^* )
&= \sum_{i\in [n], r\in [R]} \mathbb{E} \left( 
\frac{N_{i,r}(T)}{T} \left\|
\frac{y_{i,r}(t)}{N_{i,r}(T)}  
- \frac{\lambda^*_{i,r}(t)}{\Lambda^*_{i,r}(T)} 
\right\|_t^2 \right)
\end{align}
where $\lambda^*_{i,r}(t) \equiv a^*_{z_i} + \sum_{m\in[M]} S^{v^*_{i,m}+w^*_{r,m}} f^*_{z^*_{i},m}(t)$, and $\Lambda^*_{i,r}(T) \equiv \int_0^T \lambda^*_{i,r}(t) \mathrm{d}t$.
Therefore, it suffices to show that
\begin{gather}
\label{seq:E_shape_error_1}
\mathbb{E} \left(
\frac{N_{i,r}(T)}{T} \left\|
\frac{y_{i,r}(t)}{N_{i,r}(T)}  
- \frac{\lambda^*_{i,r}(t)}{\Lambda^*_{i,r}(T)} 
\right\|_t^2 
 \right)
\lessapprox (T\Delta t)^{-1}.
\end{gather}
To this end, we consider the following conditional expectation:
\begin{align}
\label{seq:E_Nir_T}
\begin{split}
&\mathbb{E} \left(
\frac{N_{i,r}(T)}{T} \left\|
\frac{y_{i,r}(t)}{N_{i,r}(T)}  
- \frac{\lambda^*_{i,r}(t)}{\Lambda^*_{i,r}(T)} 
\right\|_t^2 
\Bigg\lvert N_{i,r}(T)
 \right)
\\
&= \mathbb{E} \left(
\frac{N_{i,r}(T)}{T} \left\|
\frac{N_{i,r}(t+\Delta t) - N_{i,r}(t) }{N_{i,r}(T) \Delta t }  
- \frac{\lambda^*_{i,r}(t)}{\Lambda^*_{i,r}(T)} 
\right\|_t^2 
\Bigg\lvert N_{i,r}(T)
 \right)
\\
&= \mathbb{E} \left(
\frac{N_{i,r}(T)}{T} \left\|
\frac{ \sum_{j=1}^{N_{i,r}(T)} \textbf{1}(t < t_{i,r,j} \leq t+\Delta t) }{N_{i,r}(T) \Delta t }  
- \frac{\lambda^*_{i,r}(t)}{\Lambda^*_{i,r}(T)} 
\right\|_t^2 
\Bigg\lvert N_{i,r}(T)
 \right)
\\
&= \frac{N_{i,r}(T)}{T} 
\int_0^T \mathbb{E} \left(
 \left|
\frac{ \sum_{j=1}^{N_{i,r}(T)} \textbf{1}(t < t_{i,r,j} \leq t+\Delta t) }{N_{i,r}(T) \Delta t }  
- \frac{\lambda^*_{i,r}(t)}{\Lambda^*_{i,r}(T)} 
\right|^2 
\Bigg\lvert N_{i,r}(T)
 \right) \mathrm{d}t,
\\
&\equiv \frac{N_{i,r}(T)}{T} 
\int_0^T \mathbb{E} \left(
 \big|
X(t)
- Y(t)
\big|^2 
\big\lvert N_{i,r}(T)
 \right) \mathrm{d}t,
\end{split}
\end{align}
where the first equality follows from the definition of $y_{i,r}(t)$, the second equality follows from the definition of $N_{i,r}(t)$, and the third equality follows from the definition of $L^2$-norm.
By definitions of $X(t)$ and $Y(t)$, we know that
\begin{align}
\label{seq:E_frac_given_N}
\begin{split}
\mathbb{E} \left[ X(t) \lvert N_{i,r}(T) \right]
&= \mathbb{E} \left(
\frac{ \sum_{j=1}^{N_{i,r}(T)} \textbf{1}(t < t_{i,r,j} \leq t+\Delta t) }{N_{i,r}(T) \Delta t }
\Bigg\lvert N_{i,r}(T)
\right)
\\
&= \mathbb{E} \left(
\frac{ \textbf{1}(t < t_{i,r,1} \leq t+\Delta t) }{ \Delta t }
\Bigg\lvert N_{i,r}(T)
\right)
\\
&= \frac{ \mathbb{P} \big( t < t_{i,r,1} \leq t+\Delta t \mid N_{i,r}(T) \big)  }{ \Delta t }
\\
&\approx \frac{\lambda^*_{i,r}(t)}{\Lambda^*_{i,r}(T)} 
= Y(t)
\end{split}
\end{align}
where the first equality follows from the definition of $X(t)$. The second equality holds because, for Poisson processes, the event times $\{t_{i,r,j}: j=1,\cdots,N_{i,r}(T)\}$ can be treated as i.i.d. samples given the total event count $N_{i,r}(T)$.
The approximation in the fourth line follows from the definition of $\lambda^*_{i,r}(t)$ and $\Lambda^*_{i,r}(T)$.
Therefore we have
\begin{align}
\label{seq:E_frac_diff_given_N}
\begin{split}
& ~~~~\mathbb{E} \left(
 \big|X(t) - Y(t)\big|^2 \big\lvert N_{i,r}(T)
 \right)
\\
&= \mathrm{var} \left(
X(t) 
\big\lvert N_{i,r}(T)
 \right) 
\\ 
&= \frac{1}{|\Delta t|^2}  \left(
\frac{  \mathrm{var} \{ \textbf{1}(t < t_{i,r,j} \leq t+\Delta t) \mid N_{i,r}(T) \} }{N_{i,r}(T) } 
 \right) 
\\ 
&= \frac{1}{|\Delta t|^2}  \left(
\frac{  \mathbb{P} \{ t < t_{i,r,j} \leq t+\Delta t \mid N_{i,r}(T) \} [ 1-\mathbb{P} \{ t < t_{i,r,j} \leq t+\Delta t \mid N_{i,r}(T) \} ] }{N_{i,r}(T) } 
 \right) 
\\ 
&\approx \frac{1}{|\Delta t|^2} \frac{1}{N_{i,r}(T)} 
\left( \frac{\lambda^*_{i,r}(t)}{\Lambda^*_{i,r}(T)} \Delta t  \right) 
\left( 1 - \frac{\lambda^*_{i,r}(t)}{\Lambda^*_{i,r}(T)} \Delta t \right) ,
\end{split}
\end{align}
where the second equality follows from the definition of $X(t)$, the approximation in the fourth line follows from the definition of $\lambda^*_{i,r}(t)$ and $\Lambda^*_{i,r}(T)$.

Plugging~\eqref{seq:E_frac_diff_given_N} into~\eqref{seq:E_Nir_T}, we can derive that
\begin{align}
\label{seq:E_approx_approxeq}
\begin{split}
&\mathbb{E} \left(
\frac{N_{i,r}(T)}{T} \left\|
\frac{y_{i,r}(t)}{N_{i,r}(T)}  
- \frac{\lambda^*_{i,r}(t)}{\Lambda^*_{i,r}(T)} 
\right\|^2 
\Bigg\lvert N_{i,r}(T)
 \right)
\\
& \approx \frac{N_{i,r}(T)}{T} 
\int_0^T \frac{1}{|\Delta t|^2} \frac{1}{N_{i,r}(T)} 
\left( \frac{\lambda^*_{i,r}(t)}{\Lambda^*_{i,r}(T)} \Delta t  \right) 
\left( 1 - \frac{\lambda^*_{i,r}(t)}{\Lambda^*_{i,r}(T)} \Delta t \right)  \mathrm{d}t
\\ 
&= \frac{1}{T} \left( \frac{1}{ \Delta t} - \left\| \frac{\lambda^*_{i,r}(t)}{\Lambda^*_{i,r}(T)}  \right\|^2 \right)
\\ 
&\lessapprox (T\Delta t)^{-1}.
\end{split}
\end{align}
Finally, by taking the expectation of~\eqref{seq:E_approx_approxeq}, we obtain
\begin{align}
\begin{split}
\small 
& ~~~~ \mathbb{E} \left(
\frac{N_{i,r}(T)}{T} \left\|
\frac{y_{i,r}(t)}{N_{i,r}(T)}  
- \frac{\lambda^*_{i,r}(t)}{\Lambda^*_{i,r}(T)} 
\right\|_t^2 
 \right)
\\
& = \mathbb{E} \left[ \mathbb{E} \left(
\frac{N_{i,r}(T)}{T} \left\|
\frac{y_{i,r}(t)}{N_{i,r}(T)}  
- \frac{\lambda^*_{i,r}(t)}{\Lambda^*_{i,r}(T)} 
\right\|_t^2 
\Bigg\lvert N_{i,r}(T)
 \right) \right]
\\
& \lessapprox (T\Delta t)^{-1}.
\end{split}
\end{align}

\subsubsection*{Derivation of~\eqref{seq:E_scale_error}}
Employing the definition of $L_2$ in~(\ref*{eq:L2_zLambda}) of the main text, along with the variance expression for the Poisson distribution, we can derive that
\begin{align}
\mathbb{E}[L_2 (\mathbf{z}^*, \boldsymbol{\Lambda}^* )]
&= \mathbb{E} \left[ \sum_{i\in [n], r\in [R]} \big| N_{i,r}(T) - \Lambda^*_{i,r}(T) \big|^2 \right]
= \sum_{i\in[n],r\in[R]} \Lambda^*_{i,r}(T) 
\approx \sum_{i\in[n],r\in[R]} N_{i,r}(T) .
\end{align}

\end{document}